\documentclass[english,aps,A4,english,superscriptaddress, nofootinbib]{revtex4-1}
\usepackage[T1]{fontenc}
\usepackage[latin9]{inputenc}
\usepackage{color}
\usepackage{babel}
\usepackage{verbatim}
\usepackage{mathrsfs}
\usepackage{amstext}
\usepackage{amssymb}
\usepackage{graphicx}
\usepackage{esint}
\usepackage[pdfusetitle,
 bookmarks=true,bookmarksnumbered=false,bookmarksopen=false,
 breaklinks=false,pdfborder={0 0 0},pdfborderstyle={},backref=false,colorlinks=true]
 {hyperref}
\hypersetup{
 linkcolor=blue, citecolor=cyan}

\makeatletter
\usepackage{slashed}

\usepackage{xcolor}

\makeatother
\usepackage{soul}

\begin{document}
\title{Collisional corrections to spin polarization from quantum kinetic
theory using Chapman-Enskog expansion}
\author{Shuo Fang}
\email{fangshuo@mail.ustc.edu.cn}

\affiliation{Department of Modern Physics, University of Science and Technology
of China, Anhui 230026, China}
\author{Shi Pu}
\email{shipu@ustc.edu.cn}

\affiliation{Department of Modern Physics, University of Science and Technology
of China, Anhui 230026, China}
\affiliation{Southern Center for Nuclear-Science Theory (SCNT), Institute of Modern
Physics, Chinese Academy of Sciences, Huizhou 516000, Guangdong Province,
China}

\begin{abstract}
We have investigated the collisional corrections to the spin polarization
pseudo-vector, $\delta\mathcal{P}^{\mu}$, using quantum kinetic theory
in Chapman-Enskog expansion. We derive the spin Boltzmann equation
incorporating M{\o}ller scattering process. We further consider two distinct
scenarios using hard thermal loop approximations for simplification.
In scenario (I), the vector charge distribution function is treated
as off-equilibrium under the validity domain of gradient expansion.
Remarkably, the polarization induced by thermal vorticity and shear
viscous tensors are modified, but $\delta\mathcal{P}_{\textrm{ }}^{\mu}$
in this scenario does not depend on the coupling constant. In scenario
(II), the vector charge distribution function is assumed to be in
local thermal equilibrium. Then collisional corrections $\delta\mathcal{P}_{\textrm{ }}^{\mu}$
in this scenario are at $\mathcal{O}(\hbar^{2}\partial^{2})$. Additionally,
we evaluate the $\delta\mathcal{P}^{\mu}$ using relaxation time approach
for comparative analysis. Our results establish the theoretical framework
necessary for the future numerical investigations on the interaction
corrections to spin polarization.

\end{abstract}
\maketitle

\section{Introduction}

The spin-orbit coupling in many-body systems was first discovered
over a century ago, known as the Barnett effect \citep{PhysRev.6.239}
and the Einstein-de Haas effect \citep{1915KNAB...18..696E}. In relativistic
non-central heavy-ion collisions, a large orbital angular momentum
perpendicular to the reaction plane is produced, polarizing the hadrons
created in the collisions through spin-orbit coupling \citep{Liang:2004ph,Liang:2004xn,Gao:2007bc}.
Recently, the RHIC-STAR and LHC-ALICE collaborations have observed
the global and local polarization of $\Lambda$ and $\overline{\Lambda}$
hyperons \citep{STAR:2017ckg,STAR:2019erd,ALICE:2019aid,STAR:2020xbm}
and the spin alignment of vector mesons \citep{STAR:2022fan,ALICE:2019aid,ALICE:2023jad}. 

The RHIC-STAR collaboration has measured the spin alignment of $\phi$
mesons \citep{STAR:2022fan}. More specifically, a significant positive
deviation of the $00$-component of the spin density matrix from $1/3$
for $\phi$ mesons is observed, i.e., $\rho_{00}(\phi)>1/3$, while
$\rho_{00}$ of $K^{*0}$ is consistent with $1/3$. The spin alignment
of $\phi$ mesons can be well understood by the fluctuation of effective
strong force fields \citep{Sheng:2019kmk,Sheng:2020ghv,Sheng:2022ffb,Sheng:2022wsy,Sheng:2023urn}.
On the other hand, $\rho_{00}$ for $J/\psi$ is found to be smaller
than $1/3$ as reported by the LHC-ALICE collaboration \citep{ALICE:2023jad}.
Despite many recent developments on this topic \citep{ALICE:2023jad},
explaining the spin alignment of $J/\psi$ mesons remains challenging
\citep{Muller:2021hpe,Kumar:2023ghs}.

The global polarization of $\Lambda$ and $\overline{\Lambda}$ hyperons
is well described by various phenomenological models \citep{Karpenko:2016jyx,Becattini:2017gcx,Xie:2017upb,Pang:2016igs,Li:2017slc,Wei:2018zfb,Ryu:2021lnx,Shi:2017wpk,Fu:2020oxj,Sun:2017xhx,Wu:2022mkr,Alzhrani:2022dpi}
based on the modified Cooper-Frye formula \citep{Becattini:2013fla,Fang:2016vpj}
for spin polarization, assuming the global equilibrium condition is
fulfilled. This suggests that spin polarization in relativistic heavy
ion collisions is driven by the collective motion of the quark-gluon
plasma. See also recent reviews \citep{Becattini:2020ngo,Gao:2020vbh,Hidaka:2022dmn,Becattini:2022zvf,Becattini:2024uha}
and the references therein. On the other hand, theoretical predictions
for local spin polarization along the beam direction, which exhibit
dependence on the azimuthal angle, transverse momentum and centrality,
disagree with experimental results \citep{STAR:2019erd}, and even
display opposite trends \citep{Becattini:2017gcx,Xia:2018tes}. This
discrepancy is commonly referred to as the '\emph{sign problem}' in
local spin polarization. Recent studies have identified polarization
effects induced in local equilibrium, such as the shear viscous tensor
and the gradient of the chemical potential over temperature \citep{Liu:2021uhn,Fu:2021pok,Becattini:2021suc,Becattini:2021iol,Hidaka:2017auj,Yi:2021ryh,Fu:2022myl,Wu:2022mkr},
as crucial for understanding the data. Concurrently, polarization
along the beam direction in high transverse momentum or low multiplicity
scenarios in relativistic nucleus-nucleus collisions \citep{STAR:2023eck},
as well as in p-Pb collisions \citep{Yi:2024kwu}, remains puzzling.

Consequently, the explanation of local polarization remains an open
question, and further investigation beyond the local thermal equilibrium
is necessary. One of the macroscopic approaches involves incorporating
spin degrees of freedom into conventional relativistic viscous hydrodynamics,
termed relativistic spin hydrodynamics \citep{Florkowski:2017ruc,Florkowski:2018fap,Florkowski:2018myy,Hattori:2019lfp,Hongo:2021ona,Hongo:2022izs,Cao:2022aku,Li:2020eon,Fukushima:2020qta,Fukushima:2020ucl,Wang:2021ngp,Wang:2021wqq,Xie:2023gbo,She:2021lhe,Bhadury:2020puc,Bhadury:2020cop,Bhadury:2022ulr,Biswas:2023qsw,Garbiso:2020puw,Montenegro:2017lvf,Montenegro:2017rbu,Peng:2021ago,Weickgenannt:2022zxs,Weickgenannt:2022qvh,Weickgenannt:2022zhj,Weickgenannt:2023btk,Gallegos:2020otk,Gallegos:2021bzp,Shi:2020htn,Hu:2021lnx,Hu:2021pwh,Hu:2022azy,Becattini:2023ouz,Wagner:2024fhf}.
In spin hydrodynamics, total angular momentum conservation is considered,
modifying the energy-momentum tensor and the conserved currents from
internal symmetries due to spin interactions. Spin hydrodynamics is
developed through various methodologies including entropy current
analysis \citep{Hattori:2019lfp,Li:2020eon,Fukushima:2020ucl,Gallegos:2021bzp,She:2021lhe,Hongo:2021ona,Cao:2022aku,Biswas:2023qsw},
quantum kinetic theory \citep{Florkowski:2017ruc,Florkowski:2018myy,Bhadury:2020puc,Bhadury:2020cop,Shi:2020htn,Peng:2021ago,Weickgenannt:2022zxs,Weickgenannt:2022zhj,Weickgenannt:2022qvh,Weickgenannt:2023btk,Wagner:2024fhf},
effective field theory \citep{Montenegro:2017lvf,Montenegro:2017rbu},
and holographic method \citep{Gallegos:2020otk,Garbiso:2020puw}.
Recently, the causality and stability of canonical spin hydrodynamics
have been investigated through linear modes analysis \citep{Xie:2023gbo}
and revisited using thermodynamic stability methods \citep{Ren:2024pur}.
Meanwhile, analytic solutions for spin hydrodynamics in Bjorken \citep{Wang:2021ngp}
and Gubser flows \citep{Wang:2021wqq} have been derived, and attractors
within spin hydrodynamics \citep{Wang:2024afv} have also been reported.

As a macroscopic effective theory, hydrodynamics only contains the
evolution of several long wavelength modes, while quantum kinetic
theory (QKT) with quantum corrections, as the microscopic approaches,
can capture more off-equilibrium features. The QKT has been derived
for fermions \citep{Gao:2019znl,Hattori:2019ahi,Weickgenannt:2019dks,Weickgenannt:2020aaf,Yang:2020hri,Liu:2020flb,Weickgenannt:2021cuo,Sheng:2021kfc,Wang:2021qnt,Weickgenannt:2022jes,Wang:2019moi,Huang:2020wrr,Lin:2021mvw,Fang:2022ttm,Wang:2022yli,Lin:2022tma,Sheng:2022ssd,Wagner:2022amr,Li:2019qkf,Das:2022azr,Yang:2021fea,Fang:2023bbw}
and for spin-1 particles \citep{Huang:2020kik,Hattori:2020gqh,Mameda:2022ojk},
broadening the scope of chiral kinetic theory (CKT) \citep{Son:2012wh,Son:2012zy,Stephanov:2012ki,Gao:2012ix,Chen:2012ca,Chen:2014cla,Chen:2015gta,Hidaka:2016yjf,Hidaka:2017auj,Huang:2017tsq}
for massless fermions and bridging the gap between fundamental quantum
field theory and macroscopic theories. For more details, one can refer
to the recent review in Ref. \citep{Hidaka:2022dmn} and the references
cited therein. To better understand the experimental data, we further
explore the off-equilibrium contributions from QKT beyond the scenarios
of local equilibrium. This analysis requires the computation of collision
kernels and self-energies with quantum corrections. Early pioneering
works \citep{Chen:2014cla,Hidaka:2016yjf} discovered that collision
effects in CKT are associated with side-jump effects. It then has
since been extended to QKT with effective interactions \citep{Weickgenannt:2020aaf,Yang:2020hri,Weickgenannt:2021cuo,Sheng:2021kfc,Wang:2020pej,Wang:2021qnt,Weickgenannt:2022jes,Lin:2021mvw,Fang:2022ttm,Wang:2022yli,Lin:2022tma,Wagner:2022amr,Li:2019qkf}.
In our previous work \citep{Fang:2022ttm}, we computed the vector
and axial collision kernels of $\mathcal{O}(\hbar)$ up to the leading
logarithm in the 2-2 scattering process for real quantum electrodynamics
interactions under the hard thermal loop (HTL) approximation. We also
derived a spin Boltzmann equation (SBE) where the collision kernel
is explicitly expressed through vector and axial distribution functions.
For similar discussions, see Refs. \citep{Wang:2022yli,Lin:2022tma}.
More recently, the effects of self-energy corrections on chiral transport
and spin polarization have been investigated in CKT \citep{Yamamoto:2023okm}
and QKT \citep{Fang:2023bbw}. However, the quantitative contribution
of collisions to spin polarization remains unclear. Also see the very
recent work in Ref. \citep{Lin:2024zik} for relevant discussions.
Furthermore, the off-equilibrium contributions from collisions to
the spin polarization current have been extensively discussed in condensed
matter physics, particularly in the context of extrinsic anomalous
Hall effects and spin Hall effects in nearly perfect crystals \citep{Xiao:2009rm,RevModPhys.82.1539,sinova2015spin}.

Our main goal in this work is to derive the off-equilibrium corrections
to the axial Wigner function $\mathcal{A}^{<,\mu}$ and investigate possible new corrections to the spin polarization vector arising from off-equilibrium effects. 
We begin by deriving
both the vector Boltzmann equation for the vector Wigner function
$\mathcal{V}^{<,\mu}$ and the SBE for $\mathcal{A}^{<,\mu}$ from
QKT in 2-2 fermion elastic scattering processes. Subsequently, we
apply the Chapman-Enskog expansion \citep{DeGroot:1980dk,denicol2022microscopic}
to obtain the variation of the axial distribution function $\delta f_{{\rm A}}$
and axial Wigner function $\delta\mathcal{A}^{<,\mu}$ from a Boltzmann-type
axial transport equation with $f_{{\rm V}}$ in local equilibrium
and compute the off-equilibrium corrections to the spin polarization
pseudo-vector $\delta\mathcal{P}^{\mu}$. Additionally, we calculate
$\delta f_{{\rm A}}$ and $\delta\mathcal{P}^{\mu}$ using the relaxation
time approach (RTA) for comparison.

This paper is organized as follows. In Sec. \ref{sec:Collsional-quantum-kinetic},
we begin by briefly reviewing the collisional QKT formalism and deriving
the SBE in the M{\o}ller scattering processes. We then analyze the structure
of the axial collision kernel and discuss the matching conditions.
In Sec. \ref{sec:Chapman-Enskog-expansion}, we reexamine the matching
condition using the properties of the axial collision kernel and subsequently
solve for the off-equilibrium correction to the axial distribution
in the HTL approximation using the Chapman-Enskog expansion. In Sec.
\ref{sec:Fermion-Spin-polarization-from}, we also compute the collisional
corrections to the modified Cooper-Frye formula and the off-equilibrium
axial distribution function using the RTA. We conclude in Sec.\ref{sec:Conclusions-and-outlook}.

Throughout this paper, we set $c=k_{{\rm B}}=1$ but keep $\hbar$
explicitly. We adopt the most minus signature of Minkowski metric,
$\eta^{\mu\nu}=(+,-,-,-)$. We also define the projection operator
\begin{equation}
\Delta^{\mu\nu}=\eta^{\mu\nu}-u^{\mu}u^{\nu},
\end{equation}
which is orthogonal to the fluid velocity $u^{\mu}$, and a symmetrized
traceless operator orthogonal to $u^{\mu}$ 
\begin{equation}
\Delta^{\mu\nu\alpha\beta}=\frac{\Delta^{\mu\alpha}\Delta^{\nu\beta}+\Delta^{\mu\beta}\Delta^{\nu\alpha}}{2}-\frac{1}{3}\Delta^{\mu\nu}\Delta^{\alpha\beta}.
\end{equation}
 The rank-3 operator can be similarly introduced as in Refs. \citep{DeGroot:1980dk,denicol2022microscopic},
\begin{eqnarray}
\Delta_{\nu_{1}\nu_{2}\nu_{3}}^{\mu_{1}\mu_{2}\mu_{3}} & = & \frac{1}{6}\sum_{\mathscr{P}_{\mu}^{3}\mathscr{P}_{\nu}^{3}}\Delta_{\nu_{i}}^{\mu_{i}}\Delta_{\nu_{j}}^{\mu_{j}}\Delta_{\nu_{k}}^{\mu_{k}}-\frac{1}{15}\sum_{\mathscr{P}_{\mu}^{3}\mathscr{P}_{\nu}^{3}}\Delta^{\mu_{i}\mu_{j}}\Delta_{\nu_{i}\nu_{j}}\Delta_{\nu_{k}}^{\mu_{k}},\label{eq:Rank-3=000020symmetrized=000020tensor}
\end{eqnarray}
where the summation over the permutations $\mathscr{P}_{\mu}^{3},\mathscr{P}_{\nu}^{3}$
runs over all distinct permutations of $\mu$- and $\nu$-type indices.
For an arbitrary vector $A^{\mu}$, we define, 
\begin{equation}
A^{\langle\mu\rangle}=A_{\perp}^{\mu}\equiv\Delta^{\mu\nu}A_{\nu}.
\end{equation}
We also introduce the symmetric and anti-symmetric symbols for an
arbitrary tensor $A^{\mu\nu}$, 
\begin{equation}
A^{(\mu\nu)}=\frac{A^{\mu\nu}+A^{\nu\mu}}{2},\quad A^{[\mu\nu]}=\frac{A^{\mu\nu}-A^{\nu\mu}}{2}.
\end{equation}
and, the traceless part, 
\begin{equation}
A^{\langle\mu\nu\rangle}=\Delta_{\alpha\beta}^{\mu\nu}A^{\alpha\beta}.
\end{equation}
We also define the following hydrodynamic symbols, the expansion scalar
$\theta=\partial_{\mu}u^{\mu}$, the comoving derivative $D=u_{\mu}\partial^{\mu}$,
the spacelike derivative $\nabla_{\mu}=\Delta_{\mu}^{\nu}\partial_{\nu}$,
the shear tensor 
\begin{equation}
\sigma^{\mu\nu}=\Delta_{\alpha\beta}^{\mu\nu}\partial^{\alpha}u^{\beta},
\end{equation}
the fluid vorticity tensor 
\begin{equation}
\omega^{\mu\nu}=\Delta^{\mu\alpha}\Delta^{\nu\beta}\partial_{[\alpha}u_{\beta]}\equiv\epsilon^{\mu\nu\rho\sigma}u_{\rho}\omega_{\sigma},
\end{equation}
with the kinetic vorticity vector 
\begin{equation}
\omega_{\sigma}=\frac{1}{2}\epsilon_{\sigma\alpha\beta\gamma}u^{\alpha}\partial^{\beta}u^{\gamma}.
\end{equation}
We decompose $\partial_{\mu}u_{\nu}$ as follows, 
\begin{eqnarray}
\partial_{\mu}u_{\nu} & = & u_{\mu}Du_{\nu}+\sigma_{\mu\nu}+\frac{1}{3}\Delta_{\mu\nu}\theta+\omega_{\mu\nu}.\label{eq:Fluid=000020velocity=000020gradient=000020decomposition}
\end{eqnarray}
Another type of projection operator is defined as 
\begin{equation}
\Theta^{\mu\nu}(p)=\Delta^{\mu\nu}+\hat{p}_{\perp}^{\mu}\hat{p}_{\perp}^{\nu},
\end{equation}
with $\hat{p}_{\perp}^{\mu}=p_{\perp}^{\mu}/\sqrt{-p_{\perp}\cdot p_{\perp}}$
and we introduce $p^{\{\mu\}}=\Theta^{\mu\nu}p_{\nu}$.

\section{Quantum kinetic theory with collisions }

\label{sec:Collsional-quantum-kinetic} We begin by reviewing the
derivation of quantum kinetic theory for massless fermions in Sec.
\ref{subsec:Brief-review-of}. Next, we list the self-energies and
collision kernels, and discuss the structure of the axial vector collision
kernel and its relationship to the off-equilibrium corrections of
the axial distribution $f_{{\rm A}}$ in Sec. \ref{subsec:Self-energies-in-2-2}.
We also present the mapping from kinetic theory to the hydrodynamic
quantities in Sec. \ref{subsec:Spin-hydrodynamics-from}.

\subsection{Brief review of quantum kinetic theory for massless fermions \protect\label{subsec:Brief-review-of}}

In this work, we consider a dilute fermionic gas with fermion elastic scattering. We focus on the quantum
electrodynamics (QED) interactions rather than the quantum chromodynamics (QCD) interactions. The reason is as follows. 
First, the QED interaction shares some basic properties of the gauge theories \cite{Blaizot:2001nr}, e.g. the logarithm-type infrared divergence of the damping rate \cite{Blaizot:1996az}, same coupling dependence of the relaxation time for the transport coefficients (e.g. see Ref.~\cite{Arnold:2000dr} for the calculations on the shear viscosity and  charge conductivity) and others.
Second, full QCD calculations require incorporating quantum corrections to the gluon propagator within the QKT framework for vector fields. While the quantum Wigner function for the photon field has been studied in the Coulomb gauge \cite{Huang:2020kik, Hattori:2020gqh}, even its equilibrium form remains incompletely understood. This lack of understanding further complicates the investigation of quark spin polarization due to collisions in a QCD medium. Consequently, the spin transport involved with gauge interaction is still challenging to perform computation within the framework of QKT. Therefore, we focus on the QED type interactions for simplicity.

We start from the Lagrangian density for massless  fermions with quantum
electrodynamics interaction, 
\begin{eqnarray}
\mathcal{L} & = & \frac{1}{2}\overline{\psi}i\hbar\left(\gamma^{\mu}\overrightarrow{D}_{\mu}-\overleftarrow{D}_{\mu}\gamma^{\mu}\right)\psi-ea_{\mu}\overline{\psi}\gamma^{\mu}\psi-\frac{1}{4}F_{\mu\nu}F^{\mu\nu},\label{eq:QED=000020Lagrangian}
\end{eqnarray}
where the gauge covariant derivatives are defined as, 
\begin{eqnarray}
\overrightarrow{D}_{\mu} & = & \overrightarrow{\partial}_{\mu}+ie\hbar^{-1}A_{\mu},\quad\overleftarrow{D}_{\mu}=\overleftarrow{\partial}_{\mu}-ie\hbar^{-1}A_{\mu}.\label{eq:Gauge=000020covariant=000020derivative}
\end{eqnarray}
Here $A_{\mu},a_{\mu}$ are the classical background and quantum U(1)
gauge field, respectively, chosen in the background field gauge \citep{Peskin:1995ev,Blaizot:2001nr}.
The $a_{\mu}$ is invariant under the gauge transformation while the
classical $A_{\mu}$ transforms as the gauge field. $F_{\mu\nu}[A+a]=\partial_{\mu}(A_{\nu}+a_{\nu})-\partial_{\nu}(A_{\mu}+a_{\mu})$
is the field strength tensor.

Denoting $\{\overline{\eta},\eta,j_{\mu}\}$
the external sources coupled with $\{\psi,\overline{\psi},a^{\mu}\}$,
the generating functional along the Schwinger-Keldysh contour 'C'
can be defined as \citep{Blaizot:1999xk,Blaizot:2001nr}, 
\begin{eqnarray}
Z[j] & = & \int\mathcal{D}a_{\mu}\mathcal{D}\overline{\psi}\mathcal{D}\psi\exp\left(\frac{i}{\hbar}\int_{\mathrm{C}}\mathrm{d}^{4}z\left[\mathcal{L}-\overline{\eta}\psi-\overline{\psi}\eta-j^{\mu}a_{\mu}\right]\right).\label{eq:C=000020generating=000020functional}
\end{eqnarray}
We then introduce the connected two-point Green's function,
\begin{eqnarray}
S_{\mathrm{c}}^{(2)}(x,y) & = & \left.\frac{(i\hbar)^{2}\delta^{(2)}\ln Z[\eta,\overline{\eta},j^{\mu}]}{\delta\overline{\eta}(x)\delta\eta(y)}\right|_{j=0}\nonumber \\
 & = & \theta_{\mathrm{C}}(x_{0},y_{0})S_{{\rm c}}^{>}(x,y)-\theta_{\mathrm{C}}(y_{0},x_{0})S_{{\rm c}}^{<}(x,y)\label{eq:2-p=000020GF=000020along=000020C}
\end{eqnarray}
where $\widetilde{T}_{\mathrm{C}}$ is the time-ordering operator
and $\theta_{\mathrm{C}}(x_{0},y_{0})$ is the unit-step function
along $C$. The subscript ${\rm c}$ denotes the ``connected'' correlation
functions and operators are the ensemble averaged. We can further
define the gauge invariant two-point functions as,
\begin{eqnarray}
S_{ab}^{<}(x,y) & = & U(y,x)\langle\overline{\psi}_{b}(y)\psi_{a}(x)\rangle_{{\rm c}},\label{eq:Les=000020gauge-inv=000020GF}\\
S_{ab}^{>}(x,y) & = & U^{\dagger}(x,y)\langle\psi_{a}(x)\overline{\psi}_{b}(y)\rangle_{{\rm c}},\label{eq:Gtr=000020gauge-inv=000020GF}
\end{eqnarray}
where $U(y,x)=\mathcal{P}e^{-\frac{ie}{\hbar}\int_{y}^{x}\mathrm{d}z_{\mu}A^{\mu}(z)}$
is the gauge link with $\mathcal{P}$ the path ordering operator \citep{Elze:1986qd,Vasak:1987um}, and we emphasize such gauge link is essential to guarantee the gauge independence of the Wigner functions under background field gauge.
The Wigner function is defined as the Wigner transformation of two-point
functions,
\begin{eqnarray}
S_{ab}^{\lessgtr}(X,p) & = & \int\mathrm{d}^{4}Ye^{i\frac{p\cdot Y}{\hbar}}S_{ab}^{\lessgtr}(x,y),\label{eq:Gauge-inv=000020WF}
\end{eqnarray}
with 
\begin{equation}
Y=x-y,\qquad X=\frac{X+Y}{2}.
\end{equation}

With Eq.(\ref{eq:C=000020generating=000020functional}), one can derive
the equations of motion for the mean field and then make variation
with respect to the external currents. After taking the Wigner transformation,
we can finally derive the Kadanoff-Baym equations up to the first
order in gradients for the Wigner function $S^{<}(X,p)$ \citep{Blaizot:2001nr,Hidaka:2016yjf,Yang:2020hri,Hidaka:2022dmn,Fang:2023bbw},
\begin{eqnarray}
\Pi_{\mu}\gamma^{\mu}S^{<}+\frac{i\hbar}{2}\gamma^{\mu}\Delta_{\mu}S^{<} & = & \frac{i\hbar}{2}(\Sigma^{<}\star S^{>}-\Sigma^{>}\star S^{<}),\label{eq:KB-eq=0000201}\\
\Pi_{\mu}S^{<}\gamma^{\mu}-\frac{i\hbar}{2}\Delta_{\mu}S^{<}\gamma^{\mu} & = & -\frac{i\hbar}{2}(S^{>}\star\Sigma^{<}-S^{<}\star\Sigma^{>}),\label{eq:KB-eq=0000202}
\end{eqnarray}
where $\Sigma^{\lessgtr}$ are the Wigner-transformed gauge-independent self-energies, and the  covariant derivative and kinetic momentum are
given by,
\begin{eqnarray*}
\Delta_{\mu} & = & \partial_{X,\mu}-eF_{\mu\nu}(X)\partial_{p}^{\nu}+\mathcal{O}(\hbar^{2}),\\
\Pi_{\mu} & = & p_{\mu}+\mathcal{O}(\hbar^{2}).
\end{eqnarray*}
We also introduce the gauge invariant Moyal product,
\begin{eqnarray*}
f(p,X)\star g(p,X) & = & f(p,X)g(p,X)+\frac{i\hbar}{2}f(p,X)(\overrightarrow{\partial}_{X}\cdot\overleftarrow{\partial}_{p}-\overleftarrow{\partial}_{X}\cdot\overrightarrow{\partial}_{p}-eF_{\mu\nu}\overleftarrow{\partial}_{p}^{\mu}\overrightarrow{\partial}_{p}^{\nu})g(p,X)+\mathcal{O}(\hbar^{2}).
\end{eqnarray*}

Under the Clifford basis, we can decompose the Wigner functions as,
\begin{eqnarray}
S & = & F+i\mathcal{P}\gamma^{5}+\mathcal{V}_{\mu}\gamma^{\mu}+\mathcal{A}_{\mu}\gamma^{5}\gamma^{\mu}+\frac{1}{2}\mathcal{S}_{\mu\nu}\gamma^{\mu\nu},\label{eq:Clifford=000020decomp}
\end{eqnarray}
where $\gamma^{\mu\nu}=\frac{i}{2}[\gamma^{\mu},\gamma^{\nu}]$. In
the massless case, all the block-diagonal components vanish, the decomposition
in Eq. (\ref{eq:Clifford=000020decomp}) reduces to, 
\begin{eqnarray}
S & = & \mathcal{V}_{\mu}\gamma^{\mu}+\mathcal{A}_{\mu}\gamma^{5}\gamma^{\mu},\label{eq:Massless=000020WF=000020Clifford=000020decomp}\\
\Sigma & = & \Sigma_{\mathrm{V}}^{\mu}\gamma_{\mu}+\Sigma_{\mathrm{A}}^{\mu}\gamma^{5}\gamma_{\mu}.\label{eq:Massless=000020SE=000020Clifford=000020decomp}
\end{eqnarray}

Before proceeding further, we briefly introduce the power counting
scheme following our previous work \citep{Fang:2023bbw}. We emphasize
that the $\hbar$ expansion generally differs from the gradient expansion
denoted as spacetime derivative $\partial_{\mu}$. In quantum theory,
$\hbar$ naturally arises, for example, in loop corrections to Green's
functions \citep{Itzykson:1980rh}, while the spacetime derivative
$\partial_{\mu}$ characterizes the relative importance of off-equilibrium
corrections in a nearly equilibrium many-body system \citep{Kovtun:2012rj}.
In quantum kinetic theory, the gradient expansion of correlators in
Eq. (\ref{eq:Gauge-inv=000020WF}), naturally induces $\hbar\partial$
factors.\textcolor{blue}{{} }Consequently, we might omit $\hbar$ from
the interactions and instead use $\hbar$ to characterize the order
of gradients $\partial_{\mu}$. Given that spin polarization effects
are quantum, we expect that all axial vector components $\mathcal{A}_{\mu}$
and $\Sigma_{{\rm A}}^{\mu}$ will inherently carry an additional
$\hbar$ relative to the vector components, $\mathcal{V}_{\mu}$ and
$\Sigma_{{\rm V}}^{\mu}$, i.e., 
\begin{eqnarray}
\mathcal{A}_{\mu},\Sigma_{{\rm A}}^{\mu}\sim\mathcal{O}(\hbar^{1}) & , & \mathcal{V}_{\mu},\Sigma_{{\rm V}}^{\mu}\sim\mathcal{O}(\hbar^{0}).\label{eq:WF=000020power=000020counting}
\end{eqnarray}

When transitioning from microscopic to macroscopic theory, we encounter
a macroscopic scale, $L$, related to the characteristic size of the
system, which we can roughly regard as the inverse of gradients, $L\sim\partial^{-1}$.
Additionally, there is a microscopic scale, $\lambda$, which approximates
the mean free path of the particles. We then define the so-called
Knudsen number, $\mathrm{Kn}=\lambda L^{-1}$, which should be small
if fluid dynamics is applicable. Consequently, all hydrodynamic quantities
can be ordered by ${\rm Kn}$ \citep{DeGroot:1980dk,denicol2022microscopic}.

Under the power counting in Eq. (\ref{eq:WF=000020power=000020counting}),
we can derive the perturbative solution of $\mathcal{V}^{\mu}$ and
$\mathcal{A}^{\mu}$ up to $\mathcal{O}(\hbar^{1})$ \citep{Yang:2020hri,Fang:2022ttm},
\begin{eqnarray}
\mathcal{V}^{\lessgtr,\mu}(X,p) & = & 2\pi p^{\mu}\delta(p^{2})f_{\mathrm{V}}^{\lessgtr}(X,p),\label{eq:Formal=000020sol.=000020V}\\
\mathcal{A}^{\lessgtr,\mu}(X,p) & = & 2\pi p^{\mu}\delta(p^{2})f_{\mathrm{A}}^{\lessgtr}(X,p)+2\pi\hbar\delta(p^{2})S_{(n)}^{\mu\alpha}\left(\Delta_{\alpha}f_{\mathrm{V}}^{\lessgtr}(X,p)-C_{\mathrm{V,}\alpha}[f_{\mathrm{V}}^{\lessgtr}]\right)\nonumber\\ 
 &  & +2\pi\delta^{\prime}(p^{2})f_{{\rm V}}^{\lessgtr}(X,p)\frac{\hbar}{2}\epsilon^{\mu\nu\rho\sigma}eF_{\rho\sigma}q_{\nu}\label{eq:Formal=000020sol.=000020A}
\end{eqnarray}
where we focus on the positive-energy fermions and have omitted the
overall sign function $\epsilon(n\cdot p)$ with $n^{\mu}$ being
the frame vector from now on. The frame-dependent spin tensor is defined
as 
\begin{equation}
S_{(n)}^{\mu\nu}=\frac{\epsilon^{\mu\nu\rho\sigma}p_{\rho}n_{\sigma}}{2n\cdot p},
\end{equation}
and 
\begin{equation}
C_{\mathrm{V,}\alpha}[f_{\mathrm{V}}^{\lessgtr}]\equiv\Sigma_{\mathrm{V},\alpha}^{\lessgtr}f_{\mathrm{V}}^{\gtrless}-\Sigma_{\mathrm{V},\alpha}^{\gtrless}f_{\mathrm{V}}^{\lessgtr}.
\end{equation}
The vector-charge distribution function $f_{\mathrm{V}}^{<}$ follows
$f_{\mathrm{V}}^{<}+f_{\mathrm{V}}^{>}=1$ and is taken as Fermi-Dirac
distribution in thermal equilibrium, while the axial-charge distribution
$f_{\mathrm{A}}^{<}$ follows $f_{\mathrm{A}}^{<}=-f_{\mathrm{A}}^{>}$.
According to Eq. (\ref{eq:WF=000020power=000020counting}), we have,
\begin{equation}
f_{\mathrm{V}}^{\lessgtr}\sim\mathcal{O}(\hbar^{0}),\quad f_{\mathrm{A}}^{\lessgtr}\sim\mathcal{O}(\hbar^{1}).
\end{equation}
To make the solutions $\{f_{{\rm V}},f_{{\rm A}}\}$ mathematically
close, we also need the following two kinetic equations up to $\mathcal{O}(\hbar^{1})$
\citep{Yang:2020hri,Fang:2022ttm},
\begin{eqnarray}
\Delta_{\mu}\mathcal{V}^{<,\mu} & = & \Sigma_{V}^{<}\cdot\mathcal{V}^{>}-\Sigma_{V}^{>}\cdot\mathcal{V}^{<}\equiv2\pi\delta(p^{2})\mathcal{C}_{\mathrm{V}},\label{eq:V=000020kinetic=000020eq}\\
\Delta_{\mu}\mathcal{A}^{<,\mu} & = & \Sigma_{\mathrm{V}}^{<}\cdot\mathcal{A}^{>}-\Sigma_{\mathrm{V}}^{>}\cdot\mathcal{A}^{<}-\Sigma_{\mathrm{A}}^{<}\cdot\mathcal{V}^{>}+\Sigma_{\mathrm{A}}^{>}\cdot\mathcal{V}^{<}\equiv2\pi\delta(p^{2})\mathcal{C}_{\mathrm{A}}.\label{eq:A=000020kinetic=000020eq}
\end{eqnarray}
Eq. (\ref{eq:V=000020kinetic=000020eq}) leads to the conventional
 collisional Vlasov equation, while Eq.(\ref{eq:A=000020kinetic=000020eq})
leads to the SBE. The collision kernels $\mathcal{C}_{\mathrm{V/A}}$
depend on the practical interaction and will be explicitly calculated
in the following context.

Lastly, we discuss the thermal equilibrium conditions for the distribution
functions in the absence of external electromagnetic fields. In global equilibrium, as proven by the $H$-theorem \citep{Chen:2015gta}
and quantum kinetic theory \citep{Weickgenannt:2020aaf} , the global
equilibrium conditions, known as Killing conditions for (spin) hydrodynamics,
are \citep{Becattini:2012tc},
\begin{eqnarray}
\partial_{(\mu}\beta u_{\nu)}=0,\qquad & \omega_{\mu\nu}^{s}=\Omega_{\mu\nu}\equiv\partial_{[\mu}\beta u_{\nu]},\qquad & \nabla_{\mu}\alpha=0,\label{eq:General=000020Geq=000020condition}
\end{eqnarray}
where 
\begin{equation}
\beta=T^{-1},\quad\alpha=\beta\mu,
\end{equation}
are the inverse temperature and the thermal chemical potential, respectively.
The $\omega_{\mu\nu}^{s}$ and $\Omega_{\mu\nu}$ are called thermal
vorticity tensor and the spin chemical potential, respectively. The
first equation in Eqs.(\ref{eq:General=000020Geq=000020condition})
can be derived by solving the relativistic Boltzmann equation \citep{DeGroot:1980dk}.
From now on, unless otherwise specified, we work in the \textit{comoving frame} with 
\begin{equation}
n^{\mu}=u^{\mu}=(1,\mathbf{0}),
\end{equation}
but $\partial_{\mu}u_{\nu}\neq0$.

In local equilibrium, entropy production remains zero and all collision
kernels in the kinetic theory should vanish. However, the free-streaming
part of the (spin) Boltzmann equations (\ref{eq:V=000020kinetic=000020eq},
\ref{eq:A=000020kinetic=000020eq}) does not vanish. Note that the
"\emph{local equilibrium}" mentioned here differs from that described
in Refs. \citep{Weickgenannt:2020aaf,Weickgenannt:2021cuo,Weickgenannt:2022zxs}.
In those works, local equilibrium corresponds to the vanishing of
the local collision kernel, while the non-local part remains nonzero.
In such scenarios, the spin chemical potential emerges as a necessary
novel thermodynamic quantity. Additionally, the local equilibrium
condition for massless fermions in a 2-2 scattering process appears
to be particularly unique \citep{Hidaka:2017auj,Fang:2022ttm},
\begin{eqnarray}
\omega_{\mu\nu}^{s} & = & \Omega_{\mu\nu},\quad\partial_{(\mu}\beta u_{\nu)}\neq0,\quad\nabla_{\mu}\alpha\neq0.\label{eq:Leq_condition=000020HTL}
\end{eqnarray}
The correctness of this condition has also been verified through explicit
calculations of axial-vector collision kernels in HTL approximations
\citep{Fang:2022ttm}. We have also listed a proof for the $u$- and
$t$-channel fermion elastic scattering process in the Appendix \ref{subsec:Local-equilibrium-condition}.
In such cases, up to $\mathcal{O}(\hbar^{1})$, the distribution functions
in our local equilibrium are as follows \citep{Chen:2015gta,Hidaka:2017auj,Fang:2022ttm},

\begin{eqnarray}
f_{\mathrm{V},\mathrm{leq}}^{<}(x,p) & = & \frac{1}{e^{\beta u\cdot p-\alpha}+1},\label{eq:fV=000020local=000020eq}\\
f_{\mathrm{A},\mathrm{leq}}^{<}(x,p) & = & -\frac{\hbar}{2}f_{V,\mathrm{leq}}^{<}(x,p)f_{V,\mathrm{leq}}^{>}(x,p)\omega_{\mu\nu}^{s}S_{(u)}^{\mu\nu}(x,p),\label{eq:fA=000020local=000020eq}
\end{eqnarray}
where we have dropped the axial chemical potential $\mu_{\mathrm{A}}$
due to the damping of axial charge in local equilibrium\footnote{Alternatively, in the presence of gradient of thermal chemical potential,
the thermal axial chemical $\mu_{\mathrm{A}}/T$ depends on the choice
of the frame \citep{Chen:2015gta}, thus we simply set $\mu_{\mathrm{A}}=0$
in the comoving frame here.} \citep{Hattori:2019ahi}. Some consequences from Eq. (\ref{eq:Leq_condition=000020HTL})
have important applications in the local polarization of $\Lambda$
and $\overline{\Lambda}$ hyperons in relativistic heavy ion collisions.
Given that $\partial_{(\mu}(\beta u_{\nu)})\neq0$ and $\nabla_{\mu}\alpha\neq0$
in local equilibrium, these conditions can lead to shear-induced polarization
and a baryonic spin-Hall effect. This has been discussed in various
studies \citep{Hidaka:2017auj,Yi:2021ryh,Liu:2020dxg,Liu:2021uhn,Becattini:2021suc,Fu:2021pok,Fu:2022myl,Becattini:2021iol,Wu:2022mkr}.
We notice that these effects do not originate from collision terms,
we can thus refer to spin polarization induced by such effects as
'intrinsic', given that the effective transport
coefficient for shear-induced polarization is time-reversal even.
These effects are associated with the side-jump term $S^{\mu\nu}\partial_{\nu}f_{\mathrm{V}}$
and, further, with the Berry curvature in momentum space. Interestingly,
similar phenomena have been extensively discussed in condensed matter
physics, referred to as intrinsic spin Hall effects \citep{sinova2015spin}.
In this work, we will discuss the off-equilibrium effects beyond the
local equilibrium regime, resulting from collisions, which we will
term '\emph{extrinsic spin Hall effects}'. To investigate the off-equilibrium
corrections to the axial Wigner function $\mathcal{A}^{\mu}$ in Eq.(\ref{eq:Formal=000020sol.=000020A}),
we need to evaluate the off-equilibrium corrections to the distribution
functions $\{f_{{\rm V}},f_{{\rm A}}\}$ by solving the Eqs. (\ref{eq:V=000020kinetic=000020eq},
\ref{eq:A=000020kinetic=000020eq}). Before proceeding with these
calculations, it is necessary to discuss the collision kernels involving
specific interactions.

\subsection{Self-energies in 2-2 scattering process and structure of collision
kernels\protect\label{subsec:Self-energies-in-2-2}}

\begin{figure}
\begin{centering}
\includegraphics{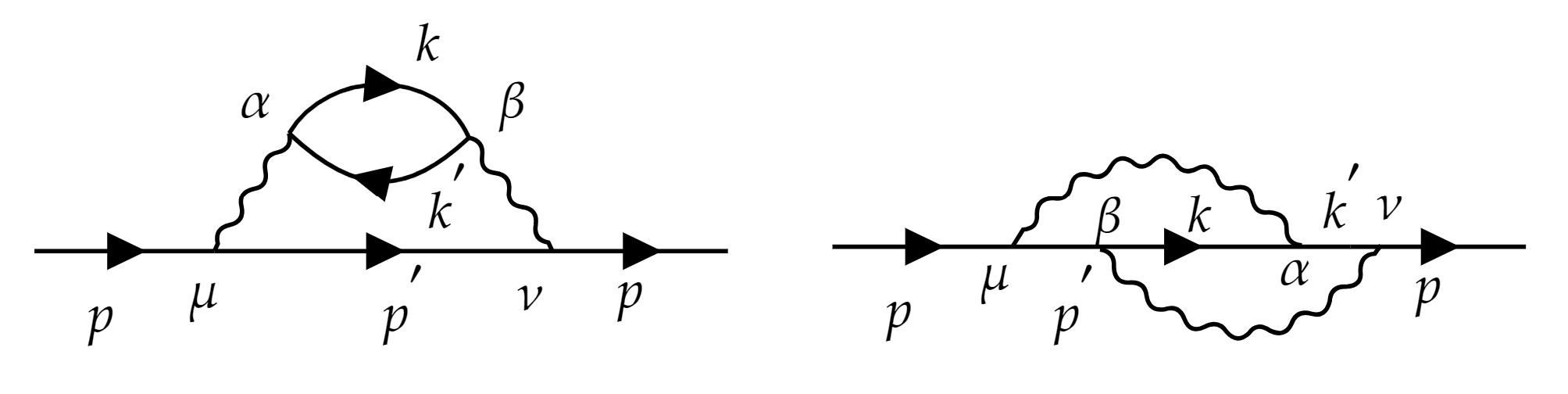}
\par\end{centering}
\caption{The 2-2 scattering process of the fermions in the $t$-channel and
$u$-channel are shown in the left and right hand panels, respectively.
The Roman indices are the $4$-momenta of the fermions while the
Greek ones represent the polarization of the photon propagators.\protect\label{fig:The-2-2-scattering}}
\end{figure}

In the current work, we consider the complete M{\o}ller scattering process
between the identical fermions at finite temperature, 
\begin{eqnarray}
f^{-}(p)+f^{-}(k) & \to & f^{-}(p^{\prime})+f^{-}(k^{\prime}).\label{eq:Electron=000020scattering}
\end{eqnarray}
instead of focusing on the scattering between a hard fermion and
a thermalized QED medium \citep{Fang:2022ttm}. In the current
work, we assume that the fermions are near-equilibrium and consider
the scattering process in both $u$- and $t$-channels. For simplicity,
we assume the photons are not thermalized. Hence, we adopt the zero-temperature
photonic Wigner function and neglect the relevant scattering processes
as discussed in Ref. \citep{Fang:2022ttm}. Consequently, we will
not include the Compton scattering process and the positron-related
processes in this analysis. The inclusion of the $u$-channel will
introduce crossing terms such as $\frac{s^{2}}{tu}$, where $s,t,u$
represent the Mandelstam variables. These crossing terms do not contribute
to the collision integrals up to the leading logarithm. However, they
can introduce corrections to the collision kernels. For more details,
see Appendix \ref{sec:Collision-integrals}. Notably, unlike our previous
work \citep{Fang:2022ttm}, the conservation of energy-momentum tensor
and U(1) currents is ensured in the current study by integrating the
vector Boltzmann equation (\ref{eq:V=000020kinetic=000020eq}) with
the collision kernel over momentum. This conservation results from
the complete contributions of both $u$- and $t$-channels. Ensuring
the conservation of these macroscopic quantities is essential for
deriving the hydrodynamics discussed in Sec. \ref{subsec:Spin-hydrodynamics-from}. 

We also remark from now on we will simply drop the background fields in the collision terms when calculating the self-energies. This is because these background fields are higher order effects in the collision terms. Moreover, such ignorance does not harm the gauge invariance of self-energies due to the properties of HTL approximation \cite{Bellac:2011kqa}. In the left-handed side, these background fields are also the leading contribution, but due to the magnitude of such external EM field is relatively small compared to the temperature and damped out very fast the in the QGP \cite{Deng:2012pc}, we can also neglect them from now on.

Using the cutting rule in Refs. \citep{Bellac:2011kqa,Mrowczynski:1992hq},
the self-energy in Fig. \ref{fig:The-2-2-scattering} is written as,
\begin{eqnarray}
 &  & -\Sigma^{<}(x_{1},x_{2})\nonumber \\
 & = & -\int\mathrm{d}^{4}x_{3}\mathrm{d}^{4}x_{4}(-ie\gamma_{\mu})S^{<}(x_{1},x_{2})(+ie\gamma_{\nu})G^{\mu\alpha}(x_{1},x_{3})G^{\beta\nu,*}(x_{4},x_{2})\nonumber \\
 &  & \qquad\qquad\times\mathrm{Tr}\left[(-ie\gamma_{\alpha})S^{<}(x_{3},x_{4})(ie\gamma_{\beta})S^{>}(x_{4},x_{3})\right]\nonumber \\
 &  & +\int\mathrm{d}^{4}x_{3}\mathrm{d}^{4}x_{4}(-ie\gamma_{\mu})S^{<}(x_{1},x_{4})(+ie\gamma_{\beta})S^{>}(x_{4},x_{3})\nonumber \\
 &  & \qquad\qquad\times(-ie\gamma_{\alpha})S^{<}(x_{3},x_{2})(+ie\gamma_{\nu})G^{\mu\alpha}(x_{1},x_{3})G^{\nu\beta,*}(x_{4},x_{2}),\label{eq:SE=000020coordinate=000020space}
\end{eqnarray}
where $G^{\mu\nu}(x,y)$ is the photonic propagator and its Hermitian
conjugate is the 2-point photonic Green's function lying in the anti-chronological
branch in Schwinger-Keldysh contour. All the vertices in the lower
contour receive an additional minus sign in Eq. (\ref{eq:SE=000020coordinate=000020space}).
Furthermore, the trace over the fermionic loop contributes an extra
$(-1)$. Taking Wigner transformation and choosing the Feynman gauge,
we obtain,
\begin{eqnarray}
 &  & -\Sigma^{<}(X,p)\nonumber \\
 & = & -e^{4}\int_{p^{\prime},k,k^{\prime}}\gamma^{\alpha}S^{<}(X,p^{\prime})\gamma^{\beta}\frac{1}{(p-p^{\prime})^{4}}\mathrm{Tr}\left[\gamma_{\alpha}S^{<}(X,k^{\prime})\gamma_{\beta}S^{>}(X,k)\right]\nonumber \\
 &  & +e^{4}\int_{p^{\prime},k,k^{\prime}}\gamma^{\alpha}S^{<}(X,p^{\prime})\gamma^{\nu}S^{>}(X,k)\gamma_{\alpha}S^{<}(X,k^{\prime})\gamma_{\nu}\frac{1}{(p-k^{\prime})^{2}}\frac{1}{(p-p^{\prime})^{2}},\label{eq:SE=000020phasespace}
\end{eqnarray}
where we have introduced the following symbols for convenience, 
\begin{eqnarray}
\int_{p^{\prime},k,k^{\prime}} & \equiv & \int\frac{\mathrm{d}^{4}k^{\prime}}{(2\pi)^{4}}\frac{\mathrm{d}^{4}p^{\prime}}{(2\pi)^{4}}\frac{\mathrm{d}^{4}k}{(2\pi)^{4}}(2\pi)^{4}\delta^{(4)}(p+k-k^{\prime}-p^{\prime}),\label{eq:Int=000020sym1}\\
\int_{k,k^{\prime}} & \equiv & \int\frac{\mathrm{d}^{4}k^{\prime}}{(2\pi)^{4}}\frac{\mathrm{d}^{4}k}{(2\pi)^{4}}(2\pi)^{4}\delta^{(4)}(p+k-k^{\prime}-p^{\prime}),\label{eq:Int=000020sym2}\\
\int_{p^{\prime}} & \equiv & \int\frac{\mathrm{d}^{4}p^{\prime}}{(2\pi)^{4}}.\label{eq:Int=000020sym3}
\end{eqnarray}

Utilizing the Clifford decomposition (\ref{eq:Massless=000020WF=000020Clifford=000020decomp},
\ref{eq:Massless=000020SE=000020Clifford=000020decomp}), the vector
and axial component of fermionic self-energies read,
\begin{eqnarray}
 &  & \Sigma_{\mathrm{V}}^{<,\mu}(X,p)\nonumber \\
 & = & 8e^{4}\int_{p^{\prime},k,k^{\prime}}\frac{1}{(p-p^{\prime})^{4}}\left[\mathcal{V}^{<,\mu}(k^{\prime})\mathcal{V}^{>}(k)\cdot\mathcal{V}^{<}(p^{\prime})+\mathcal{V}^{>,\mu}(k)\mathcal{V}^{<}(k^{\prime})\cdot\mathcal{V}^{<}(p^{\prime})\right]\nonumber \\
 &  & +8e^{4}\int_{p^{\prime},k,k^{\prime}}\frac{1}{(p-k^{\prime})^{2}}\frac{1}{(p-p^{\prime})^{2}}\mathcal{V}^{>,\mu}(k)\mathcal{V}^{<}(p^{\prime})\cdot\mathcal{V}^{<}(k^{\prime})\nonumber \\
 & = & 8e^{4}\int_{p^{\prime},k,k^{\prime}}w_{1,pk\to p^{\prime}k^{\prime}}^{\mu}(2\pi)^{3}\delta(k^{\prime,2})\delta(k^{2})\delta(p^{\prime,2})f_{\mathrm{V}}^{<}(k^{\prime})f_{\mathrm{V}}^{<}(p^{\prime})f_{\mathrm{V}}^{>}(k),\label{eq:Vector=000020SE}
\end{eqnarray}
and
\begin{eqnarray}
 &  & \Sigma_{\mathrm{A}}^{<,\mu}(X,p)\nonumber \\
 & = & -8e^{4}\int_{p^{\prime},k,k^{\prime}}\frac{1}{(p-p^{\prime})^{4}}\left[\mathcal{V}^{<,\mu}(k^{\prime})\mathcal{V}^{>}(k)\cdot\mathcal{A}^{<}(p^{\prime})+\mathcal{V}^{>,\mu}(k)\mathcal{V}^{<}(k^{\prime})\cdot\mathcal{A}^{<}(p^{\prime})\right.\nonumber \\
 &  & \qquad\qquad+\mathcal{A}^{>,\mu}(k)\mathcal{V}^{<}(p^{\prime})\cdot\mathcal{V}^{<}(k^{\prime})-\mathcal{V}^{<,\mu}(k^{\prime})\mathcal{V}^{<}(p^{\prime})\cdot\mathcal{A}^{>}(k)\nonumber \\
 &  & \qquad\qquad\left.-\mathcal{A}^{<,\mu}(k^{\prime})\mathcal{V}^{<}(p^{\prime})\cdot\mathcal{V}^{>}(k)+\mathcal{V}^{>,\mu}(k)\mathcal{V}^{<}(p^{\prime})\cdot\mathcal{A}^{<}(k^{\prime})\right]\nonumber \\
 &  & -8e^{4}\int_{p^{\prime},k,k^{\prime}}\frac{1}{(p-k^{\prime})^{2}}\frac{1}{(p-p^{\prime})^{2}}\left[\mathcal{V}^{>,\mu}(k)\mathcal{V}^{<}(p^{\prime})\cdot\mathcal{A}^{<}(k^{\prime})\right.\nonumber \\
 &  & \qquad\qquad\left.+\mathcal{A}^{>,\mu}(k)\mathcal{V}^{<}(p^{\prime})\cdot\mathcal{V}^{<}(k^{\prime})+\mathcal{V}^{>,\mu}(k)\mathcal{A}^{<}(p^{\prime})\cdot\mathcal{V}^{<}(k^{\prime})\right],\label{eq:Axial=000020SE}
\end{eqnarray}
where we introduce the auxiliary scattering matrix element, 
\begin{eqnarray}
w_{1,pk\to p^{\prime}k^{\prime}}^{\mu} & = & \frac{k^{\prime,\mu}k\cdot p^{\prime}+k^{\prime}\cdot p^{\prime}k^{\mu}}{(p-p^{\prime})^{4}}+\frac{k^{\mu}p^{\prime}\cdot k^{\prime}}{(p-k^{\prime})^{2}(p-p^{\prime})^{2}}.\label{eq:Auxillary_scattering_element_=00005Cmu1}
\end{eqnarray}

Inserting Eqs.(\ref{eq:Formal=000020sol.=000020V}, \ref{eq:Vector=000020SE})
into Eq.(\ref{eq:V=000020kinetic=000020eq}), yields, 
\begin{eqnarray}
p\cdot\partial f_{\mathrm{V}}^{<}(p) & = & \mathcal{C}_{\mathrm{V}}\nonumber \\
 & \equiv & \frac{1}{4}\int_{p^{\prime},k,k^{\prime}}\mathcal{W}_{pk\to p^{\prime}k^{\prime}}(2\pi)^{3}\delta(k^{2})\delta(p^{\prime,2})\delta(k^{\prime,2})\nonumber \\
 &  & \times\left(f_{\mathrm{V}}^{<}(k^{\prime})f_{\mathrm{V}}^{<}(p^{\prime})f_{\mathrm{V}}^{>}(p)f_{\mathrm{V}}^{>}(k)-f_{\mathrm{V}}^{>}(k^{\prime})f_{\mathrm{V}}^{>}(p^{\prime})f_{\mathrm{V}}^{<}(p)f_{\mathrm{V}}^{<}(k)\right),\label{eq:V=000020BE=000020Moller=000020process}
\end{eqnarray}
where $\mathcal{W}_{pk\to p^{\prime}k^{\prime}}$ is the scattering
matrix elements for the M{\o}ller process, 
\begin{eqnarray}
\mathcal{W}_{pk\to p^{\prime}k^{\prime}} & = & 4e^{4}\left[\frac{u^{2}+s^{2}}{t^{2}}+\frac{2s^{2}}{ut}+\frac{t^{2}+s^{2}}{u^{2}}\right],\label{eq:Moller=000020scattering=000020matrix=000020elements}
\end{eqnarray}
with the Mandelstam variables being $t=(p-p^{\prime})^{2}$, $u=(p-k^{\prime})^{2}$
and $s=(p+k)^{2}$. Here, the spins of the particles in final states
are summed over, while those in the initial states are not averaged.
We have also extracted a factor $1/2$ to correct the double counting
of initial states. It is easy to check that $\mathcal{W}_{pk\to p^{\prime}k^{\prime}}$
is symmetric under the exchange of $p^{\prime}\leftrightarrow k^{\prime}$
and $p\leftrightarrow k$, and the collision kernel in Eq. (\ref{eq:V=000020BE=000020Moller=000020process})
vanishes when $f_{\mathrm{V}}^{<}(p)=f_{\mathrm{V,leq}}^{<}(p)$ as
a result of detailed balance. Another remark on Eq.(\ref{eq:Moller=000020scattering=000020matrix=000020elements})
is that the evolution of vector charge distribution $f_{\mathrm{V}}^{<}$
is decoupled from the evolution of spin up to $\mathcal{O}(\hbar^{1})$.
This indicates that the spin does not influence the properties of
conventional quantities up to $\mathcal{O}(\partial^{1})$. A similar
observation can also be found in the hydrodynamics derived from Winger
functions, as shown in Sec. \ref{subsec:Spin-hydrodynamics-from}.

Similarly, using Eqs.(\ref{eq:V=000020kinetic=000020eq}, \ref{eq:A=000020kinetic=000020eq},
\ref{eq:Vector=000020SE}, \ref{eq:Axial=000020SE}), we derive SBE,
\begin{eqnarray}
p^{\mu}\partial_{\mu}f_{\mathrm{A}}^{<}(p)+\hbar\partial_{\mu}S^{(u),\mu\alpha}(p)\partial_{\alpha}f_{\mathrm{V}}^{<}(p) & = & \mathcal{C}_{\mathrm{A}}+\hbar\partial_{\mu}\left(S_{(u)}^{\mu\alpha}C_{\mathrm{V,}\alpha}[f_{\mathrm{V}}^{<}]\right)\nonumber \\
 & \equiv & \mathcal{C}_{\mathrm{A}}^{\mathrm{B}}[f_{\mathrm{V}},f_{\mathrm{A}}]+\mathcal{C}_{\mathrm{A}}^{\partial}[f_{\mathrm{V}}]+\mathcal{C}_{\mathrm{A}}^{\Sigma}[f_{\mathrm{V}}].\label{eq:SBE=000020Moller=000020process}
\end{eqnarray}
The term $\mathcal{C}_{\mathrm{A}}^{\mathrm{B}}[f_{\mathrm{V}},f_{\mathrm{A}}]$
has a very similar structure to the collision kernel in the vector
Boltzmann equation (\ref{eq:V=000020BE=000020Moller=000020process}),
and all dependence on $f_{\mathrm{A}}$ are collected there. Its expression
is as follows,
\begin{eqnarray}
 &  & \mathcal{C}_{\mathrm{A}}^{\mathrm{B}}[f_{\mathrm{V}},f_{\mathrm{A}}]\nonumber \\
 & = & 8e^{4}\int_{p^{\prime},k,k^{\prime}}w_{pk\to p^{\prime}k^{\prime}}^{1}(2\pi)^{3}\delta(k^{\prime,2})\delta(k^{2})\delta(p^{\prime,2})\nonumber \\
 &  & \qquad\times\left[f_{\mathrm{V}}^{<}(k^{\prime})f_{\mathrm{V}}^{<}(p^{\prime})f_{\mathrm{V}}^{>}(k)f_{\mathrm{A}}^{>}(p)-f_{\mathrm{V}}^{>}(k^{\prime})f_{\mathrm{V}}^{>}(p^{\prime})f_{\mathrm{V}}^{<}(k)f_{\mathrm{A}}^{<}(p)\right.\nonumber \\
 &  & \qquad\qquad\left.+f_{\mathrm{V}}^{>}(p)f_{\mathrm{V}}^{>}(k)f_{\mathrm{V}}^{<}(k^{\prime})f_{\mathrm{A}}^{<}(p^{\prime})-f_{\mathrm{V}}^{<}(k)f_{\mathrm{V}}^{<}(p)f_{\mathrm{V}}^{>}(k^{\prime})f_{\mathrm{A}}^{>}(p^{\prime})\right]\nonumber \\
 &  & +8e^{4}\int_{p^{\prime},k,k^{\prime}}w_{pk\to p^{\prime}k^{\prime}}^{2}(2\pi)^{3}\delta(k^{2})\delta(p^{\prime,2})\delta(k^{\prime,2})\nonumber \\
 &  & \qquad\times\left[f_{\mathrm{V}}^{<}(p^{\prime})f_{\mathrm{V}}^{<}(k^{\prime})f_{\mathrm{V}}^{>}(p)f_{\mathrm{A}}^{>}(k)-f_{\mathrm{V}}^{>}(p^{\prime})f_{\mathrm{V}}^{>}(k^{\prime})f_{\mathrm{V}}^{<}(p)f_{\mathrm{A}}^{<}(k)\right.\nonumber \\
 &  & \qquad\qquad\left.+f_{\mathrm{V}}^{<}(p^{\prime})f_{\mathrm{A}}^{<}(k^{\prime})f_{\mathrm{V}}^{>}(p)f_{\mathrm{V}}^{>}(k)-f_{\mathrm{V}}^{<}(p)f_{\mathrm{V}}^{<}(k)f_{\mathrm{V}}^{>}(p^{\prime})f_{\mathrm{A}}^{>}(k^{\prime})\right],\label{eq:Axial=000020CK=000020Boltzmann-type}
\end{eqnarray}
where, 
\begin{eqnarray}
w_{pk\to p^{\prime}k^{\prime}}^{1} & = & \frac{(p\cdot k^{\prime})^{2}+(p\cdot k)^{2}}{(p-p^{\prime})^{4}}+\frac{(p\cdot k)^{2}}{(p-k^{\prime})^{2}(p-p^{\prime})^{2}},\label{eq:Auxillary_scattering_element_2}\\
w_{pk\to p^{\prime}k^{\prime}}^{2} & = & \frac{(p\cdot k)^{2}-(p\cdot k^{\prime})^{2}}{(p-p^{\prime})^{4}}+\frac{(p\cdot k)^{2}}{(p-k^{\prime})^{2}(p-p^{\prime})^{2}}.\label{eq:Auxillary_scattering_element_3}
\end{eqnarray}
The terms $\mathcal{C}_{\mathrm{A}}^{\partial}[f_{\mathrm{V}}]$ and
$\mathcal{C}_{\mathrm{A}}^{\Sigma}[f_{\mathrm{V}}]$ function solely
as functionals of $f_{\mathrm{V}}$. The term $\mathcal{C}_{\mathrm{A}}^{\partial}[f_{\mathrm{V}}]$
is related to $\partial f_{\mathrm{V}}$. While the term $\mathcal{C}_{\mathrm{A}}^{\Sigma}[f_{\mathrm{V}}]$
includes terms like $e^{4}C_{\mathrm{V,\alpha}}[f_{\mathrm{V}}^{<}]$,
which are higher order in coupling but leading order in gradient expansion.
The explicit form of $\mathcal{C}_{\mathrm{A}}^{\partial}$ and $\mathcal{C}_{\mathrm{A}}^{\Sigma}$
are presented in Appendix. \ref{sec:Axial-collisional-kernels}. If
we include the coupling constant in our power counting, we find that
these three types of collision kernels are 
\begin{eqnarray}
\mathcal{C}_{\mathrm{A}}^{\mathrm{B}}[f_{\mathrm{V}},f_{\mathrm{A}}]\sim\mathcal{C}_{\mathrm{A}}^{\partial}[f_{\mathrm{V}}]\sim\mathcal{O}(e^{4}\partial^{1}) & ,\qquad & \mathcal{C}_{\mathrm{A}}^{\Sigma}[f_{\mathrm{V}}]\sim\mathcal{O}(e^{8}\partial^{0}).\label{eq:ACK=000020General=000020power=000020counting=000020scheme}
\end{eqnarray}

In the current work, we focus on the scenarios close to local equilibrium,
where the distribution functions $f_{\mathrm{V}},f_{\mathrm{A}}$
slightly deviate from their local equilibrium forms, 
\begin{eqnarray}
f_{\mathrm{V}}=f_{\mathrm{V,leq}}+\delta f_{\mathrm{V}} & ,\quad & f_{\mathrm{A}}=f_{\mathrm{A,leq}}+\delta f_{\mathrm{A}},\quad\delta f_{\mathrm{V(A})}\ll f_{\mathrm{V(A),leq}},\label{eq:Perturbation=000020of=000020f}
\end{eqnarray}
where $f_{\mathrm{V,leq}}$ and $f_{\mathrm{A,leq}}$ are given by
Eqs. (\ref{eq:fV=000020local=000020eq}, \ref{eq:fA=000020local=000020eq}),
respectively. When $f_{\mathrm{V}}=f_{\mathrm{V,leq}}$ and $f_{\mathrm{A}}=f_{\mathrm{A,leq}}$
, it has been proven in our previous work \citep{Fang:2022ttm} that
$\mathcal{C}_{\mathrm{A}}[f_{\mathrm{V,leq}},f_{\mathrm{A,leq}}]=0$.
Furthermore, since $C_{\mathrm{V},\alpha}[f_{\mathrm{V,leq}}^{<}]\propto h_{1}u_{\alpha}+h_{2}p_{\perp,\alpha}$\footnote{We can actually prove this relation using the HTL approximation by
making the angle projection between different momenta.}, we can get $S^{(u),\mu\alpha}C_{\mathrm{V,}\alpha}[f_{\mathrm{V,leq}}^{<}]=0$,
i.e. $\mathcal{C}_{\mathrm{V}}[f_{\mathrm{V,leq}}]=0$. When $f_{\mathrm{V}}=f_{\mathrm{V,leq}}$
and $f_{\mathrm{A}}\neq f_{\mathrm{A,leq}}$, we have $\mathcal{C}_{\mathrm{V}}[f_{\mathrm{V,leq}}]=0$
but $\mathcal{C}_{\mathrm{A}}[f_{\mathrm{V,leq}},f_{\mathrm{A}}]\neq0$.
This indicates that the ordinary quasi-particle modes reach their
equilibrium states faster than the spin modes \citep{Fang:2022ttm}.
Using Eqs. (\ref{eq:Perturbation=000020of=000020f}), we can linearize
both the collision kernels $\mathcal{C}_{\mathrm{V}}$ and $\mathcal{C}_{\mathrm{A}}$
up to the first order of $\delta f_{\mathrm{V}}$, 
\begin{eqnarray}
\mathcal{C}_{\mathrm{V}}[\delta f_{\mathrm{V}}]\sim\mathcal{O}(e^{4}\delta f_{{\rm V}}),\quad \mathcal{C}_{\mathrm{A}}^{\partial}[\delta f_{\mathrm{V}}]\sim\mathcal{O}(e^{4}\partial\delta f_{{\rm V}}),\quad \mathcal{C}_{\mathrm{A}}^{\Sigma}[\delta f_{\mathrm{V}}]\sim\mathcal{O}(e^{8}\delta f_{{\rm V}}).\label{eq:near_Leq=000020ACK=000020power=000020counting=000020scheme}
\end{eqnarray}
Comparison both sides of Eq. (\ref{eq:V=000020BE=000020Moller=000020process}),
we formally obtain 
\begin{equation}
\delta f_{\mathrm{V}}\sim\mathcal{O}(e^{-4}\partial^{1}).
\end{equation}
 The Boltzmann type axial collision kernel $\mathcal{C}_{\mathrm{A}}^{\mathrm{B}}$
can be divided into two parts, 
\begin{equation}
\mathcal{C}_{\mathrm{A}}^{\mathrm{B}}[\delta f_{\mathrm{V}},\delta f_{\mathrm{A}}]=g_{1}[\delta f_{\mathrm{V}}]+g_{2}[\delta f_{\mathrm{A}}],\quad g_{1}[\delta f_{\mathrm{V}}]\sim\mathcal{O}(e^{4}\partial\delta f_{{\rm V}}),\quad g_{2}[\delta f_{\mathrm{A}}]\sim\mathcal{O}(e^{4}\delta f_{{\rm A}}).
\end{equation}
Recalling the gradient expansion, where $\partial\sim\lambda^{-1}\mathrm{Kn}\ll\lambda^{-1}$,
and noting that $\lambda$ is related to the cross section of microscopic
interactions, we usually have $\lambda^{-1}\sim\sigma_{col}n\sim Te^{4}$.
Here, $\sigma_{col}$ represents the typical cross section for quantum
electrodynamics type interactions and $n$ denotes the number density.
We, therefore, only need to keep axial collision kernel at $\mathcal{O}(e^{8}\delta f_{\mathrm{V}})$
in Eq.(\ref{eq:SBE=000020Moller=000020process}). Comparing with both
sides of Eq. (\ref{eq:SBE=000020Moller=000020process}), when $\delta f_{{\rm V}}\neq0$,
we get,
\begin{eqnarray}
\delta f_{\mathrm{A}} & \sim & \mathcal{O}(e^{4}\delta f_{{\rm V}})+\mathcal{O}(e^{-4}\partial^{2}).\label{eq:delta_fA=000020CaseI}
\end{eqnarray}

The above discussion shows that, two possible scenarios to derive
the off equilibrium correction to $\mathcal{A}^{\mu}$, $\delta\mathcal{A}^{\mu}$,
can exist.
\begin{itemize}
\item Scenario (I): we consider that the vector charge distribution function
is off-equilibrium, i.e. $f_{{\rm V}}\neq f_{\mathrm{V,leq}}^{<}$
, and assume that $\mathcal{O}(\partial)\ll\mathcal{O}(Te^{4})$.
Eq. (\ref{eq:Formal=000020sol.=000020A}) gives, 
\begin{eqnarray}
\delta\mathcal{A}_{\textrm{(I)}}^{<,\mu}(X,p) & = & 2\pi p^{\mu}\delta(p^{2})\delta f_{\mathrm{A}}^{<}(X,p)+2\pi\hbar\delta(p^{2})S_{(u)}^{\mu\alpha}\left(\partial_{X,\alpha}\delta f_{\mathrm{V}}^{<}(X,p)-C_{\mathrm{V,}\alpha}[\delta f_{\mathrm{V}}^{<}]\right)\nonumber \\
 & = & 2\pi p^{\mu}\delta(p^{2})\delta f_{\mathrm{A}}^{<}(X,p)-2\pi\hbar\delta(p^{2})S_{(u)}^{\mu\alpha}C_{\mathrm{V,}\alpha}[\delta f_{\mathrm{V}}^{<}]+\mathcal{O}(\partial^{2}).\label{eq:Off-eq-A-Case1}
\end{eqnarray}
Here, $\delta f_{{\rm V}}^{<}$ can be derived by solving the conventional
Boltzmann equation using the standard methods \citep{Arnold:2000dr,Arnold:2002zm,Arnold:2003zc}.
For $\delta f_{A}$, such off-equilibrium corrections to $f_{{\rm A}}$
automatically vanish up to $\mathcal{O}(\partial^{1})$ due to our
matching conditions, as discussed in Sec. \ref{subsec:A-revisiting-of}, 
\item Scenario (II): we consider that the vector charge distribution function
is in local equilibrium, i.e. $f_{{\rm V}}=f_{\mathrm{V,leq}}^{<}$.
Since $S_{(u)}^{\mu\alpha}C_{\mathrm{V,}\alpha}[f_{\mathrm{V,leq}}^{<}]=0$,
Eq. (\ref{eq:Formal=000020sol.=000020A}) reduces to, 
\begin{eqnarray}
\delta\mathcal{A}_{\textrm{(II)}}^{<,\mu}(X,p) & = & 2\pi p^{\mu}\delta(p^{2})\delta f_{\mathrm{A}}^{<}(X,p).\label{eq:Off-eq-A-Case2}
\end{eqnarray}
In this scenario, from Eq.(\ref{eq:SBE=000020Moller=000020process}),
only the Boltzmann-type axial collision kernel $\mathcal{C}_{\mathrm{A}}^{\mathrm{B}}$
survives. By linearizing the collision kernels, $\mathcal{C}_{\mathrm{A}}^{\mathrm{B}}[f_{\mathrm{V,leq}},\delta f_{\mathrm{A}}]\sim\mathcal{O}(e^{4}\delta f_{{\rm A}}),$
we find,
\begin{eqnarray}
\delta f_{\mathrm{A}} & \sim & \mathcal{O}(e^{-4}\partial^{2}).\label{eq:delta_fA_order}
\end{eqnarray}
This scenario corresponds to the case in which system is close to
the chemical freeze-out. In the chemical freeze-out hypersurface,
particle emission spectrum is well described by the Cooper-Frye formula
\citep{Cooper:1974mv} with thermal equilibrium distribution functions
for particles, as computed in the statistical hadronization model
\citep{Andronic:2017pug}.
\end{itemize}
Lastly, we briefly explain how to calculate the self-energies (\ref{eq:Vector=000020SE},
\ref{eq:Axial=000020SE}) and collision kernels on the right hand
side of Eq. (\ref{eq:V=000020BE=000020Moller=000020process}, \ref{eq:SBE=000020Moller=000020process})
using the HTL approximation. Following Refs. \citep{Arnold:2000dr,Arnold:2002zm,Arnold:2003zc},
we assume the transferred momentum $q=p-p^{\prime}$ is much larger
than the soft (electric) scale of the interaction system, denoted
as $eT$, but significantly smaller than the hard scale, such as the
characteristic temperature of the system $T$ and the momentum of
the hard particles $p,k,p^{\prime},k^{\prime}$, i.e.
\begin{eqnarray}
eT\ll & q & \ll T,p,k,p^{\prime},k^{\prime}.\label{eq:Momentum=000020hierachy}
\end{eqnarray}
This hierarchy corresponds to the leading logarithm result, and it
physically describes the small angle multi-scattering process \citep{Arnold:2007pg}.
Generally, the large angle scattering and collinear splitting process
are also important. But, we will not discuss such complete leading
order process for simplicity. Eq. (\ref{eq:Momentum=000020hierachy})
is sufficient to capture the transport properties we are interested
in, especially since the system consists of massless components. Finally,
all the collision integrals will be reduced to integrals over the
transferred momentum $q$, and we only need to keep the results proportional
to $\int_{eT}^{T}|\mathbf{q}|^{-1}\mathrm{d}|\mathbf{q}|$ in the
leading-logarithm contributions. The integrals often involves the
thermal chemical potential $\alpha$, but for simplicity, we will
omit them and only retain their gradients to derive analytical expressions.
For more details, one is referred to see Appendix. \ref{sec:Collision-integrals}
or our previous work \citep{Fang:2022ttm}.

\subsection{Matching conditions from QKT to relativistic hydrodynamics \protect\label{subsec:Spin-hydrodynamics-from}}

Before considering the higher order corrections from QKT, we need
to establish the matching conditions to match the macroscopic effective
theory, e.g. the relativistic hydrodynamics, to our microscopic theory.
The purpose of these matching conditions is to close the equations
of variables by defining the off-equilibrium energy density, particle
number density, and spacelike currents. For example, by integrating
the momentum, one can align the moments of microscopic distributions
with macroscopic quantities \citep{DeGroot:1980dk,denicol2022microscopic}.
This matching process eliminates any extra degrees of freedom that
originate from the symmetry of the theory\footnote{It reminds us the gauge ``symmetry'' is also a redundancy and needs
a constraint (the gauge fixing).}.

According to the N\"oether theorem, the canonical energy-momentum
and spin angular momentum current tensor, and conserved U(1) current
operators for fermions corresponding to the Lagrangian in Eq. (\ref{eq:QED=000020Lagrangian})
read,
\begin{eqnarray}
\hat{T}^{\mu\nu} & = & \frac{1}{2}\overline{\psi}i\hbar(\gamma^{\mu}\overrightarrow{\partial}^{\nu}-\overleftarrow{\partial}^{\nu}\gamma^{\mu})\psi,\label{eq:Canonical=000020EMT}\\
\hat{S}^{\mu\rho\sigma} & = & \frac{\hbar}{2}\epsilon^{\alpha\mu\rho\sigma}\overline{\psi}\gamma_{\alpha}\gamma^{5}\psi,\label{eq:Canonical=000020SAMT}\\
\hat{j}^{\mu} & = & \overline{\psi}\gamma^{\mu}\psi.\label{eq:U(1)=000020current}
\end{eqnarray}
where we have used the on-shell condition to isolate the fermion energy-momentum
tensor in Eq. (\ref{eq:Canonical=000020EMT}) from the total one.
Taking the ensemble average over the operators, we can express the
expectations of these operators in terms of Wigner functions \citep{Hidaka:2017auj,Yang:2018lew},
\begin{eqnarray}
T^{\mu\nu}(x) & = & 4\int_{p}p^{\nu}\mathcal{V}^{<,\mu}(x,p),\quad S^{\mu\rho\sigma}(x)=4\epsilon^{\alpha\mu\rho\sigma}\frac{\hbar}{2}\int_{p}\mathcal{A}_{\alpha}^{<}(x,p),\label{eq:EMT=000020=000026=000020SAMT=000020in=000020WF}\\
j^{\mu}(x) & = & 4\int_{p}\mathcal{V}^{<,\mu}(x,p).\label{eq:U(1)=000020current=000020in=000020WF}
\end{eqnarray}
We can further insert $\mathcal{V}^{<,\mu}(x,p)$ in Eq. (\ref{eq:Formal=000020sol.=000020V})
up to $\mathcal{O}(\hbar^{1})$ to Eqs.(\ref{eq:EMT=000020=000026=000020SAMT=000020in=000020WF},
\ref{eq:U(1)=000020current=000020in=000020WF}),
\begin{eqnarray}
T^{\mu\nu} & = & 4\int_{p}2\pi\delta(p^{2})p^{\nu}p^{\mu}f_{\mathrm{V}}^{<}(x,p)=\langle p^{\nu}p^{\mu}\rangle,\label{eq:EMT=000020in=000020distribution=000020f}\\
j^{\mu} & = & 4\int_{p}2\pi\delta(p^{2})p^{\mu}f_{\mathrm{V}}^{<}(x,p)=\langle p^{\mu}\rangle,\label{eq:U(1)=000020current=000020in=000020distribution=000020f}
\end{eqnarray}
where we have introduced the symbol 
\begin{equation}
\langle a\rangle=4\int_{p}2\pi\delta(p^{2})f_{\mathrm{V}}^{<}(x,p)a.
\end{equation}
We notice that up to $\mathcal{O}(\hbar^{1})$, the energy-momentum
tensor and U(1) current do not receive any corrections from the collective
spin degree of freedom and $T^{[\mu\nu]}\sim\mathcal{O}(\hbar^{2})$.
The macroscopic effects for spin, such as spin boost current or spin
rotational tensor in the energy-momentum tensor of spin hydrodynamics
\citep{Hattori:2019lfp,Fukushima:2020ucl,Hongo:2021ona} do not appear
due to our special power counting in Eqs. (\ref{eq:fV=000020local=000020eq},
\ref{eq:fA=000020local=000020eq}) with vanishing $\mu_{\mathrm{A}}$. 

In general, the $\mu_{\mathrm{A}}$ can be as large as the $\mu$,
such as in systems containing only right handed fermions. In such
cases, $\mathcal{A}_{\alpha}^{<}$ can have an additional term of
order $\mathcal{O}(\mu_{\mathrm{A}})\sim\mathcal{O}(\partial^{0})$.
Consequently, $\mathcal{V}^{<,\mu}$ can receive extra contributions
related to spin, e.g. see the Refs. \citep{Gao:2012ix,Chen:2012ca}.
If one can solve the off-equilibrium QKT up to $\mathcal{O}(\hbar^{2})$,
it becomes possible to derive a thermodynamic relation that includes
the spin chemical potential and spin density. The equations of motion
for hydrodynamics should then also take account of currents related
to spin. Under these conditions, the equilibrium condition would be
modified, e.g. the conventional Killing condition and $\nabla^{\mu}\alpha=0$
may no longer be valid \citep{Chen:2015gta}. 

Despite the absence of anti-symmetric part of energy-momentum tensor,
we can still construct the spin hydrodynamics using the conservation
law of angular momentum, 
\begin{eqnarray}
\partial_{\mu}S^{\mu\rho\sigma} & = & -2T^{[\rho\sigma]}.\label{eq:Conservation=000020of=000020AMT}
\end{eqnarray}
We regard Eq.(\ref{eq:Conservation=000020of=000020AMT}) as the definition
of anti-symmetric energy-momentum tensor. The expression of $T^{[\rho\sigma]}$
can be derived once we have the explicit expression for $S^{\mu\rho\sigma}$
from kinetic theory, e.g. see Refs. \citep{Weickgenannt:2020aaf,Weickgenannt:2021cuo,Weickgenannt:2022zxs}.
However, it is beyond the scope of the current work and will present
it somewhere else.

Here, we directly apply the thermodynamic relations and the equations
of motion for conventional viscous hydrodynamics. Following the conventional
relativistic hydrodynamics \citep{landau2003fluid,denicol2022microscopic},
we decompose the energy-momentum tensor and U(1) current as, 
\begin{eqnarray}
T^{\mu\nu} & = & \varepsilon u^{\mu}u^{\nu}-P\Delta^{\mu\nu}+2h^{(\mu}u^{\nu)}+\pi^{\mu\nu},\label{eq:EMT=000020decomposition}\\
j^{\mu} & = & nu^{\mu}+\nu^{\mu},\label{eq:U(1)=000020current=000020decomposition}
\end{eqnarray}
where the bulk viscous pressure $\Pi=0$ in the massless theory due
to the conformal symmetry. The energy density $\varepsilon$, the
pressure $P$, the heat current $h^{\mu}$, the shear viscous tensor
$\pi^{\mu\nu}$, the particle number density $n$ and the particle
diffusion current $\nu^{\mu}$ can be derived from the kinetic theory,
\begin{eqnarray}
\varepsilon=\langle E_{\mathbf{p}}^{2}\rangle,\quad & P=\frac{1}{3}\langle E_{\mathbf{p}}^{2}\rangle,\quad & \pi^{\mu\nu}=\langle p^{\langle\nu}p^{\mu\rangle}\rangle,\quad h^{\mu}=\langle E_{\mathbf{p}}p^{\langle\mu\rangle}\rangle,\nonumber \\
n=\langle E_{\mathbf{p}}\rangle,\quad & \nu^{\mu}=\langle p^{\langle\mu\rangle}\rangle,
\end{eqnarray}
where $E_{\mathbf{p}}=u_{\mu}p^{\mu}$. We find 
\begin{eqnarray}
\varepsilon,P,n,u^{\mu},...\sim\mathcal{O}(\partial^{0}) & ,\quad & \pi^{\mu\nu},\nu^{\mu},\partial\cdot u,\sigma_{\mu\nu},\nabla^{\mu}\alpha,...\sim\mathcal{O}(\partial),\label{eq:Normal=000020hydro=000020power=000020counting}
\end{eqnarray}
which is the same as those in conventional relativistic hydrodynamics.
For the quantity with quantum characteristics, we simply add an extra
$\hbar$, i.e. $S^{\mu\rho\sigma}\sim\mathcal{O}(\hbar^{1}\partial^{1})$.
We find that the thermal vorticity and the shear tensor are of the
same order, $\Omega_{\mu\nu}\sim\sigma_{\mu\nu}\sim\mathcal{O}(\partial^{1})$
based on our local equilibrium condition (\ref{eq:Leq_condition=000020HTL}).
This differs from the power counting scheme in Refs. \citep{Li:2020eon,Weickgenannt:2022zxs}.

We then discuss the matching conditions. First, we define the hydrodynamic
quantities in the local equilibrium, 
\begin{eqnarray}
\varepsilon_{0}=\langle E_{\mathbf{p}}^{2}\rangle_{0},\quad & P_{0}=\frac{1}{3}\langle E_{\mathbf{p}}^{2}\rangle_{0},\quad n_{0}=\langle E_{\mathbf{p}}\rangle_{0},\quad & \pi_{0}^{\mu\nu}=0,\quad h_{0}^{\mu}=\nu_{0}^{\mu}=0,
\end{eqnarray}
and, 
\begin{eqnarray}
\pi^{\mu\nu}=\langle p^{\langle\nu}p^{\mu\rangle}\rangle_{\delta},\quad & h^{\mu}=\langle E_{\mathbf{p}}p^{\langle\mu\rangle}\rangle_{\delta},\quad & \nu^{\mu}=\langle p^{\langle\mu\rangle}\rangle_{\delta},\label{eq:Viscous=000020current=000020from=000020kinetic=000020theory}
\end{eqnarray}
where 
\begin{equation}
\langle a\rangle_{0}=4\int_{p}2\pi\delta(p^{2})f_{\mathrm{V,leq}}^{<}(x,p)a,\quad\langle a\rangle_{\delta}=\langle a\rangle-\langle a\rangle_{0}.
\end{equation}
For convenience, we work in the Landau frame where the heat current
$h^{\mu}$ is zero, 
\begin{eqnarray}
u_{\mu}T^{\mu\nu} & = & \varepsilon u^{\nu},\quad\langle E_{\mathbf{p}}p^{\langle\mu\rangle}\rangle_{\delta}=0.\label{eq:Heat=000020flow=000020matching=000020condition}
\end{eqnarray}
Considering a fictitious equilibrium state described by 
\begin{equation}
\alpha_{0}=\alpha,\quad\beta_{0}=\beta,
\end{equation}
 and thus, and we thus have \citep{Israel:1979wp}, 
\begin{eqnarray}
\varepsilon=\varepsilon_{0}(\alpha_{0},\beta_{0}) & = & \langle E_{\mathbf{p}}^{2}\rangle_{0},\quad n=n_{0}(\alpha_{0},\beta_{0})=\langle E_{\mathbf{p}}\rangle_{0},\label{eq:e,n=000020matching=000020condition_origin}
\end{eqnarray}
which gives, 
\begin{eqnarray}
\langle E_{\mathbf{p}}\rangle_{\delta} & = & \langle E_{\mathbf{p}}^{2}\rangle_{\delta}=0.\label{eq:e,n=000020matching=000020condition}
\end{eqnarray}
Eqs. (\ref{eq:Heat=000020flow=000020matching=000020condition}, \ref{eq:e,n=000020matching=000020condition})
are the matching conditions in conventional relativistic hydrodynamics
\citep{denicol2022microscopic}. In the kinetic theory with quantum
corrections, we also need one extra matching condition to determine
hydrodynamic quantities from $f_{\mathrm{A}}$. In local equilibrium
with $\mu_{\mathrm{A}}=0$, we find the axial charge density $n_{\mathrm{A}}$
satisfies,
\begin{eqnarray}
n_{\mathrm{A},0}(x) & = & u_{\mu}\mathcal{J}_{5,\mathrm{leq}}^{\mu}(x)=4\int_{p}2\pi\delta(p^{2})E_{\mathbf{p}}f_{\mathrm{A,leq}}^{<}(x,p)=0,\label{eq:n5=000020leq}
\end{eqnarray}
which agrees with Refs. \citep{Gao:2012ix,Chen:2012ca} in the absence
of $\mu_{\mathrm{A}}$. The number of left and right handed fermions
must be conserved in the absence of external gauge fields. Following
the same arguments in Eqs. (\ref{eq:e,n=000020matching=000020condition_origin}),
it leads to
\begin{equation}
n_{\mathrm{A}}(x)=n_{\mathrm{A},0}(\alpha_{0},\beta_{0}),
\end{equation}
which provides us another matching condition, 
\begin{eqnarray}
\int_{p}2\pi\delta(p^{2})(f_{\mathrm{A}}^{<}-f_{\mathrm{A,leq}}^{<})E_{\mathbf{p}} & = & 0.\label{eq:Axial=000020matching=000020condition}
\end{eqnarray}
We regard Eq. (\ref{eq:Axial=000020matching=000020condition}) as
a result of the conservation of axial charge from the perspective
of QKT. Later, by using the property of the axial collision kernel,
we will demonstrate in Sec.\ref{subsec:A-revisiting-of} that the
matching condition in Eq.(\ref{eq:Axial=000020matching=000020condition})
represents the only possible additional matching condition that can
be derived from microscopic theory. 

Before ending this subsection, we emphasize that Eq.(\ref{eq:Axial=000020matching=000020condition})
is applicable only to massless particles in the absence of external
gauge fields. Generally, if the spin chemical potential is considered
as a dynamical quantity, an additional matching condition is required.
This condition can be established by defining the total angular momentum
density in local equilibrium, as discussed in Ref. \citep{Weickgenannt:2022zxs}. 

\section{SBE in Chapman-Enskog expansion }

\label{sec:Chapman-Enskog-expansion} In this section, we apply the
Chapman-Enskog expansion \citep{chapman1990mathematical,Kockel:1959auz,israel1963relativistic}
to QKT and derive the off-equilibrium corrections to the distribution
functions $\{\delta f_{\mathrm{V}},\delta f_{{\rm A}}\}$. The collision
kernels are calculated using the HTL approximations up to leading
logarithm explicitly. For $\delta f_{{\rm V}}$, our derivation adheres
to the standard procedure in the textbooks \citep{DeGroot:1980dk,cercignani2002relativistic,denicol2022microscopic}.
While, it is challenging to derive the off-equilibrium correction
to axial distribution, $\delta f_{{\rm A}}$, due to the complexity
of axial collision kernel in Eq.(\ref{eq:Moller=000020scattering=000020matrix=000020elements})
and Appendix. \ref{sec:Axial-collisional-kernels}. We consider two
scenarios as mentioned in Sec. \ref{subsec:Self-energies-in-2-2}
to simplify this derivation. We will revisit the matching condition
(\ref{eq:Axial=000020matching=000020condition}) for SBE and demonstrate
that $\delta f_{{\rm A}}$ must be of $\mathcal{O}(\partial^{2})$
in Sec.\ref{subsec:A-revisiting-of}. Subsequently, we solve the SBE
using the Chapman-Enskog expansion with $f_{{\rm V}}=f_{{\rm V,leq}}$
in Sec.\ref{subsec:The--correction}.A summary of this section is
presented in Sec.\ref{subsec:Collisional-corrections-to}.

\subsection{Revisited matching condition from SBE and simplification in scenario
(I) \protect\label{subsec:A-revisiting-of}}

Multiplying the both sides of SBE (\ref{eq:SBE=000020Moller=000020process})
by $2\pi\delta(p^{2})G_{p}$, where $G_{p}$ is an arbitrary function
of momentum, and integrating over momentum, yields
\begin{eqnarray}
\partial_{\mu}\int_{p}2\pi\delta(p^{2})G_{p}\mathcal{A}^{\mu}(x,p) & = & \int_{p}2\pi\delta(p^{2})G_{p}\mathcal{C}_{\mathrm{A}}.\label{eq:SBE_temp1}
\end{eqnarray}
Inserting Eq. (\ref{eq:Formal=000020sol.=000020V}) into above equation,
we get, 
\begin{eqnarray}
 &  & \int_{p}2\pi\delta(p^{2})G_{p}\mathcal{C}_{\mathrm{A}}\nonumber \\
 & = & 8e^{4}\int_{p,p^{\prime},k,k^{\prime}}\left[\frac{(G_{p}-G_{k}-G_{p^{\prime}}+G_{k^{\prime}})k_{\mu}^{\prime}(k\cdot p^{\prime})+(G_{p}+G_{k}-G_{p^{\prime}}-G_{k^{\prime}})k_{\mu}(p^{\prime}\cdot k^{\prime})}{(p-p^{\prime})^{4}}\right.\nonumber \\
 &  & \quad\qquad\left.+(G_{p}+G_{k}-G_{p^{\prime}}-G_{k^{\prime}})\frac{p^{\prime}\cdot k^{\prime}k_{\mu}}{(p-k^{\prime})^{2}(p-p^{\prime})^{2}}\right](2\pi)^{3}\delta(k^{\prime,2})\delta(k^{2})\delta(p^{\prime,2})\nonumber \\
 &  & \quad\times\left[f_{V}^{<}(k^{\prime})f_{V}^{>}(k)f_{V}^{<}(p^{\prime})\mathcal{A}^{>,\mu}(p)-f_{V}^{>}(k^{\prime})f_{V}^{<}(k)f_{V}^{>}(p^{\prime})\mathcal{A}^{<,\mu}(p)\right].\label{eq:GpCA=000020Integral}
\end{eqnarray}
We observe that the integral vanishes only when $G_{p}$ is a constant
since the $\delta^{(4)}(p+k-p^{\prime}-k^{\prime})$ in the integral
measure is not symmetric under $k\leftrightarrow k^{\prime}$. In
that case, Eq. (\ref{eq:GpCA=000020Integral}) becomes, 
\begin{eqnarray}
\partial_{\mu}\mathcal{J}_{5}^{\mu}(x) & = & 0,\label{eq:J5=000020conservation}
\end{eqnarray}
where 
\begin{equation}
\mathcal{J}_{5}^{\mu}(x)=\int_{p}2\pi\delta(p^{2})\mathcal{A}^{<,\mu}(x,p).
\end{equation}
is the axial vector current. It is consistent with the fact that the
axial current is conserved at both classical and quantum levels in
the absence of electromagnetic or color electromagnetic fields. Referring
to Eq.(\ref{eq:J5=000020conservation}), we need a matching condition
to define the axial charge density $n_{\mathrm{A}}$ in local equilibrium.
Note that this axial charge density has been set to zero in Eq.(\ref{eq:n5=000020leq}).

We now suppose $\delta f_{{\rm A}}$is of first order in gradients
and use the matching condition in Eq.(\ref{eq:Axial=000020matching=000020condition})
to show its vanish. Assuming $\mathcal{O}(\partial)\ll\mathcal{O}(Te^{4})\ll\mathcal{O}(T)$,
Eq. (\ref{eq:near_Leq=000020ACK=000020power=000020counting=000020scheme})
gives $\mathcal{C}_{\mathrm{A}}^{\Sigma}[\delta f_{\mathrm{V}}]\gg\mathcal{C}_{\mathrm{A}}^{\partial}[\delta f_{\mathrm{V}}]$.
Therefore, the collision kernel of SBE in Eq. (\ref{eq:SBE=000020Moller=000020process})
reduces to, 
\begin{eqnarray}
\mathcal{C}_{\mathrm{A}}^{\mathrm{B}}[\delta f_{\mathrm{A}},f_{\mathrm{V,leq}}]+\mathcal{C}_{\mathrm{A}}^{\Sigma}[\delta f_{\mathrm{V}}]+\mathcal{O}(e^{4}\partial^{2}) & = & 0,\label{eq:SBE=0000201st_scenario}
\end{eqnarray}
where we have dropped the left-hand side of Eq.(\ref{eq:SBE=000020Moller=000020process})
which is of $\mathcal{O}(\partial^{2})$ and used
\begin{eqnarray}
\mathcal{C}_{\mathrm{A}}^{\mathrm{B}}[\delta f_{\mathrm{A}},\delta f_{\mathrm{V}}] & = & \mathcal{C}_{\mathrm{A}}^{\mathrm{B}}[f_{\mathrm{A,leq}},\delta f_{\mathrm{V}}]+\mathcal{C}_{\mathrm{A}}^{\mathrm{B}}[\delta f_{\mathrm{A}},f_{\mathrm{V,leq}}]+\mathcal{O}(\delta f_{\mathrm{A}}\delta f_{\mathrm{V}})\nonumber \\
 & = & \mathcal{C}_{\mathrm{A}}^{\mathrm{B}}[\delta f_{\mathrm{A}},f_{\mathrm{V,leq}}]+\mathcal{O}(e^{4}\delta f_{{\rm V}}^{2}).\label{eq:ACK=000020Boltzmann_type}
\end{eqnarray}
The $\mathcal{C}_{\mathrm{A}}^{\mathrm{B}}[\delta f_{\mathrm{A}},f_{\mathrm{V,leq}}]$
is given by, up to the leading logarithm,
\begin{eqnarray}
 &  & \mathcal{C}_{\mathrm{A}}^{\mathrm{B}}[\delta f_{\mathrm{A}},f_{\mathrm{V,leq}}]\nonumber \\
 & = & \int_{p^{\prime},k,k^{\prime}}W_{pk\to p^{\prime}k^{\prime}}(2\pi)^{4}\delta(k^{\prime,2})\delta(k^{2})\delta(p^{\prime,2})\delta(p^{2})f_{\mathrm{V,leq}}^{<}(k^{\prime})f_{\mathrm{V,leq}}^{<}(p^{\prime})f_{\mathrm{V,leq}}^{>}(k)f_{\mathrm{V,leq}}^{>}(p)\nonumber \\
 &  & \qquad\qquad\times\left[\frac{\delta f_{\mathrm{A}}^{<}(p^{\prime})}{f_{\mathrm{V,leq}}^{<}(p^{\prime})f_{\mathrm{V,leq}}^{>}(p^{\prime})}-\frac{\delta f_{\mathrm{A}}^{<}(p)}{f_{\mathrm{V,leq}}^{<}(p)f_{\mathrm{V,leq}}^{>}(p)}\right],\label{eq:ACK=000020Boltmann_type=000020fV_leq}
\end{eqnarray}
with $W_{pk\to p^{\prime}k^{\prime}}=8e^{4}\frac{(p\cdot k^{\prime})^{2}+(p\cdot k)^{2}}{(p-p^{\prime})^{4}}=8e^{4}\frac{u^{2}+s^{2}}{4t^{2}}$
being the $t$-channel scattering matrix element. In Eq.(\ref{eq:ACK=000020Boltmann_type=000020fV_leq})
the symmetry of the scattering matrix element and the distribution
function, such as the invariance of replacing $p^{\prime}\leftrightarrow k^{\prime}$,
is broken explicitly. It indicates  that the SBE  (\ref{eq:SBE=000020Moller=000020process})
does not contain most symmetries of the vector Boltzmann equation
(\ref{eq:V=000020BE=000020Moller=000020process}). Nevertheless,
we still find that when $\frac{\delta f_{\mathrm{A}}^{<}(p)}{f_{\mathrm{V,leq}}^{<}(p)f_{\mathrm{V,leq}}^{>}(p)}\sim a$
with $a$ being a constant, Eq.(\ref{eq:ACK=000020Boltmann_type=000020fV_leq})
vanishes automatically. It indicates the conservation for number of
left and right handed particles in the gauge interactions. Eq. (\ref{eq:CA=000020Sigma})
gives,
\begin{eqnarray}
\mathcal{C}_{\mathrm{A}}^{\Sigma}[\delta f_{\mathrm{V}}] & = & 8e^{4}\int_{p^{\prime},k,k^{\prime}}w_{pk\to p^{\prime}k^{\prime},\mu}^{1}(2\pi)^{3}\delta(k^{\prime,2})\delta(k^{2})\delta(p^{\prime,2})\hbar S_{(u)}^{\mu\alpha}(p)\delta C_{\mathrm{V,\alpha}}[f_{\mathrm{V}}^{<}(p)]\nonumber \\
 &  & \qquad\qquad\times\left[f_{\mathrm{V,leq}}^{<}(k^{\prime})f_{\mathrm{V,leq}}^{>}(k)f_{\mathrm{V,leq}}^{<}(p^{\prime})+f_{\mathrm{V,leq}}^{>}(k^{\prime})f_{\mathrm{V,leq}}^{<}(k)f_{\mathrm{V,leq}}^{>}(p^{\prime})\right]\nonumber \\
 &  & +...\label{eq:CA=000020Sigma=00005Bdelta_fV=00005D}
\end{eqnarray}
where ... denotes other three similar terms. Using the angle projections
discussed in Appendix. \ref{sec:Collision-integrals} and expression
of $\delta f_{{\rm V}}$ in Eq. (\ref{eq:SE=000020kernel=000020HTL})
derived from vector Boltzmann equation, we find that $\mathcal{C}_{\mathrm{A}}^{\Sigma}[\delta f_{\mathrm{V}}]=0$.
Eq. (\ref{eq:SBE=0000201st_scenario}) now becomes,
\begin{eqnarray}
\mathcal{C}_{\mathrm{A}}^{\mathrm{B}}[\delta f_{\mathrm{A}},f_{\mathrm{V,leq}}]+\mathcal{O}(e^{4}\partial^{2}) & = & 0,
\end{eqnarray}
whose solution is, 
\begin{eqnarray}
\frac{\delta f_{\mathrm{A}}^{<}(p)}{f_{\mathrm{V,leq}}^{<}(p)f_{\mathrm{V,leq}}^{>}(p)} & = & a_{0}^{\mathrm{A}},
\end{eqnarray}
with $a_{0}^{\mathrm{A}}$ being a constant. The axial matching condition
in Eq. (\ref{eq:Axial=000020matching=000020condition}) gives $a_{0}^{\mathrm{A}}=0$.

To conclude, $f_{\mathrm{A}}$ does not receive any off-equilibrium
correction up to $\mathcal{O}(\partial^{1})$, and its off-equilibrium
part $\delta f_{A}$ is at least of $\mathcal{O}(\partial^{2})$.
In the scenario (I) , we find that $\delta\mathcal{A}_{\textrm{(I)}}^{<,\mu}(X,p)$
in Eq. (\ref{eq:Off-eq-A-Case1}) reduces to
\begin{eqnarray}
\delta\mathcal{A}_{\textrm{(I)}}^{<,\mu}(X,p) & = & -2\pi\hbar\delta(p^{2})S^{(u),\mu\alpha}C_{\mathrm{V,}\alpha}[\delta f_{\mathrm{V}}^{<}]+\mathcal{O}(\partial^{2}).\label{eq:delta_A_sI_01}
\end{eqnarray}
Let us comment on the above results. The vanishing $\mathcal{C}_{\mathrm{A}}^{\Sigma}[\delta f_{\mathrm{V}}]$
and $\delta f_{{\rm A}}$ up to $\mathcal{O}(\partial^{1})$ strongly
depends on the explicit expression of scattering matrix element. Generally,
from the point view of effective theory, $\mathcal{C}_{\mathrm{A}}^{\Sigma}[\delta f_{\mathrm{V}}]$
is a linear combination of $\nabla_{\nu}\alpha_{0}$ and $\sigma_{\mu\nu}$,
with the coefficients that can be functions of $p_{\perp,\mu},u_{\mu},E_{\mathbf{p}}$
and other variables. In different scattering processes or interactions,
one may encounter additional terms, potentially leading to extra contributions
from collisions to spin polarization up to $\mathcal{O}(\partial^{1})$.
In Refs. \citep{Weickgenannt:2022zxs,Weickgenannt:2022qvh}, the shear-induced
polarization, of $\mathcal{O}(\partial^{1})$, is shown to possess
an interaction-dependent coefficient. We leave a systematic discussion
of other types of interactions for future work.

\subsection{Corrections to $\delta f_{\mathrm{A}}$ up to $\mathcal{O}(\partial^{2})$
in scenario (II) \protect\label{subsec:The--correction}}

Now we focus on the scenario (II) with Eq. (\ref{eq:Off-eq-A-Case2})
and set $f_{\mathrm{V}}=f_{\mathrm{V,leq}}$ in this subsection. In
this case, the axial collision kernel of SBE in Eq. (\ref{eq:SBE=000020Moller=000020process})
reduces to $\mathcal{C}_{\mathrm{A}}^{\mathrm{B}}[\delta f_{\mathrm{A}},f_{\mathrm{V,leq}}]$
in Eq. (\ref{eq:ACK=000020Boltmann_type=000020fV_leq}). 

We follow the procedure in Ref. \citep{denicol2022microscopic} to
derive $\delta f_{\mathrm{A}}$ using the the Chapman-Enskog expansion.
We expand the axial distribution function and the axial collision
kernel in Eq. (\ref{eq:ACK=000020Boltmann_type=000020fV_leq}) in
the power series of $\mathrm{Kn}$,
\begin{eqnarray}
f_{\mathrm{A}} & = & f_{\mathrm{A,leq}}+\delta f_{\mathrm{A}}=\mathrm{Kn}f_{\mathrm{A}}^{(1)}+\mathrm{Kn}^{2}f_{\mathrm{A}}^{(2)}+\mathcal{O}(\mathrm{Kn}^{3}),
\end{eqnarray}
and
\begin{eqnarray}
\mathcal{C}_{\mathrm{A}}[\delta f_{\mathrm{A}},f_{\mathrm{V,leq}}] & = & \mathrm{Kn}\mathcal{C}_{\mathrm{A}}^{(1)}[\delta f_{\mathrm{A}},f_{\mathrm{V,leq}}]+\mathrm{Kn}^{2}\mathcal{C}_{\mathrm{A}}^{(2)}[\delta f_{\mathrm{A}},f_{\mathrm{V,leq}}]+\mathcal{O}(\mathrm{Kn}^{3})\nonumber \\
 & = & \int_{p^{\prime},k,k^{\prime}}W_{pk\to p^{\prime}k^{\prime}}(2\pi)^{4}\delta(k^{\prime,2})\delta(k^{2})\delta(p^{\prime,2})\delta(p^{2})f_{\mathrm{V,leq}}^{<}(k^{\prime})f_{\mathrm{V,leq}}^{<}(p^{\prime})f_{\mathrm{V,leq}}^{>}(k)f_{\mathrm{V,leq}}^{>}(p)\nonumber \\
 &  & \qquad\qquad\times\left[\frac{\delta f_{\mathrm{A}}^{<}(p^{\prime})}{f_{\mathrm{V,leq}}^{<}(p^{\prime})f_{\mathrm{V,leq}}^{>}(p^{\prime})}-\frac{\delta f_{\mathrm{A}}^{<}(p)}{f_{\mathrm{V,leq}}^{<}(p)f_{\mathrm{V,leq}}^{>}(p)}\right].
\end{eqnarray}
Note that $f_{\mathrm{A,leq}}=\mathrm{Kn}f_{\mathrm{A}}^{(1)}$. By
matching the orders of the Knudsen number $\mathrm{Kn}=\lambda L^{-1}$
on both sides of Eq. (\ref{eq:SBE=000020Moller=000020process}), we
get,
\begin{eqnarray}
\mathcal{C}_{\mathrm{A}}^{(1)}[\delta f_{\mathrm{A}},f_{\mathrm{V,leq}}] & = & 0,\label{eq:SBE_gradient=0000201st_order}\\
\hbar\frac{\hat{\partial}_{\mu}S_{(u)}^{\mu\alpha}(p)}{E_{\mathbf{p}}}u_{\alpha}\left(\hat{D}f_{\mathrm{V}}^{<}(p)\right)^{(1)}+\hbar\frac{\hat{\partial}_{\mu}S_{(u)}^{\mu\alpha}(p)}{E_{\mathbf{p}}}\hat{\nabla}_{\alpha}f_{\mathrm{V}}^{<,(0)}(p)\nonumber \\
+L\left(\hat{D}f_{\mathrm{A}}^{<}(p)\right)^{(2)}+L\frac{p_{\perp}^{\mu}}{E_{\mathbf{p}}}\hat{\nabla}_{\mu}f_{\mathrm{A}}^{<,(1)} & = & \frac{\lambda^{2}}{E_{\mathbf{p}}}\mathcal{C}_{\mathrm{A}}^{(2)}[\delta f_{\mathrm{A}},f_{\mathrm{V,leq}}],\label{eq:SBE_gradient=0000202st_order}
\end{eqnarray}
where we denote 
\begin{equation}
\hat{\partial}_{\mu}=L\partial_{\mu},
\end{equation}
with $L$ being the characteristic macroscopic length scale of system
and $\lambda$ being particles' mean free path. Eq.(\ref{eq:SBE_gradient=0000201st_order})
is consistent with result of detailed balance for the spin evolution,
i.e. $\mathcal{C}_{\mathrm{A}}[f_{\mathrm{V,leq}},f_{\mathrm{A,leq}}]=0$.
We can then derive $\delta f_{\mathrm{A}}$ by solving Eq. (\ref{eq:SBE_gradient=0000202st_order}).
We notice that the $u\cdot\partial$ contains higher order terms in
gradients as detailed in Appendix \ref{subsec:Thermodynamical-functions-and}
and needs to be separated out. For example, we have $p\cdot\hat{\partial}f_{\mathrm{V}}=E_{\mathbf{p}}\hat{D}f_{\mathrm{V}}+p^{\mu}\hat{\nabla}_{\mu}f_{\mathrm{V}}$,
where we introduce 
\begin{equation}
\hat{D}=L^{-1}D\equiv L^{-1}u\cdot\hat{\partial},\quad\hat{\nabla}_{\mu}=L^{-1}\nabla_{\mu}\equiv L^{-1}\Delta_{\mu\alpha}\hat{\partial}^{\alpha}.
\end{equation}
Furthermore, we need to replace $\hat{D}f_{\mathrm{V}}^{(n-1)}$ by
$(\hat{D}f_{\mathrm{V}})^{(n)}$ when we extract the terms of $\mathcal{O}(\mathrm{Kn}^{n})$.

Following a lengthy yet straightforward calculation, we obtain the
left hand side of Eq.(\ref{eq:SBE_gradient=0000202st_order}),
\begin{eqnarray}
 &  & L\left(\hat{D}f_{A}^{<}(p)+\frac{p_{\perp}^{\mu}}{E_{\mathbf{p}}}\hat{\partial}_{\mu}f_{A}^{<}(p)\right)^{(2)}+\hbar\left(\frac{\hat{\partial}_{\mu}S_{(u)}^{\mu\alpha}(p)}{E_{\mathbf{p}}}\hat{\partial}_{\alpha}(\beta_{0}u\cdot p-\alpha_{0})\right)^{(2)}(-f_{V,\mathrm{leq}}^{<}f_{\mathrm{V,leq}}^{>})\nonumber \\
 & = & -\frac{\hbar}{2}f_{\mathrm{V,leq}}^{<}(p)f_{\mathrm{V,leq}}^{>}(p)\left[A_{\mathbf{p}}+B_{\mathbf{p}}^{\alpha}p_{\langle\alpha\rangle}+C_{\mathbf{p}}^{\alpha\rho}p_{\langle\alpha}p_{\rho\rangle}+D_{\mathbf{p}}^{\mu\alpha\lambda}p_{\langle\mu}p_{\alpha}p_{\lambda\rangle}\right],\label{eq:l.h.s.=000020SBE=000020in=000020CE_expansion}
\end{eqnarray}
where 
\begin{eqnarray}
A_{\mathbf{p}} & = & \left[\frac{2}{3}h_{0}^{-1}-\frac{4}{3}E_{\mathbf{p}}^{-1}+\frac{\beta_{0}}{3}(1-E_{\mathbf{p}}h_{0}^{-1})\left(1-2f_{\mathrm{V,leq}}^{>}(p)\right)\right]\hat{\omega}^{\alpha}\hat{\nabla}_{\alpha}\alpha_{0}\nonumber \\
 &  & \quad+\frac{1}{3}\hat{\omega}^{\alpha}\hat{\nabla}_{\alpha}\beta_{0}-\frac{1}{3}\beta_{0}\hat{\nabla}^{\rho}\hat{\omega}_{\rho},\label{eq:Ap}\\
B_{\mathbf{p}}^{\alpha} & = & \left[\frac{12}{5}\beta_{0}E_{\mathbf{p}}^{-1}-\frac{2}{5}\beta_{0}^{2}\left(1-2f_{\mathrm{V,leq}}^{>}(p)\right)\right]\hat{\omega}^{\mu}\hat{\sigma}_{\mu}^{\alpha}\nonumber \\
 &  & \quad+\frac{1}{\beta_{0}}\frac{\epsilon^{\mu\nu\alpha\beta}u_{\beta}\hat{\nabla}_{\nu}\alpha_{0}\hat{\nabla}_{\mu}\beta_{0}}{E_{\mathbf{p}}}\left(-E_{\mathbf{p}}^{-1}+h_{0}^{-1}\right),\label{eq:Bp}\\
C_{\mathbf{p}}^{\alpha\rho} & = & 2\frac{\hat{\omega}^{\langle\alpha}\hat{\nabla}^{\rho\rangle}\beta_{0}}{E_{\mathbf{p}}^{2}}+\frac{\beta_{0}\hat{\nabla}^{\langle\alpha}\hat{\omega}^{\rho\rangle}}{E_{\mathbf{p}}^{2}}+\frac{2\epsilon^{\mu\nu\sigma\langle\rho}\hat{\sigma}_{\mu}^{\alpha\rangle}u_{\sigma}\hat{\nabla}_{\nu}\beta_{0}}{E_{\mathbf{p}}^{2}}-\frac{\epsilon^{\mu\nu\sigma\langle\rho}\hat{\sigma}_{\mu}^{\alpha\rangle}u_{\sigma}\hat{\nabla}_{\nu}\alpha_{0}}{E_{\mathbf{p}}^{2}}(h_{0}^{-1}+E_{\mathbf{p}}^{-1})\nonumber \\
 &  & \quad+\left[E_{\mathbf{p}}^{-3}-2E_{\mathbf{p}}^{-2}h_{0}^{-1}+\beta_{0}E_{\mathbf{p}}^{-1}(h_{0}^{-1}-E_{\mathbf{p}}^{-1})\left(1-2f_{\mathrm{V,leq}}^{>}(p)\right)\right]\hat{\omega}^{\langle\alpha}\hat{\nabla}^{\rho\rangle}\alpha_{0},\label{eq:Cp}\\
D_{\mathbf{p}}^{\mu\alpha\lambda} & = & \left[\left(1-2f_{\mathrm{V,leq}}^{>}(p)\right)\beta_{0}^{2}-E_{\mathbf{p}}^{-1}\beta_{0}\right]\frac{\hat{\omega}^{\langle\mu}\hat{\sigma}^{\alpha\lambda\rangle}}{E_{\mathbf{p}}^{2}}.\label{eq:Dp}
\end{eqnarray}
From Eq.(\ref{eq:SBE_gradient=0000202st_order}), we assume the expression
of $f_{\mathrm{A}}^{(2)}$ takes the following form, 
\begin{eqnarray}
f_{\mathrm{A}}^{<,(2)}(p) & = & -\frac{\hbar}{2}f_{\mathrm{V,leq}}^{<}(p)f_{\mathrm{V,leq}}^{>}(p)\Bigg\{\varphi_{\mathrm{A},\mathbf{p}}^{\mathrm{s,1}}\frac{1}{\beta_{0}}\hat{\omega}^{\alpha}\hat{\nabla}_{\alpha}\beta_{0}+\varphi_{\mathrm{A},\mathbf{p}}^{\mathrm{s,2}}\hat{\nabla}^{\alpha}\hat{\omega}_{\alpha}+\varphi_{\mathrm{A},\mathbf{p}}^{\mathrm{s,3}}\hat{\omega}^{\alpha}\hat{\nabla}_{\alpha}\alpha_{0}\nonumber \\
 &  & \quad+\left[\varphi_{\mathrm{A},\mathbf{p}}^{\mathrm{v,1}}\beta_{0}\hat{\omega}^{\mu}\hat{\sigma}_{\mu}^{\alpha}+\varphi_{\mathrm{A},\mathbf{p}}^{\mathrm{v,2}}\epsilon^{\mu\nu\alpha\beta}u_{\beta}\hat{\nabla}_{\nu}\alpha_{0}\hat{\nabla}_{\mu}\beta_{0}\right]p_{\langle\alpha\rangle}\nonumber \\
 &  & \quad+\left[\varphi_{\mathrm{A},\mathbf{p}}^{\mathrm{ts},1}\beta_{0}\hat{\omega}^{\langle\alpha}\hat{\nabla}^{\rho\rangle}\beta_{0}+\varphi_{\mathrm{A},\mathbf{p}}^{\mathrm{ts},2}\beta_{0}^{2}\hat{\nabla}^{\langle\alpha}\hat{\omega}^{\rho\rangle}+\varphi_{\mathrm{A},\mathbf{p}}^{\mathrm{ts},3}\beta_{0}\epsilon^{\mu\nu\sigma\langle\rho}\hat{\sigma}_{\mu}^{\alpha\rangle}u_{\sigma}\hat{\nabla}_{\nu}\beta_{0}\right.\nonumber \\
 &  & \quad\qquad\left.+\varphi_{\mathrm{A},\mathbf{p}}^{\mathrm{ts},4}\epsilon^{\mu\nu\sigma\langle\rho}\hat{\sigma}_{\mu}^{\alpha\rangle}u_{\sigma}\hat{\nabla}_{\nu}\alpha_{0}+\varphi_{\mathrm{A},\mathbf{p}}^{\mathrm{ts},5}\beta_{0}^{2}\hat{\omega}^{\langle\alpha}\hat{\nabla}^{\rho\rangle}\alpha_{0}\right]p_{\langle\alpha}p_{\rho\rangle}\nonumber \\
 &  & \quad+\varphi_{\mathrm{A},\mathbf{p}}^{\mathrm{tt}}\beta_{0}^{3}\hat{\omega}^{\langle\mu}\hat{\sigma}^{\alpha\lambda\rangle}p_{\langle\mu}p_{\alpha}p_{\lambda\rangle}+a_{1}^{\mathrm{A}}\Bigg\},\label{eq:delta=00007BfA=00007D=000020CE_Assumption}
\end{eqnarray}
where $\varphi_{\mathrm{A},\mathbf{p}}^{i}$ are the dimensionless
functions of $E_{\mathbf{p}}$ and constant $a_{1}^{\mathrm{A}}$
needs to be determined by the axial matching condition in Eq. (\ref{eq:Axial=000020matching=000020condition}).
Inserting Eq.(\ref{eq:delta=00007BfA=00007D=000020CE_Assumption}),
the axial matching condition in Eq. (\ref{eq:Axial=000020matching=000020condition})
becomes, 
\begin{eqnarray}
4\int_{p}2\pi\delta(p^{2})f_{\mathrm{V,leq}}^{<}(p)f_{\mathrm{V,leq}}^{>}(p)\varphi_{\mathrm{A},\mathbf{p}}^{\mathrm{s,}i} & = & 0,\quad(i=1,2,3)\label{eq:Axial=000020MC_phi_si}\\
a_{1}^{\mathrm{A}} & = & 0.\label{eq:Axial=000020MC_a1}
\end{eqnarray}
where we have used the fact that $\hat{\omega}^{\alpha}\hat{\nabla}_{\alpha}\beta_{0},\hat{\nabla}^{\alpha}\hat{\omega}_{\alpha},\hat{\omega}^{\alpha}\hat{\nabla}_{\alpha}\alpha_{0}$
are independent sources. Using Eq.(\ref{eq:delta=00007BfA=00007D=000020CE_Assumption}),
Eq.(\ref{eq:SBE_gradient=0000202st_order}) becomes,
\begin{eqnarray}
 &  & -f_{\mathrm{V,leq}}^{<}(p)f_{\mathrm{V,leq}}^{>}(p)\left[A_{\mathbf{p}}+B_{\mathbf{p}}^{\alpha}p_{\langle\alpha\rangle}+C_{\mathbf{p}}^{\alpha\rho}p_{\langle\alpha}p_{\rho\rangle}+D_{\mathbf{p}}^{\mu\alpha\lambda}p_{\langle\mu}p_{\alpha}p_{\lambda\rangle}\right]\nonumber \\
 & = & \int_{p^{\prime},k,k^{\prime}}W_{pk\to p^{\prime}k^{\prime}}(2\pi)^{4}\delta(k^{\prime,2})\delta(k^{2})\delta(p^{\prime,2})\delta(p^{2})f_{\mathrm{V,leq}}^{<}(k^{\prime})f_{\mathrm{V,leq}}^{<}(p^{\prime})f_{\mathrm{V,leq}}^{>}(k)f_{\mathrm{V,leq}}^{>}(p)\nonumber \\
 &  & \quad\times\Bigg\{(\varphi_{\mathrm{A},\mathbf{p}}^{\mathrm{s,1}}-\varphi_{\mathrm{A},\mathbf{p}^{\prime}}^{\mathrm{s,1}})\frac{1}{\beta_{0}}\hat{\omega}^{\alpha}\hat{\nabla}_{\alpha}\beta_{0}+(\varphi_{\mathrm{A},\mathbf{p}}^{\mathrm{s,2}}-\varphi_{\mathrm{A},\mathbf{p}^{\prime}}^{\mathrm{s,2}})\hat{\nabla}^{\alpha}\hat{\omega}_{\alpha}+(\varphi_{\mathrm{A},\mathbf{p}}^{\mathrm{s,3}}-\varphi_{\mathrm{A},\mathbf{p}^{\prime}}^{\mathrm{s,3}})\hat{\omega}^{\alpha}\hat{\nabla}_{\alpha}\alpha_{0}\nonumber \\
 &  & \quad\qquad+(\varphi_{\mathrm{A},\mathbf{p}}^{\mathrm{v,1}}p_{\langle\alpha\rangle}-\varphi_{\mathrm{A},\mathbf{p}^{\prime}}^{\mathrm{v,1}}p_{\langle\alpha\rangle}^{\prime})\beta_{0}\hat{\omega}^{\mu}\hat{\sigma}_{\mu}^{\alpha}+(\varphi_{\mathrm{A},\mathbf{p}}^{\mathrm{v,2}}p_{\langle\alpha\rangle}-\varphi_{\mathrm{A},\mathbf{p}^{\prime}}^{\mathrm{v,2}}p_{\langle\alpha\rangle}^{\prime})\epsilon^{\mu\nu\alpha\beta}u_{\beta}\hat{\nabla}_{\nu}\alpha_{0}\hat{\nabla}_{\mu}\beta_{0}\nonumber \\
 &  & \quad\qquad+(\varphi_{\mathrm{A},\mathbf{p}}^{\mathrm{ts},1}p_{\langle\alpha}p_{\rho\rangle}-\varphi_{\mathrm{A},\mathbf{p}^{\prime}}^{\mathrm{ts},1}p_{\langle\alpha}^{\prime}p_{\rho\rangle}^{\prime})\beta_{0}\hat{\omega}^{\langle\alpha}\hat{\nabla}^{\rho\rangle}\beta_{0}+(\varphi_{\mathrm{A},\mathbf{p}}^{\mathrm{ts},2}p_{\langle\alpha}p_{\rho\rangle}-\varphi_{\mathrm{A},\mathbf{p}^{\prime}}^{\mathrm{ts},2}p_{\langle\alpha}^{\prime}p_{\rho\rangle}^{\prime})\beta_{0}^{2}\hat{\nabla}^{\langle\alpha}\hat{\omega}^{\rho\rangle}\nonumber \\
 &  & \quad\qquad+(\varphi_{\mathrm{A},\mathbf{p}}^{\mathrm{ts},3}p_{\langle\alpha}p_{\rho\rangle}-\varphi_{\mathrm{A},\mathbf{p}^{\prime}}^{\mathrm{ts},3}p_{\langle\alpha}^{\prime}p_{\rho\rangle}^{\prime})\beta_{0}\epsilon^{\mu\nu\sigma\langle\rho}\hat{\sigma}_{\mu}^{\alpha\rangle}u_{\sigma}\hat{\nabla}_{\nu}\beta_{0}\nonumber \\
 &  & \quad\qquad+(\varphi_{\mathrm{A},\mathbf{p}}^{\mathrm{ts},4}p_{\langle\alpha}p_{\rho\rangle}-\varphi_{\mathrm{A},\mathbf{p}^{\prime}}^{\mathrm{ts},4}p_{\langle\alpha}^{\prime}p_{\rho\rangle}^{\prime})\epsilon^{\mu\nu\sigma\langle\rho}\hat{\sigma}_{\mu}^{\alpha\rangle}u_{\sigma}\hat{\nabla}_{\nu}\alpha_{0}+(\varphi_{\mathrm{A},\mathbf{p}}^{\mathrm{ts},5}p_{\langle\alpha}p_{\rho\rangle}-\varphi_{\mathrm{A},\mathbf{p}^{\prime}}^{\mathrm{ts},5}p_{\langle\alpha}^{\prime}p_{\rho\rangle}^{\prime})\beta_{0}^{2}\hat{\omega}^{\langle\alpha}\hat{\nabla}^{\rho\rangle}\alpha_{0}\nonumber \\
 &  & \quad\qquad+(\varphi_{\mathrm{A},\mathbf{p}}^{\mathrm{tt}}p_{\langle\mu}p_{\alpha}p_{\lambda\rangle}-\varphi_{\mathrm{A},\mathbf{p}^{\prime}}^{\mathrm{tt}}p_{\langle\mu}^{\prime}p_{\alpha}^{\prime}p_{\lambda\rangle}^{\prime})\beta_{0}^{3}\hat{\omega}^{\langle\mu}\hat{\sigma}^{\alpha\lambda\rangle}\Bigg\}.\label{eq:Intermediate_CE_SBE}
\end{eqnarray}
Following Ref. \citep{denicol2022microscopic}, we need to solve $\varphi_{\mathrm{A},\mathbf{p}}^{i}$
from the functional equation in Eq. (\ref{eq:Intermediate_CE_SBE}).
We first multiply both sides of Eq. (\ref{eq:Intermediate_CE_SBE})
by $2\pi\delta(p^{2})\xi^{(r)}(E_{\mathbf{p}})$, where $\xi^{(r)}(E_{\mathbf{p}})$
is an arbitrary function of $E_{\mathbf{p}}$, and irreducible basis
$p^{\langle\mu_{1}}...p^{\mu_{m}\rangle}$. Subsequently, we integrate
both sides of Eq. (\ref{eq:Intermediate_CE_SBE}) over momentum. Our
objective is to derive distinct, disentangled equations for $\varphi_{\mathrm{A},\mathbf{p}}^{i}$
by utilizing the following orthogonal relations \citep{Denicol:2012cn,denicol2022microscopic},
\begin{eqnarray}
 &  & \int_{p,k,k^{\prime},p^{\prime}}(2\pi)^{4}\delta(k^{2})\delta(p^{\prime2})\delta(k^{\prime2})\delta(p^{2})\mathcal{W}_{pk\to p^{\prime}k^{\prime}}f_{\mathrm{V,leq}}^{>}(k)f_{\mathrm{V,leq}}^{>}(p)f_{\mathrm{V,leq}}^{<}(k^{\prime})f_{\mathrm{V,leq}}^{<}(p^{\prime})\nonumber \\
 &  & \times E_{\mathbf{p}}^{r-1}p^{\langle\mu_{1}}...p^{\mu_{m}\rangle}\left(-\mathrm{H}_{\mathbf{p}}p_{\langle\nu_{1}...}p_{\nu_{n}\rangle}-\mathrm{H}_{\mathbf{k}}k_{\langle\nu_{1}...}k_{\nu_{n}\rangle}+\mathrm{H}_{\mathbf{p}^{\prime}}p_{\langle\nu_{1}...}^{\prime}p_{\nu_{n}\rangle}^{\prime}+\mathrm{H}_{\mathbf{k}^{\prime}}k_{\langle\nu_{1}...}^{\prime}k_{\nu_{n}\rangle}^{\prime}\right)\nonumber \\
 & = & \Delta_{\nu_{1}...\nu_{n}}^{\mu_{1}...\mu_{m}}\frac{\delta_{mn}}{2n+1}\int_{k,k^{\prime},p,p^{\prime}}(2\pi)^{4}\delta(k^{2})\delta(p^{\prime2})\delta(k^{\prime2})\delta(p^{2})\mathcal{W}_{pk\to p^{\prime}k^{\prime}}f_{\mathrm{V,leq}}^{>}(k)f_{\mathrm{V,leq}}^{>}(p)f_{\mathrm{V,leq}}^{<}(k^{\prime})f_{\mathrm{V,leq}}^{<}(p^{\prime})\nonumber \\
 &  & \times E_{\mathbf{p}}^{r-1}p^{\langle\alpha_{1}}...p^{\alpha_{n}\rangle}\left(-\mathrm{H}_{\mathbf{p}}p_{\langle\alpha_{1}...}p_{\alpha_{n}\rangle}-\mathrm{H}_{\mathbf{k}}k_{\langle\alpha_{1}...}k_{\alpha_{n}\rangle}+\mathrm{H}_{\mathbf{p}^{\prime}}p_{\langle\alpha_{1}...}^{\prime}p_{\alpha_{n}\rangle}^{\prime}+\mathrm{H}_{\mathbf{k}^{\prime}}k_{\langle\alpha_{1}...}^{\prime}k_{\alpha_{n}\rangle}^{\prime}\right),\label{eq:moment_Orthogonal=000020relation_1}
\end{eqnarray}
and
\begin{eqnarray}
\int_{p}\mathrm{H}(E_{\mathbf{p}})p^{\langle\mu_{1}}...p^{\mu_{m}\rangle}p_{\langle\nu_{1}}...p_{\nu_{n}\rangle} & = & \frac{n!\delta_{mn}}{(2n+1)!!}\Delta_{\nu_{1}...\nu_{n}}^{\mu_{1}...\mu_{m}}\int_{p}\mathrm{H}(E_{\mathbf{p}})(-E_{\mathbf{p}}^{2})^{n}.\label{eq:moment_Orthogonal=000020relation_2}
\end{eqnarray}
Note that the $\xi^{(r)}(E_{\mathbf{p}})$ is a rotation-invariant
function. We can also opt to set $\xi^{i,(r)}(E_{\mathbf{p}})=\varphi_{\mathrm{A},\mathbf{p}}^{i}$
and apply the variational method numerically to solve differential
equations for $\varphi$ as discussed in Refs. \citep{Arnold:2000dr,Arnold:2002zm,Arnold:2003zc}.
Alternatively, following the approach in Ref. \citep{Arnold:2000dr},
we adopt a different method by using a set of basis functions. Here,
we choose $\xi^{(r)}(E_{\mathbf{p}})$ as these basis functions, allowing
$\varphi_{\mathrm{A},\mathbf{p}}^{i}$ to be expanded using $\xi^{(r)}(E_{\mathbf{p}})$,
\begin{eqnarray}
\varphi_{\mathrm{A},\mathbf{p}}^{i} & = & \sum_{n=0}^{N_{\mathrm{A}}^{i}}\varepsilon_{\mathrm{A},n}^{i}\xi_{\mathbf{p}}^{(n)}(E_{\mathbf{p}}).\label{eq:Expansion_phi_Ap}
\end{eqnarray}
Here, $\varepsilon_{\mathrm{A},n}^{i}$ represents thermodynamical
functions, where $i=\mathrm{s},\mathrm{v},\mathrm{ts},\mathrm{tt}$
denotes scalar, vector, second rank tensor, and third rank tensor,
respectively. By explicitly calculating the collision kernel, we can
determine all the coefficients $\varepsilon_{\mathrm{V,n}}^{i}$ by
appropriately truncating the functional space and inverting the collisional
integral matrix.

In the current work, we adopt $\xi^{(r)}(E_{\mathbf{p}})=E_{\mathbf{p}}^{r}$
\citep{Arnold:2000dr}, which has been demonstrated to behave well
in classical limit within the conventional kinetic theory \citep{denicol2022microscopic}.
Consequently, Eq.(\ref{eq:Expansion_phi_Ap}) simplifies as follows,
\begin{eqnarray}
\varphi_{\mathrm{A},\mathbf{p}}^{i} & = & \sum_{n=0}^{N_{\mathrm{A}}^{i}}\varepsilon_{\mathrm{A},n}^{i}E_{\mathbf{p}}^{n}.\label{eq:Expansion_scheme:=000020phi_Ap}
\end{eqnarray}
Using Eqs.(\ref{eq:moment_Orthogonal=000020relation_1}, \ref{eq:moment_Orthogonal=000020relation_2}),
we multiply both sides of Eq.(\ref{eq:Intermediate_CE_SBE}) by $2\pi\delta(p^{2})\mathcal{B}$,
where $\mathcal{B}$ represents the set $\{E_{\mathbf{p}}^{r}$, $E_{\mathbf{p}}^{r}p^{\langle\alpha\rangle}$, $E_{\mathbf{p}}^{r}p^{\langle\alpha}p^{\beta\rangle}$, $E_{\mathbf{p}}^{r}p^{\langle\alpha}p^{\beta\rangle}$, $E_{\mathbf{p}}^{r}p^{\langle\alpha}p^{\beta}p^{\gamma\rangle}\}$.
After integrating over momentum $p$, Eq.(\ref{eq:Intermediate_CE_SBE})
is reduced to the following arrays of linear equations, 
\begin{eqnarray}
\sum_{n=0}^{N_{\mathrm{A}}^{\mathrm{s},i}}\mathcal{A}_{\mathrm{A},rn}^{\mathrm{s}}\varepsilon_{\mathrm{A},n}^{\mathrm{s},i}=\alpha_{\mathrm{A},r}^{\mathrm{s},i} & ,\quad & \sum_{n=0}^{N_{\mathrm{A}}^{\mathrm{v},j}}\mathcal{A}_{\mathrm{A},rn}^{\mathrm{v}}\varepsilon_{\mathrm{A},n}^{\mathrm{v},j}=\alpha_{\mathrm{A},r}^{\mathrm{v},j},\label{eq:epsilon_A:=000020eqs1}\\
\sum_{n=0}^{N_{\mathrm{A}}^{\mathrm{ts},i}}\mathcal{A}_{\mathrm{A},rn}^{\mathrm{ts}}\varepsilon_{\mathrm{A},n}^{\mathrm{ts},k}=\alpha_{\mathrm{A},r}^{\mathrm{ts},k} & ,\quad & \sum_{n=0}^{N_{\mathrm{A}}^{\mathrm{tt}}}\mathcal{A}_{\mathrm{A},rn}^{\mathrm{tt}}\varepsilon_{\mathrm{A},n}^{\mathrm{tt}}=\alpha_{\mathrm{A},r}^{\mathrm{tt}},\label{eq:epsilon_A:=000020eqs2}
\end{eqnarray}
with $i=1,2,3$, $j=1,2$ and $k=1,...,5$. The expression of axial
thermodynamic functions $\alpha_{\mathrm{A},r}^{i}$ are presented
in Appendix. \ref{subsec:Definitions-of-thermodynamical}. In the
aforementioned equation, the collision integrals $\mathcal{A}_{\mathrm{A},rn}^{i}$
are independent of thermodynamic sources, such as $\hat{\omega}^{\alpha}\hat{\nabla}_{\alpha}\beta_{0}$,
\begin{eqnarray}
\mathcal{A}_{\mathrm{A},rn}^{\mathrm{s}} & = & \lambda^{2}\int_{p,p^{\prime},k,k^{\prime}}W_{pk\to p^{\prime}k^{\prime}}(2\pi)^{4}\delta(p^{2})\delta(k^{\prime,2})\delta(k^{2})\delta(p^{\prime,2})\nonumber \\
 &  & \quad\times f_{\mathrm{V,leq}}^{<}(k^{\prime})f_{\mathrm{V,leq}}^{<}(p^{\prime})f_{\mathrm{V,leq}}^{>}(k)f_{\mathrm{V,leq}}^{>}(p)E_{\mathbf{p}}^{r-1}(E_{\mathbf{p}}^{n}-E_{\mathbf{p}^{\prime}}^{n}),\label{eq:Axial=000020collision=000020int=0000201}\\
\mathcal{A}_{\mathrm{A},rn}^{\mathrm{v}} & = & \frac{\lambda^{2}}{3}\int_{p,p^{\prime},k,k^{\prime}}W_{pk\to p^{\prime}k^{\prime}}(2\pi)^{4}\delta(p^{2})\delta(k^{\prime,2})\delta(k^{2})\delta(p^{\prime,2})\nonumber \\
 &  & \quad\times f_{\mathrm{V,leq}}^{<}(k^{\prime})f_{\mathrm{V,leq}}^{<}(p^{\prime})f_{\mathrm{V,leq}}^{>}(k)f_{\mathrm{V,leq}}^{>}(p)E_{\mathbf{p}}^{r-1}p^{\langle\beta\rangle}(E_{\mathbf{p}}^{n}p_{\langle\beta\rangle}-E_{\mathbf{p}^{\prime}}^{n}p_{\langle\beta\rangle}^{\prime}),\label{eq:Axial=000020collision=000020int=0000202}\\
\mathcal{A}_{\mathrm{A},rn}^{\mathrm{ts}} & = & \frac{\lambda^{2}}{5}\int_{p,p^{\prime},k,k^{\prime}}W_{pk\to p^{\prime}k^{\prime}}(2\pi)^{4}\delta(p^{2})\delta(k^{\prime,2})\delta(k^{2})\delta(p^{\prime,2})\nonumber \\
 &  & \quad\times f_{\mathrm{V,leq}}^{<}(k^{\prime})f_{\mathrm{V,leq}}^{<}(p^{\prime})f_{\mathrm{V,leq}}^{>}(k)f_{\mathrm{V,leq}}^{>}(p)E_{\mathbf{p}}^{r-1}p^{\langle\alpha}p^{\rho\rangle}(E_{\mathbf{p}}^{n}p_{\langle\alpha}p_{\rho\rangle}-E_{\mathbf{p}^{\prime}}^{n}p_{\langle\alpha}^{\prime}p_{\rho\rangle}^{\prime}),\label{eq:Axial=000020collision=000020int=0000203}\\
\mathcal{A}_{\mathrm{A},rn}^{\mathrm{tt}} & = & \frac{\lambda^{2}}{7}\int_{p,p^{\prime},k,k^{\prime}}W_{pk\to p^{\prime}k^{\prime}}(2\pi)^{4}\delta(p^{2})\delta(k^{\prime,2})\delta(k^{2})\delta(p^{\prime,2})f_{\mathrm{V,leq}}^{<}(k^{\prime})f_{\mathrm{V,leq}}^{<}(p^{\prime})\nonumber \\
 &  & \quad\times f_{\mathrm{V,leq}}^{>}(k)f_{\mathrm{V,leq}}^{>}(p)E_{\mathbf{p}}^{r-1}p^{\langle\mu}p^{\alpha}p^{\lambda\rangle}(E_{\mathbf{p}}^{n}p_{\langle\mu}p_{\alpha}p_{\lambda\rangle}-E_{\mathbf{p}^{\prime}}^{n}p_{\langle\mu}^{\prime}p_{\alpha}^{\prime}p_{\lambda\rangle}^{\prime}),\label{eq:Axial=000020collision=000020int=0000204}
\end{eqnarray}
Interestingly, we find $\mathcal{A}_{\mathrm{A},r0}^{\mathrm{s}}=0$
from Eq.(\ref{eq:Axial=000020collision=000020int=0000201}). We therefore
need the matching condition (\ref{eq:Axial=000020MC_phi_si}) to determine
$\varepsilon_{\mathrm{A},0}^{\mathrm{s},i}$.

To solve the arrays of linear equation (\ref{eq:epsilon_A:=000020eqs1},
\ref{eq:epsilon_A:=000020eqs2}), we can adopt the minimal truncation
scheme \citep{denicol2022microscopic},
\begin{eqnarray}
N_{\mathrm{A}}^{\mathrm{s},i}=1 & , & N_{\mathrm{A}}^{\mathrm{v},i}=N_{\mathrm{A}}^{\mathrm{ts},i}=N_{\mathrm{A}}^{\mathrm{tt}}=0.
\end{eqnarray}
From Eqs. (\ref{eq:epsilon_A:=000020eqs1}, \ref{eq:epsilon_A:=000020eqs2},
\ref{eq:Axial=000020MC_phi_si}), we get,
\begin{eqnarray}
\varphi_{\mathrm{A},\mathbf{p}}^{\mathrm{s},i}=\frac{\alpha_{\mathrm{A},0}^{\mathrm{s},i}}{\mathcal{A}_{\mathrm{A},01}^{\mathrm{s}}}\left(E_{\mathbf{p}}-\frac{J_{20}}{J_{10}}\right) & ,\quad & \varphi_{\mathrm{A},\mathbf{p}}^{\mathrm{v},i}=\frac{\alpha_{\mathrm{A},0}^{\mathrm{v},i}}{\mathcal{A}_{\mathrm{A},00}^{\mathrm{v}}},\label{eq:phi_Ap_s,v}\\
\varphi_{\mathrm{A},\mathbf{p}}^{\mathrm{ts},i}=\frac{\alpha_{\mathrm{A},0}^{\mathrm{ts},i}}{\mathcal{A}_{\mathrm{A},00}^{\mathrm{ts}}} & ,\quad & \varphi_{\mathrm{A},\mathbf{p}}^{\mathrm{tt}}=\frac{\alpha_{\mathrm{A},0}^{\mathrm{tt}}}{\mathcal{A}_{\mathrm{A},00}^{\mathrm{tt}}},\label{eq:phi_Ap_t}
\end{eqnarray}
where $J_{nq}$ are the moments defined in Appendix. \ref{subsec:Thermodynamical-functions-and}.
Details for the calculations of the axial collisional integrals and
functions $\alpha_{\mathrm{A},0}^{i}$ are also shown in Appendix.
\ref{subsec:Axial-collision-integrals} and \ref{subsec:Some-thermodynamical-integrals},
respectively.

Inserting Eqs.(\ref{eq:Alpha_s123}, \ref{eq:Alpha_v12}, \ref{eq:Alpha_ts1-4},
\ref{eq:Alpha_ts5-tt}\ref{eq:A_01=000020s}, \ref{eq:A_00=000020v,ts},
\ref{eq:A_00=000020t}) into Eqs. (\ref{eq:phi_Ap_s,v},\ref{eq:phi_Ap_t}),
we eventually derive the off-equilibrium corrections to $f_{\mathrm{A}}$,
\begin{eqnarray}
\delta f_{\mathrm{A}}^{<}(p) & = & -\frac{\hbar}{2}f_{\mathrm{V,leq}}^{<}(p)f_{\mathrm{V,leq}}^{>}(p)\frac{\beta_{0}^{2}}{e^{4}\ln e^{-1}}\mathcal{F},\label{eq:=00005Cdelta_fA}
\end{eqnarray}
where 
\begin{eqnarray}
\mathcal{F} & = & \left(E_{\mathbf{p}}-d_{1}\frac{1}{\beta_{0}}\right)\left[d_{2}(\omega^{\alpha}\nabla_{\alpha}\beta_{0}-\beta_{0}\nabla^{\alpha}\omega_{\alpha})-d_{3}\beta_{0}\omega^{\alpha}\nabla_{\alpha}\alpha_{0}\right]\nonumber \\
 &  & \quad-\left[d_{4}\beta_{0}\omega^{\mu}\sigma_{\mu}^{\alpha}+d_{5}\epsilon^{\mu\nu\alpha\beta}u_{\beta}\nabla_{\nu}\alpha_{0}\nabla_{\mu}\beta_{0}\right]p_{\langle\alpha\rangle}\nonumber \\
 &  & \quad+\Big[-d_{6}\beta_{0}\left(\omega^{\langle\alpha}\nabla^{\rho\rangle}\beta_{0}+\frac{1}{2}\beta_{0}\nabla^{\langle\alpha}\omega^{\rho\rangle}+\epsilon^{\mu\nu\sigma\langle\rho}\sigma_{\mu}^{\alpha\rangle}u_{\sigma}\nabla_{\nu}\beta_{0}\right)+d_{7}\beta_{0}^{2}\epsilon^{\mu\nu\sigma\langle\rho}\sigma_{\mu}^{\alpha\rangle}u_{\sigma}\nabla_{\nu}\alpha_{0}\nonumber \\
 &  & \quad\qquad+d_{8}\beta_{0}^{2}\omega^{\langle\alpha}\nabla^{\rho\rangle}\alpha_{0}\Big]p_{\langle\alpha}p_{\rho\rangle}+d_{9}\beta_{0}^{3}\omega^{\langle\mu}\sigma^{\alpha\lambda\rangle}p_{\langle\mu}p_{\alpha}p_{\lambda\rangle},\label{eq:def_temp_F}
\end{eqnarray}
the $d_{i}$ are dimensionless numbers, 
\begin{eqnarray}
 &  & d_{1}=\frac{27\zeta(3)}{\pi^{2}},\quad d_{2}=32\pi\ln2,\quad d_{3}=\frac{560\pi^{4}-17280\zeta(3)\ln2}{7\pi^{3}},\quad d_{4}=\frac{72\pi^{3}}{5\ln2},\label{eq:d1234}\\
 &  & d_{5}=\frac{540\zeta(3)-168\pi^{2}\ln2}{7\pi\ln2},\quad d_{6}=12\pi,\label{eq:d56}\\
 &  & d_{7}=\frac{7290\zeta^{2}(3)+14\pi^{6}}{63\pi^{3}\zeta(3)},\quad d_{8}=\frac{43740\zeta^{2}(3)-56\pi^{6}}{63\pi^{3}\zeta(3)},\quad d_{9}=\frac{112\pi^{5}}{3375\zeta(5)}.\label{eq:d789}
\end{eqnarray}

\subsection{Collisional corrections to $\delta\mathcal{A}^{<,\mu}$ in two scenarios\protect\label{subsec:Collisional-corrections-to}}

In this subsection, we derive the collisional corrections to $\delta\mathcal{A}^{<,\mu}$
in two scenarios.

Similar to Sec. \ref{subsec:The--correction}, we derive $\delta f_{{\rm V}}$
from the vector Boltzmann equation (\ref{eq:V=000020BE=000020Moller=000020process}),
\begin{eqnarray}
\delta f_{\mathrm{V}}^{<}(x,p) & = & f_{\mathrm{V,leq}}^{<}(p)f_{\mathrm{V,leq}}^{>}(p)\frac{\beta_{0}^{3}}{e^{4}\ln e^{-1}}\Bigg[c_{2}p^{\langle\alpha}p^{\beta\rangle}\sigma_{\alpha\beta}+(-c_{3}\frac{1}{\beta_{0}}+E_{\mathbf{p}})c_{1}p^{\langle\alpha\rangle}\nabla_{\alpha}\alpha_{0}\Bigg],\label{eq:=00005Cdelta_fV}
\end{eqnarray}
where 
\begin{eqnarray}
c_{1}=(2\pi)^{3}\left[\frac{1215\zeta^{2}(3)}{14\pi^{4}}-\frac{\pi^{2}}{6}\right]\left[\frac{\pi^{4}}{18}-12\zeta(3)\ln2\right]^{-1},\quad & c_{2}=(2\pi)^{3}\frac{7\pi^{2}}{180\zeta(3)},\quad & c_{3}=\frac{3375\zeta(5)}{7\pi^{4}}.\label{eq:c123}
\end{eqnarray}
Using Eq. (\ref{eq:=00005Cdelta_fV}), we obtain
\begin{eqnarray}
S_{(u)}^{\mu\nu}(p)C_{\mathrm{V,\nu}}[\delta f_{\mathrm{V}}^{<}(p)] & = & S_{(u)}^{\mu\nu}(p)\beta_{0}\left[g_{1}(E_{\mathbf{p}})\nabla_{\nu}\alpha_{0}+g_{2}(E_{\mathbf{p}})\sigma_{\nu\alpha}p_{\perp}^{\alpha}\right],\label{eq:SE=000020kernel=000020HTL}
\end{eqnarray}
where
\begin{eqnarray}
g_{1}(E_{\mathbf{p}}) & = & \frac{1}{(2\pi)^{3}}f_{\mathrm{V,leq}}^{<}(p)f_{\mathrm{V,leq}}^{>}(p)c_{1}\left[\frac{1}{|\mathbf{p}|}\frac{9\zeta(3)}{\beta_{0}^{2}}-\frac{\pi^{2}}{3\beta_{0}}+\left(\frac{21\zeta(3)}{4\beta_{0}}-|\mathbf{p}|\frac{\pi^{2}}{12}\right)\left(1-2f_{\mathrm{V,leq}}^{>}(p)\right)\right],\label{eq:g1(E)}\\
g_{2}(E_{\mathbf{p}}) & = & \frac{1}{(2\pi)^{3}}f_{\mathrm{V,leq}}^{<}(p)f_{\mathrm{V,leq}}^{>}(p)c_{2}\frac{1}{|\mathbf{p}|}\left[\frac{1}{|\mathbf{p}|}\frac{6\zeta(3)}{\beta_{0}^{2}}-\frac{2\pi^{2}}{3\beta_{0}}+\left(\frac{15\zeta(3)}{2\beta_{0}}-|\mathbf{p}|\frac{\pi^{2}}{6}\right)\left(1-2f_{\mathrm{V,leq}}^{>}(p)\right)\right].\label{eq:g2(E)}
\end{eqnarray}

We eventually obtain the $\delta\mathcal{A}^{<,\mu}(p)$ in two scenarios.
In scenario (I), inserting Eq. (\ref{eq:SE=000020kernel=000020HTL})
into Eq. (\ref{eq:delta_A_sI_01}) yields, 
\begin{eqnarray}
\delta\mathcal{A}_{\textrm{(I)}}^{<,\mu}(X,p) & = & 2\pi\hbar\delta(p^{2})\beta_{0}\left[-g_{1}(E_{\mathbf{p}})\frac{\epsilon^{\mu\nu\rho\sigma}p_{\rho}u_{\sigma}}{2E_{\mathbf{p}}}\nabla_{\nu}\alpha_{0}-g_{2}(E_{\mathbf{p}})\frac{\epsilon^{\mu\nu\rho\sigma}p_{\rho}u_{\sigma}}{2E_{\mathbf{p}}}\sigma_{\nu\alpha}p^{\alpha}\right].\label{eq:delta_A_sI}
\end{eqnarray}
We comment on Eqs. (\ref{eq:SE=000020kernel=000020HTL}, \ref{eq:delta_A_sI}).
Remarkably, $\delta\mathcal{A}_{\textrm{(I)}}^{<,\mu}(p)$ appears
to be independent of the coupling constant $e$ and persists even
as $e\rightarrow0$. When setting $e\rightarrow0$, i.e. considering
the free-streaming scenario where the particles' mean free path $\lambda$
is comparable to the characteristic macroscopic length scale of system
$L$, the gradient expansion becomes invalid in such limits.

In scenario (II), we observe that $\delta f_{\mathrm{A}}$ in Eq.
(\ref{eq:=00005Cdelta_fA}) is of $\mathcal{O}\left(\frac{\partial^{2}}{e^{4}\ln e^{-1}}\right)$,
and consequently, its correction to the axial Wigner function $\mathcal{A}^{\mu}$
in Eq. (\ref{eq:Off-eq-A-Case2}) should also be of the same order.
We summarize the expression for $\delta\mathcal{A}^{<,\mu}(X,p)$
in scenario (II) as follows,
\begin{eqnarray}
\delta\mathcal{A}_{\textrm{(II)}}^{<,\mu}(X,p) & = & -\hbar\pi p^{\mu}\delta(p^{2})f_{\mathrm{V,leq}}^{<}(p)f_{\mathrm{V,leq}}^{>}(p)\frac{\beta^{2}}{e^{4}\ln e^{-1}}\mathcal{F},\label{eq:delta_A_sII}
\end{eqnarray}
where $\mathcal{F}$ is given by Eq. (\ref{eq:def_temp_F}). On the
other hand, when numerically solving the SBE in Eq. (\ref{eq:SBE=000020Moller=000020process}),
the evolution of $f_{\mathrm{A}}$ effectively incorporates all orders
of the gradient expansion, suggesting that $\mathcal{A}^{\mu}$ may
receive collisional corrections of $\mathcal{O}(\partial^{1})$ even
when $f_{{\rm V}}=f_{{\rm V,leq}}$.

Similar effects related to the collisional corrections found in Eqs.(\ref{eq:delta_A_sI},
\ref{eq:delta_A_sII}) have also been widely discussed in condensed
matter physics for anomalous and spin Hall effects \citep{RevModPhys.82.1539,sinova2015spin,dyakonov2017spin}.

\section{Collisional corrections to spin polarization pseudo-vector}

\label{sec:Fermion-Spin-polarization-from} In the previous sections,
we have solved the Boltzmann equation array (\ref{eq:V=000020BE=000020Moller=000020process},\ref{eq:SBE=000020Moller=000020process})
near local equilibrium using systematical gradient expansion and derived
the collisional corrections to $\mathcal{A}^{\mu}$ in two scenarios.
In this section, we further implement our results to spin polarization
of  $\Lambda$ hyperons and discuss the corrections to the modified Cooper-Frye formula \citep{Becattini:2013fla,Fang:2016vpj}.

\subsection{Modified Cooper-Frye formula and spin polarization pseudo-vector
in local equilibrium}

We briefly review the derivation of the Pauli-Lubanski pseudo-vector
and modified Cooper-Frye formula from QKT following Refs. \citep{Fang:2016vpj,Yang:2017sdk,Yi:2021ryh}.
The Pauli-Lubanski vector density in phase space is defined as,
\begin{eqnarray}
\mathcal{W}^{\mu}(x,p) & = & -\frac{1}{2}\epsilon^{\mu\nu\rho\sigma}\mathcal{J}_{\nu\rho}(x,p)p_{\sigma},\label{eq:Pauli-Lubanski=000020vector}
\end{eqnarray}
where $\mathcal{J}_{\nu\rho}(x,p)=u^{\lambda}J_{\lambda\nu\rho}(x,p)$
is the total angular momentum density in phase space with $J_{\lambda\nu\rho}(x,p)$
being the angular momentum current. $J_{\lambda\nu\rho}(x,p)$ can be decomposed as the orbital and spin
parts, 
\begin{eqnarray}
J_{\lambda\nu\rho}(x,p) & = & x_{\nu}T_{\lambda\rho}(x,p)-x_{\rho}T_{\lambda\nu}(x,p)+S_{\lambda\nu\rho}(x,p).
\end{eqnarray}
Using Eq. (\ref{eq:EMT=000020=000026=000020SAMT=000020in=000020WF}) within the canonical pseudo-gauge,
 the energy momentum and spin tensors in phase space are, 
\begin{eqnarray}
T_{\mu\nu}(x,p)=4p_{\nu}\mathcal{V}_{\mu}^{<}(x,p) & ,\quad & S_{\lambda\nu\rho}(x,p)=2\hbar\epsilon_{\alpha\lambda\nu\rho}\mathcal{A}^{<,\alpha}(x,p),
\end{eqnarray}
where the factor $2$ for spin current tensor, $S_{\lambda\nu\rho}$,
comes from the degeneracy of spin. Now Eq.(\ref{eq:Pauli-Lubanski=000020vector})
becomes,
\begin{eqnarray}
\mathcal{W}^{\mu}(x,p) & = & -\hbar\epsilon^{\mu\nu\rho\sigma}\epsilon_{\alpha\lambda\nu\rho}u^{\lambda}\mathcal{A}^{<,\alpha}(x,p)p_{\sigma}=2\hbar p_0 \mathcal{A}^{<,\mu}(x,p),\label{eq:PL-vector-canonical-pg}
\end{eqnarray}
where we have used $p\cdot\mathcal{A}^{<}(x,p)=0$ from Eq.(\ref{eq:Formal=000020sol.=000020A}).
We emphasize that Eq.(\ref{eq:PL-vector-canonical-pg}) is, by definition,
pseudo-gauge dependent, also see e.g. Refs.~\cite{Speranza:2020ilk,Weickgenannt:2022zxs} for related discussions.

Assuming the fermions with mass $m$ reach the local equilibrium in
the freeze-out hypersurface $\Sigma^{\mu}$, the particle number element
in phase space is given by,
\begin{eqnarray}
\mathrm{d}\mathcal{N}(t;\mathbf{x},\mathbf{p}) & = & \frac{2}{(2\pi)^{3}}\theta(p_{0})\delta(p^{2}-m^{2})f(x,p)p_{\mu}\mathrm{d}\Sigma^{\mu}\mathrm{d}p_{0}\mathrm{d}^{3}\mathbf{p},\label{eq:Particle-number=000020phase=000020space=000020element}
\end{eqnarray}
where $f(x,p)=2f_{\mathrm{FD}}(x,p)$ with $f_{\mathrm{FD}}$ defined
in Eq.(\ref{eq:fV=000020local=000020eq}) and factor $2$ denoting
the degeneracy coefficient, $\mathrm{d}\Sigma^{\mu}$ denotes the
time-like hypersurface element. For simplicity, we only consider the
particles and exclude the anti-particles. In terms of the Wigner function,
Eq.(\ref{eq:Particle-number=000020phase=000020space=000020element})
can be expressed using Eq. (\ref{eq:Formal=000020sol.=000020V}) as
follows,
\begin{eqnarray}
\mathrm{d}\mathcal{N}(t;\mathbf{x},\mathbf{p}) & = & \frac{1}{(2\pi)^{4}}4\mathcal{V}_{\mu,\mathrm{leq}}^{<}(x,p)\mathrm{d}\Sigma^{\mu}\mathrm{d}p_{0}\mathrm{d}^{3}\mathbf{p}=\frac{1}{(2\pi)^{4}}4\mathcal{V}_{\mu,\mathrm{leq}}^{<}(x,p)\hat{t}^{\mu}\mathrm{d}\sigma\mathrm{d}p_{0}\mathrm{d}^{3}\mathbf{p}\nonumber \\
 & = & \frac{1}{(2\pi)^{4}}4\theta(p_{0})2\pi\delta(p^{2}-m^{2})f_{\mathrm{V,leq}}^{<}(x,p)p_0\mathrm{d}\sigma\mathrm{d}p_{0}\mathrm{d}^{3}\mathbf{p},
\end{eqnarray}
where $\hat{t}^{\mu}=(1,\mathbf{0})$ is a the timelike normal vector
of the freeze-out hypersurface and we have adopt $\mathrm{d}\Sigma_{\mu}=(\mathrm{d}\sigma,\mathbf{0})=(\mathrm{d}^{3}\mathbf{x},\mathbf{0})$
temporarily. Then the Lorentz-invariant momentum spectrum is given
by,
\begin{eqnarray}
(2\pi)^{3} p_0 \frac{\mathrm{d}\mathcal{N}(t;\mathbf{p})}{\mathrm{d}^{3}\mathbf{p}} & = & \int_{\Sigma}f(x,p)p_{\mu}\mathrm{d}\Sigma^{\mu}=4p_0\int_{\Sigma}\mathrm{d}\Sigma^{\mu}\int\frac{\mathrm{d}p_{0}}{2\pi}\mathcal{V}_{\mu,\mathrm{leq}}^{<}(x,p).\label{eq:C-F=000020formula=000020in=000020Wigner=000020function}
\end{eqnarray}

This procedure can usually be generalized to derive the Lorentz-covariant
Cooper-Frye formula for spin near the freeze-out hypersurface. 
The
Pauli-Lubanski pseudo-vector element can be written as,
\begin{eqnarray}
\mathrm{d}\mathcal{W}^{\mu}(t;\mathbf{x},\mathbf{p}) & = & \frac{1}{(2\pi)^{4}}2\hbar p_0\mathcal{A}^{<,\mu}(x,p)\mathrm{d}\sigma\mathrm{d}p_{0}\mathrm{d}^{3}\mathbf{p}=\frac{1}{(2\pi)^{4}}2\hbar\mathcal{A}^{<,\mu}(x,p)p_{\alpha}\mathrm{d}\Sigma^{\alpha}\mathrm{d}p_{0}\mathrm{d}^{3}\mathbf{p}.
\end{eqnarray}
The Lorentz-covariant Cooper-Frye formula for spin is then given by,
\begin{eqnarray}
(2\pi)^{3}p_0\frac{\mathrm{d}\mathcal{W}^{\mu}(t;\mathbf{p})}{\mathrm{d}^{3}\mathbf{p}} & = & 2p_0\hbar\int_{\Sigma}\mathrm{d}\Sigma^{\alpha}p_{\alpha}\int\frac{\mathrm{d}p_{0}}{2\pi}\mathcal{A}^{<,\mu}(x,p).
\end{eqnarray}
From now on, we introduce the 
on-shell energy for fermions, 
$E_{\mathbf{p}}\equiv u^{\mu}p_{\mu}$ 
defined outside the $p_{0}$ integrals. We proceed to define the normalized Lorentz-covariant spin
polarization pseudo-vector for massive fermions,
\begin{equation}
\mathcal{P}^{\mu}(t;\mathbf{p})=\frac{1}{m}\frac{(2\pi)^{3}E_{\mathbf{p}}\frac{\mathrm{d}\mathcal{W}^{\mu}(t;\mathbf{p})}{\mathrm{d}^{3}\mathbf{p}}}{(2\pi)^{3}E_{\mathbf{p}}\frac{\mathrm{d}\mathcal{N}(t;\mathbf{p})}{\mathrm{d}^{3}\mathbf{p}}}=\hbar\frac{\int_{\Sigma}\mathrm{d}\Sigma\cdot p\int\frac{\mathrm{d}p_{0}}{2\pi}\mathcal{A}^{<,\mu}(x,p)}{2m\int_{\Sigma}\int\frac{\mathrm{d}p_{0}}{2\pi}\mathrm{d}\Sigma\cdot\mathcal{V}^{<}(x,p)}.\label{eq:Modified_CF_massive}
\end{equation}
Here, given the Pauli-Lubanski vector $\mathcal{W}^{\mu}(t;\mathbf{p})$
has the dimension of mass, we introduce an additional fermion mass
$m$, which serves as eigenvalue of Poincare group Casimir operator
$\hat{P}^{2}$. This mass is included in the denominator on the right-hand
side of Eq. (\ref{eq:Modified_CF_massive}), ensuring the spin polarization
pseudo-vector $\mathcal{P}^{\mu}(t;\mathbf{p})$ is dimensionless. This formulation is referred to as the modified Cooper-Frye formula
\citep{Becattini:2013fla,Fang:2016vpj,Liu:2021nyg}.

The above discussions hold for general massive fermions, including our interested $\Lambda$ hyperons. In our previous calculations, we worked out the spin polarization of light fermions (quarks) in a QED medium by solving (spin) Boltzmann equations. These light fermions finally freeze out in a timelike hypersurface and the hyperons directly inherit the polarization of $s$-quarks \cite{Liang:2004ph}, namely, following the $s$-quark scenario \cite{Fu:2021pok}. So we might directly use the axial Wigner function of the in-medium light fermions in Eq.(\ref{eq:Modified_CF_massive}). Strictly speaking, we should use the axial Wigner functions of valence $s$ quarks which are massive. 

Next, we compute $\mathcal{P}^{\mu}(t;\mathbf{p})$ in local equilibrium.
Inserting Eqs.(\ref{eq:fV=000020local=000020eq}, \ref{eq:fA=000020local=000020eq})
into Eqs.(\ref{eq:Formal=000020sol.=000020V}, \ref{eq:Formal=000020sol.=000020A})
and using the tensor decomposition (\ref{eq:Fluid=000020velocity=000020gradient=000020decomposition})
yields,
\begin{eqnarray}
\mathcal{A}_{\mathrm{leq}}^{<,\mu}(x,p) & = & 2\pi\delta(p^{2})\hbar f_{\mathrm{V,leq}}^{<}(x,p)f_{\mathrm{V,leq}}^{>}(x,p)\frac{1}{2}\left[\frac{1}{2}\epsilon^{\mu\nu\alpha\beta}p_{\nu}\partial_{\alpha}(\beta_{0}u_{\beta})-\frac{1}{E_{\mathbf{p}}}\beta_{0}\epsilon^{\mu\nu\rho\sigma}p_{\rho}u_{\sigma}p^{\alpha}\sigma_{\nu\alpha}\right.\nonumber \\
 &  & \quad\left.-\frac{1}{2}\epsilon^{\mu\nu\rho\sigma}p_{\rho}u_{\sigma}(\beta_{0}Du_{\nu}+\nabla_{\nu}\beta_{0})+\frac{\epsilon^{\mu\nu\rho\sigma}p_{\rho}u_{\sigma}}{E_{\mathbf{p}}}\nabla_{\nu}\alpha_{0}\right].\label{eq:Wigner-A=000020Leq}
\end{eqnarray}
The polarization pseudo-vector in Eq. (\ref{eq:Modified_CF_massive})
becomes \citep{Yi:2021ryh}, 
\begin{eqnarray}
\mathcal{P}_{\mathrm{leq}}^{\mu}(\mathbf{p}) & = & \hbar\frac{\int_{\Sigma}\mathrm{d}\Sigma\cdot p\mathcal{A}_{\mathrm{leq}}^{<,\mu}(x,\mathbf{p})}{2\int_{\Sigma}\mathrm{d}\Sigma\cdot\mathcal{V}_{\mathrm{leq}}^{<}(x,\mathbf{p})}=\mathcal{P}_{\mathrm{leq},\mathrm{thermal}}^{\mu}+\mathcal{P}_{\mathrm{leq},\mathrm{shear}}^{\mu}+\mathcal{P}_{\mathrm{leq},\mathrm{accT}}^{\mu}+\mathcal{P}_{\mathrm{leq},\mathrm{chem}}^{\mu}.\label{eq:P-L=000020vector=000020Leq}
\end{eqnarray}
The polarization induced by thermal vorticity, shear tensor, fluid
acceleration and $\nabla\alpha_{0}$, are 
\begin{eqnarray}
\mathcal{P}_{\mathrm{leq},\mathrm{thermal}}^{\mu} & = & \frac{\hbar^{2}}{8N}\int_{\Sigma}\mathrm{d}\Sigma\cdot pf_{\mathrm{V,leq}}^{<}(p)f_{\mathrm{V,leq}}^{>}(p)\epsilon^{\mu\nu\alpha\beta}p_{\nu}\partial_{\alpha}(\beta_{0}u_{\beta}),\label{eq:Leq=000020thermal=000020contributioin}\\
\mathcal{P}_{\mathrm{leq},\mathrm{shear}}^{\mu} & = & -\frac{\hbar^{2}}{4N}\int_{\Sigma}\mathrm{d}\Sigma\cdot pf_{\mathrm{V,leq}}^{<}(p)f_{\mathrm{V,leq}}^{>}(p)\frac{\epsilon^{\mu\nu\rho\sigma}p_{\rho}u_{\sigma}}{E_{\mathbf{p}}}\beta_{0}p^{\alpha}\sigma_{\nu\alpha},\label{eq:Leq=000020shear=000020contributioin}\\
\mathcal{P}_{\mathrm{leq},\mathrm{accT}}^{\mu} & = & -\frac{\hbar^{2}}{8N}\int_{\Sigma}\mathrm{d}\Sigma\cdot pf_{\mathrm{V,leq}}^{<}(p)f_{\mathrm{V,leq}}^{>}(p)\epsilon^{\mu\nu\rho\sigma}p_{\rho}u_{\sigma}(\beta_{0}Du_{\nu}+\nabla_{\nu}\beta_{0}),\label{eq:Leq=000020accelaration=000020contributioin}\\
\mathcal{P}_{\mathrm{leq},\mathrm{chem}}^{\mu} & = & \frac{\hbar^{2}}{4N}\int_{\Sigma}\mathrm{d}\Sigma\cdot pf_{\mathrm{V,leq}}^{<}(p)f_{\mathrm{V,leq}}^{>}(p)\frac{\epsilon^{\mu\nu\rho\sigma}p_{\rho}u_{\sigma}}{E_{\mathbf{p}}}\nabla_{\nu}\alpha_{0},\label{eq:Leq=000020chemical=000020contributioin}
\end{eqnarray}
respectively, where 
\begin{equation}
N\equiv m\int_{\Sigma}\mathrm{d}\Sigma_{\mu}p^{\mu}f_{\mathrm{V},\mathrm{leq}}^{<}(x,p).
\end{equation}

We emphasize that we have extended the discussion to massive fermions by simply substituting the $\mathcal{W}^\mu$ and $\mathcal{A}^{<,\mu}$ for massless fermions into the massive case. The $\mathcal{P}^\mu$ in Eqs.~(\ref{eq:Leq=000020thermal=000020contributioin}-\ref{eq:Leq=000020chemical=000020contributioin}) are perpendicular to $p^\mu$ after integrating over the freeze-out hypersurface. These results are consistent with those derived from linear response theory \cite{Liu:2021uhn} for massive fermions, suggesting that our simple extension is effective. However, a very recent study on non-interacting massless fermions \cite{Palermo:2023cup} suggests that $\mathcal{P}^\mu$ is exclusively proportional to $p^\mu$. 
Currently, we do not have a clear understanding of the discrepancies between our findings and those in Ref.~\cite{Palermo:2023cup}. This question will be addressed in future studies.

\subsection{Collisional corrections}

Now, we compute the collisional corrections to the spin polarization
pseudo-vector in scenario (I). By inserting Eq. (\ref{eq:delta_A_sI})
into Eq. (\ref{eq:Modified_CF_massive}), we find that this substitution
simply yields, 
\begin{eqnarray}
\delta\mathcal{P}_{\textrm{(I)}}^{\mu}(\mathbf{p}) & = & \delta\mathcal{P}_{\textrm{(I)},\mathrm{shear}}^{\mu}+\delta\mathcal{P}_{\mathrm{\textrm{(I)}},\mathrm{chem}}^{\mu}+\mathcal{O}(\hbar^{2}\partial^{2}),\label{eq:Case1-CF-formula}
\end{eqnarray}
with 
\begin{eqnarray}
\delta\mathcal{P}_{\mathrm{\textrm{(I)}},\mathrm{shear}}^{\mu} & = & -\frac{\hbar^{2}}{4N}\int_{\Sigma}\frac{\mathrm{d}\Sigma\cdot p}{E_{\mathbf{p}}}\beta_{0}g_{2}(E_{\mathbf{p}})\epsilon^{\mu\nu\rho\sigma}p_{\rho}u_{\sigma}\sigma_{\nu\alpha}p^{\alpha},\label{eq:P=000020Formal_correction=000020shear}\\
\delta\mathcal{P}_{\textrm{(I)},\mathrm{chem}}^{\mu} & = & -\frac{\hbar^{2}}{4N}\int_{\Sigma}\frac{\mathrm{d}\Sigma\cdot p}{E_{\mathbf{p}}}\beta_{0}g_{1}(E_{\mathbf{p}})\epsilon^{\mu\nu\rho\sigma}p_{\rho}u_{\sigma}\nabla_{\nu}\alpha_{0}.\label{eq:P=000020Formal_correction=000020chem}
\end{eqnarray}
We find that collision corrections from $\delta f_{\mathrm{V}}$ solely
alter the magnitude of spin polarization induced by thermal vorticity
and shear tensor in Eqs. (\ref{eq:Leq=000020shear=000020contributioin},
\ref{eq:Leq=000020chemical=000020contributioin}). Although Eqs. (\ref{eq:P=000020Formal_correction=000020shear},
\ref{eq:P=000020Formal_correction=000020chem}) appear to be coupling
constant independent, they indeed originate from collisional corrections,
known as ``side-jump'' contribution as discussed in Sec. \ref{subsec:The--correction}.
Moreover, inclusion of the complete leading order contribution, such
as collinear splitting \citep{Arnold:2007pg}, could explicitly reveal
the coupling dependence. It is important to emphasize that such collisional
corrections were not considered in previous studies on the shear-induced
polarization \citep{Liu:2021uhn,Becattini:2021suc,Yi:2021ryh}. For
further discussion of similar findings, refer to the recent work in
Ref. \citep{Lin:2024zik}. 

In scenario (II), using Eq. (\ref{eq:delta_A_sII}), the spin polarization
pseudo-vector in Eq. (\ref{eq:Modified_CF_massive}) is given by,
\begin{eqnarray}
\delta\mathcal{P}_{\textrm{(II)}}^{\mu}(\mathbf{p}) & = & \frac{\hbar}{2N}\int_{\Sigma}\mathrm{d}\Sigma\cdot pp^{\mu}\delta f_{\mathrm{A}}^{<}(x,p)+\mathcal{O}(\hbar^{2}\partial^{3})\nonumber \\
 & = & \mathcal{P}_{\mathrm{\textrm{(II)}},\omega-\nabla T}^{\mu}+\mathcal{P}_{\mathrm{\textrm{(II)}},\nabla\omega}^{\mu}+\mathcal{P}_{\mathrm{\textrm{(II)},\omega-\mathrm{chem}}}^{\mu}+\mathcal{P}_{\mathrm{\textrm{(II)},\omega-\mathrm{shear}}}^{\mu}\nonumber \\
 &  & +\mathcal{P}_{\mathrm{\textrm{(II)},chem-\nabla T}}^{\mu}+\mathcal{P}_{\mathrm{\textrm{(II)},shear-\nabla T}}^{\mu}+\mathcal{P}_{\mathrm{\textrm{(II)},shear-chem}}^{\mu},\label{eq:Off-eq=000020correction=000020C-F=000020formula}
\end{eqnarray}
where 
\begin{eqnarray}
\mathcal{P}_{\mathrm{\textrm{(II)}},\omega-\nabla T}^{\mu} & = & -\hbar^{2}\int_{\Sigma}\mathrm{d}\Sigma\cdot pa_{\textrm{(II)}}^{\mu}\left[d_{2}\left(E_{\mathbf{p}}-d_{1}\frac{1}{\beta_{0}}\right)\omega^{\alpha}\nabla_{\alpha}\beta_{0}-d_{6}\beta_{0}p_{\langle\alpha}p_{\rho\rangle}\omega^{\alpha}\nabla^{\rho}\beta_{0}\right],\nonumber \\
\mathcal{P}_{\mathrm{\textrm{(II)}},\nabla\omega}^{\mu} & = & \hbar^{2}\int_{\Sigma}\mathrm{d}\Sigma\cdot pa_{\textrm{(II)}}^{\mu}\left[\left(E_{\mathbf{p}}-d_{1}\frac{1}{\beta_{0}}\right)d_{2}\beta_{0}\nabla^{\alpha}\omega_{\alpha}+d_{6}\frac{1}{2}\beta_{0}^{2}\nabla^{\alpha}\omega^{\rho}p_{\langle\alpha}p_{\rho\rangle}\right],\nonumber \\
\mathcal{P}_{\mathrm{\textrm{(II)},\omega-\mathrm{chem}}}^{\mu} & = & \hbar^{2}\int_{\Sigma}\mathrm{d}\Sigma\cdot pa_{\textrm{(II)}}^{\mu}\left[\left(E_{\mathbf{p}}-d_{1}\frac{1}{\beta_{0}}\right)d_{3}\beta_{0}\omega^{\alpha}\nabla_{\alpha}\alpha_{0}-d_{8}\beta_{0}^{2}\omega^{\alpha}\nabla^{\rho}\alpha_{0}p_{\langle\alpha}p_{\rho\rangle}\right],\nonumber \\
\mathcal{P}_{\mathrm{\textrm{(II)},\omega-\mathrm{shear}}}^{\mu} & = & -\hbar^{2}\beta_{0}\int_{\Sigma}\mathrm{d}\Sigma\cdot pa_{\textrm{(II)}}^{\mu}\left[-d_{4}\omega^{\rho}\sigma_{\rho}^{\alpha}p_{\langle\alpha\rangle}+d_{9}\beta_{0}^{2}\omega^{\beta}\sigma^{\alpha\lambda}p_{\langle\beta}p_{\alpha}p_{\lambda\rangle}\right]\nonumber \\
\mathcal{P}_{\mathrm{\textrm{(II)},chem-\nabla T}}^{\mu} & = & \hbar^{2}\int_{\Sigma}\mathrm{d}\Sigma\cdot pa_{\textrm{(II)}}^{\mu}d_{5}\epsilon^{\rho\nu\alpha\beta}u_{\beta}\nabla_{\nu}\alpha_{0}\nabla_{\rho}\beta_{0}p_{\langle\alpha\rangle},\nonumber \\
\mathcal{P}_{\mathrm{\textrm{(II)},shear-\nabla T}}^{\mu} & = & \hbar^{2}\beta_{0}\int_{\Sigma}\mathrm{d}\Sigma\cdot pa_{\textrm{(II)}}^{\mu}d_{6}\epsilon^{\beta\nu\sigma\rho}\sigma_{\beta}^{\alpha}u_{\sigma}\nabla_{\nu}\beta_{0}p_{\langle\alpha}p_{\rho\rangle},\nonumber \\
\mathcal{P}_{\mathrm{\textrm{(II)},shear-chem}}^{\mu} & = & -\hbar^{2}\beta_{0}^{2}\int_{\Sigma}\mathrm{d}\Sigma\cdot pa_{\textrm{(II)}}^{\mu}d_{7}\epsilon^{\mu\nu\sigma\rho}\sigma_{\mu}^{\alpha}u_{\sigma}\nabla_{\nu}\alpha_{0}p_{\langle\alpha}p_{\rho\rangle},
\end{eqnarray}
with $a_{\textrm{(II)}}^{\mu}=\frac{\beta_{0}^{2}}{4Ne^{4}\ln e^{-1}}f_{\mathrm{V,leq}}^{<}(p)f_{\mathrm{V,leq}}^{>}(p)p^{\mu}$.
Here the $\{c_{1},c_{2},c_{3}\}$ and $\{d_{1},...,d_{9}\}$ are
the dimensionless constants defined in Eqs. (\ref{eq:c123}, \ref{eq:d1234},
\ref{eq:d56}, \ref{eq:d789}). We find that all of above corrections
are of $\mathcal{O}(\hbar^{2}\partial^{2})$ and proportional to the
first order of spin relaxation time. They differ from those obtained
in Eq. (\ref{eq:Case1-CF-formula}) for scenario (I). Interestingly,
in scenario (II), many of collisional corrections are found to be
proportional solely to kinetic vorticity $\omega^{\mu}$ instead of
thermal vorticity.

\section{Relaxation time approximation to SBE}

\label{sec:Relaxation-time-approximation} In this section, we apply
the RTA to SBE (\ref{eq:SBE=000020Moller=000020process}) with $f_{\mathrm{V}}=f_{\mathrm{V,leq}}^{<}$,
\begin{eqnarray}
p^{\mu}\partial_{\mu}f_{\mathrm{A}}^{<}(p)+\hbar\partial_{\mu}S_{(u)}^{\mu\alpha}(p)\partial_{\alpha}f_{\mathrm{V,leq}}^{<}(p) & = & -E_{\mathbf{p}}\frac{f_{\mathrm{A}}^{<}-f_{\mathrm{A,leq}}^{<}}{\tau_{\mathrm{A}}},\label{eq:SBE_fVleq}
\end{eqnarray}
where we parameterize the collision kernel $\mathcal{C}_{\mathrm{A}}^{\mathrm{B}}[f_{\mathrm{V,leq}},\delta f_{\mathrm{A}}]$
by using the Anderson-Witting relaxation time approximation \citep{ANDERSON1974466}
and $\tau_{\mathrm{A}}$ is the relaxation time of axial distribution
function. We introduce the effective Green function $G(x,p)$ for
$f_{\mathrm{A}}^{<}(x,p)$ in Eq. (\ref{eq:SBE_fVleq}), 
\begin{eqnarray}
\left(p^{\mu}\partial_{\mu}+\frac{E_{\mathbf{p}}}{\tau_{\mathrm{A}}}\right)f_{\mathrm{A}}^{<}(x,p) & = & E_{\mathbf{p}}\frac{f_{\mathrm{A,leq}}^{<}(x,p)}{\tau_{\mathrm{A}}}-\hbar\partial_{\mu}S_{(u)}^{\mu\alpha}(x,p)\partial_{\alpha}f_{\mathrm{V,leq}}^{<}(x,p)\equiv G(x,p).\label{eq:SBE_RTA}
\end{eqnarray}
We take the Fourier transform to both sides of Eq. (\ref{eq:SBE_RTA})
and obtain, 
\begin{eqnarray}
f_{\mathrm{A}}^{<}(x,p) & = & \int\frac{\mathrm{d}^{4}k}{(2\pi)^{4}}\int\mathrm{d}^{4}x^{\prime}e^{-ik\cdot(x-x^{\prime})}\frac{iG(x^{\prime},p)}{k_{\mu}p^{\mu}+iE_{\mathbf{p}}\tau_{\mathrm{A}}^{-1}}.\label{eq:fA=000020init=000020RTA=000020sol}
\end{eqnarray}

In this work, we consider the system expands as a Bjorken flow \citep{Bjorken:1982qr},
i.e. the system is homogeneous in transverse plane and boost-invariant
in longitudinal direction. We refer to Refs. \citep{BAYM198418} and
\citep{Ebihara:2017suq} for the discussion on the conventional and
chiral kinetic theory in a Bjorken flow. In the current work, we take
$p^{\mu}=(E_{\mathbf{p}},0,0,p^{z})$ in Eq. (\ref{eq:fA=000020init=000020RTA=000020sol})
and assume that $G(x,p)$ is independent on $\mathbf{x}_{\mathrm{T}}=(x,y)$.
Then, Eq. (\ref{eq:fA=000020init=000020RTA=000020sol}) reduces to,
\begin{eqnarray}
f_{\mathrm{A}}^{<}(x,p) & = & \frac{1}{E_{\mathbf{p}}}\int\frac{\mathrm{d}k_{0}\mathrm{d}k^{z}\mathrm{d}t^{\prime}\mathrm{d}z^{\prime}}{(2\pi)^{2}}\frac{ie^{-ik^{0}(t-t^{\prime})+ik^{z}(z-z^{\prime})}}{k^{0}-k^{z}p^{z}E_{\mathbf{p}}^{-1}+i\tau_{\mathrm{A}}^{-1}}G(x^{\prime},p)\Big|_{\mathbf{x}_{\mathrm{T}}^{\prime}=\mathbf{x}_{\mathrm{T}}}\nonumber \\
 & = & \frac{1}{E_{\mathbf{p}}}\int_{-\infty}^{+\infty}\mathrm{d}t^{\prime}\theta(t-t^{\prime})\exp\left(-\frac{t-t^{\prime}}{\tau_{\mathrm{A}}}\right)G(x^{\prime},p)\Big|_{\mathbf{x}_{\mathrm{T}}^{\prime}=\mathbf{x}_{\mathrm{T}},z-z^{\prime}=p^{z}E_{\mathbf{p}}^{-1}(t-t^{\prime})}.\label{eq:fA=000020intermediate=000020RTA_sol}
\end{eqnarray}
which has the same structure as that in the solution of $\widetilde{a}^{\mu}$
in Ref. \citep{Kumar:2023ghs}. Here, $\theta(x)$ is the unit-step
function.

As mentioned in Sec. \ref{sec:Fermion-Spin-polarization-from}, to
apply the modified Cooper-Frye formula, it is essential to know the
axial distribution function $f_{\mathrm{A}}^{<}(x,p)$ at the (chemical)
freeze out hypersurface. We assume that the freeze out occurs at $t_{f}$
and focus on the interactions corrections near this hypersurface.
For analytical convenience, we further assume that $G(x,p)$ defined
in Eq. (\ref{eq:SBE_RTA}) is nonzero for $t>t_{i}$, and negligible
for $t<t_{i}$, effectively $G(x,p)\rightarrow\theta(t-t_{i})G(x,p)$.
We retain the leading order of $G(x,p)$ in Eq. (\ref{eq:SBE_RTA}),
which is of $\mathcal{O}(\hbar\partial^{1})$, and derive the expression
for $f_{\mathrm{A}}^{<}(x,p)$ in Eq.(\ref{eq:fA=000020intermediate=000020RTA_sol}),
\begin{equation}
f_{\mathrm{A}}^{<}(x,p)=\frac{1}{\tau_{\mathrm{A}}}\int_{t_{i}}^{t_{f}}\mathrm{d}t^{\prime}\exp\left(-\frac{t-t^{\prime}}{\tau_{\mathrm{A}}}\right)\left.f_{\mathrm{A,leq}}^{<}(t^{\prime},\mathbf{x}_{\mathrm{T}},z^{\prime};p)\right|_{z^{\prime}=z-p^{z}E_{\mathbf{p}}^{-1}(t-t^{\prime})}+\mathcal{O}(\hbar\partial^{2}).
\end{equation}
For simplicity, we assume that $t_{i}$ is very close to $t_{f}$
and neglect the variation of $f_{\mathrm{A,leq}}^{<}(x,p)$ during
this time interval, as these are of $\mathcal{O}(\hbar\partial^{2})$.
Then,  we obtain,
\begin{eqnarray}
f_{\mathrm{A}}^{<}(t=t_{f},\mathbf{x},p) & \approx & \Gamma_{\mathrm{A}}f_{\mathrm{A,leq}}^{<}(x,p)+\mathcal{O}\left(\hbar\partial^{2}\right),\quad\Gamma_{\mathrm{A}}\equiv\frac{t_{f}-t_{i}}{\tau_{\mathrm{A}}},\label{eq:fA=000020near_initial=000020RTA=000020sol}
\end{eqnarray}
Here, $\Gamma_{\mathrm{A}}$ describes the ratio of the time interval
over relaxation time for spin. As reported in our previous work \citep{Fang:2022ttm},
the relaxation time for spin can exceed the conventional thermalization
time, making it reasonable to assume that $\Gamma_{\mathrm{A}}<1.$
The off-equilibrium correction to $f_{\mathrm{A}}^{<}$ is directly
proportional to $f_{\mathrm{A,leq}}^{<}$ in our approximation, particularly
when only interactions near the freeze-out hypersurface are considered.
The relaxation time of the axial distribution function, $\tau_{\mathrm{A}}$,
can also be derived using HTL approximation at the operator level,
e.g. see the discussion in Refs. \citep{Yang:2020hri,Fang:2022ttm}.
For the purposes of this discussion, we simply treat $\Gamma_{\mathrm{A}}$
as a parameter.

The axial Wigner function then reduces to,%
\begin{eqnarray}
\mathcal{A}^{<,\mu}(X,p) & = & \mathcal{A}_{\mathrm{leq}}^{<,\mu}(X,p)+\delta\mathcal{A}_{\mathrm{RTA}}^{<,\mu}(X,p),\label{eq:Off-Leq=000020Axial=000020WF=000020RTA}
\end{eqnarray}
where 
\begin{eqnarray}
\delta\mathcal{A}_{\mathrm{RTA}}^{<,\mu}(X,p) & \approx & -2\pi\hbar\delta(p^{2})(\Gamma_{\mathrm{A}}-1)f_{\mathrm{V,leq}}^{<}(p)f_{\mathrm{V,leq}}^{>}(p)\left(\frac{u^{\mu}}{2}\beta p_{\rho}\omega^{\rho}+\frac{\beta p^{\langle\mu\rangle}}{2E_{\mathbf{p}}}p_{\rho}\omega^{\rho}\right).
\end{eqnarray}
The spin polarization pseudo-vector in Eq.(\ref{eq:Modified_CF_massive})
becomes,
\begin{eqnarray}
\mathcal{P}_{\mathrm{RTA}}^{\mu} & = & \mathcal{P}_{\mathrm{leq}}^{\mu}+\delta\mathcal{P}_{\mathrm{RTA}}^{\mu},\label{eq:RTA=000020C-F=000020formula}
\end{eqnarray}
where $\mathcal{P}_{\mathrm{leq}}^{\mu}$ is given by Eq. (\ref{eq:P-L=000020vector=000020Leq})
and 
\begin{eqnarray}
\delta\mathcal{P}_{\mathrm{RTA}}^{\mu} & = & \frac{\hbar^{2}}{4N}\int_{\Sigma}\mathrm{d}\Sigma\cdot pf_{\mathrm{V,leq}}^{<}(p)f_{\mathrm{V,leq}}^{>}(p)(1-\Gamma_{\mathrm{A}})\left(\frac{u^{\mu}}{2}\beta p_{\rho}\omega^{\rho}+\frac{\beta p^{\langle\mu\rangle}}{2E_{\mathbf{p}}}p_{\rho}\omega^{\rho}\right).\label{eq:RTA_vorticity=000020C-F=000020formula}
\end{eqnarray}
Interestingly, within the framework of RTA, the collisional corrections
to the spin polarization are related solely to the kinetic vorticity.
For more realistic cases, it might be reasonable to assume that $\tau_{\mathrm{A}}\propto\frac{\beta_{0}}{e^{4}\ln e^{-1}}$
by connecting Eq.(\ref{eq:SBE_RTA}) to the evolution equation for
spin density, and parameterize $\Gamma_{\mathrm{A}}=c/(e^{4}\ln e^{-1})$
with $c$ being a positive constant. We will present such result elsewhere
in the future.

\section{Conclusions }

\label{sec:Conclusions-and-outlook} In this work, we have computed
the collisional corrections to the SBE using Chapman-Enskog expansion
and derived the collisional corrections to spin polarization pseudo-vector.
We first derive the SBE for $f_{{\rm A}}$ in Eq. (\ref{eq:SBE=000020Moller=000020process}),
considering the M{\o}ller scattering process in both $u$- and $t$-channels.
For simplification, we consider two distinct scenarios.

In scenario (I), we consider the vector distribution function is off-equilibrium,
$f_{{\rm V}}\neq f_{\mathrm{V,leq}}^{<}$, and assume the applicability
of gradient expansion, i.e. $\mathcal{O}(\partial)\ll\mathcal{O}(Te^{4})$.
We revisit the matching condition in Eq. (\ref{eq:Axial=000020matching=000020condition})
in Sec. \ref{subsec:A-revisiting-of} and derive $\delta\mathcal{A}_{\textrm{(I)}}^{<,\mu}(X,p)$
shown in Eq. (\ref{eq:delta_A_sI}). Remarkably, we find that collisional
corrections to $\delta\mathcal{A}_{\textrm{(I)}}^{<,\mu}(p)$ in this
scenario do not explicitly depend on the coupling constant, similar
to the side-jump mechanism in the conductivities of anomalous and
spin Hall effects in condensed matter physics \citep{RevModPhys.82.1539,sinova2015spin}.
Consequently, the collisional corrections to the spin polarization
pseudo-vector $\delta\mathcal{P}_{\textrm{(I)}}^{\mu}(\mathbf{p})$
as shown in Eq. (\ref{eq:Case1-CF-formula}) , exhibit a similar property.

In scenario (II), we consider vector distribution function is in local
equilibrium $f_{{\rm V}}=f_{\mathrm{V,leq}}^{<}$. In such case, the
axial collision kernel in spin Boltzmann equation is similar to the
scalar-type Boltzmann equation as shown in Eq.(\ref{eq:ACK=000020Boltmann_type=000020fV_leq}).
The collisional correction to the axial Wigner function, $\delta\mathcal{A}_{\textrm{(II)}}^{<,\mu}(X,p)$,
and the spin polarization pseudo-vector, $\delta\mathcal{P}_{\textrm{(II)}}^{\mu}(\mathbf{p})$,
are detailed in Eqs. (\ref{eq:delta_A_sI}) and (\ref{eq:Off-eq=000020correction=000020C-F=000020formula}),
respectively. Interestingly, we observe that there are many new dissipative
corrections to $\delta\mathcal{P}_{\textrm{(II)}}^{\mu}(\mathbf{p})$
at $\mathcal{O}(\hbar^{2}\partial^{2})$ proportional to the spin
relaxation time $\tau_{{\rm A}}$.

As a comparison, we also studied the SBE using RTA and derive $\delta\mathcal{A}_{\mathrm{RTA}}^{<,\mu}(x,p)$
and $\delta\mathcal{P}_{\mathrm{RTA}}^{\mu}$, shown in Eqs. (\ref{eq:RTA_vorticity=000020C-F=000020formula})
and (\ref{eq:RTA_vorticity=000020C-F=000020formula}), respectively.
In this case, we observe that the collisional corrections are solely
induced by kinetic vorticity affecting the spin polarization.

Before concluding this work, we would like to address the limitations
of our current study. First, we emphasize that the local equilibrium
condition outlined in Eq. (\ref{eq:Leq_condition=000020HTL}) may
be influenced by the specifics of interactions and approximation schemes.
Secondly, our formalism for fermion elastic scattering in the HTL
approximation is limited to the leading-logarithm order. A complete
leading-order correction, including radiative corrections \citep{Arnold:2003zc}
and contributions from the photon Wigner function up to $\mathcal{O}(\hbar^{1})$
\citep{Huang:2020kik,Hattori:2020gqh}, are also crucial. Furthermore,
it is necessary to extend our techniques to real quantum chromodynamical
processes. Thirdly, future numerical simulations based on our results
are essential to quantitatively study the collisional corrections
to spin polarization. 
\begin{acknowledgments}
The authors would like to thank Di-Lun Yang and Yi Yin for useful discussions.
This work is supported in part by the National Key Research and Development
Program of China under Contract No. 2022YFA1605500, by the Chinese
Academy of Sciences (CAS) under Grants No. YSBR-088 and by National
Natural Science Foundation of China (NSFC) under Grants No. 12075235
and No. 12135011.
\end{acknowledgments}

\appendix

\section{Axial collision kernels\protect\label{sec:Axial-collisional-kernels}}

Here, we list parts of axial collision kernels in Eq. (\ref{eq:SBE=000020Moller=000020process}).
The $\mathcal{C}_{\mathrm{A}}^{\partial}[f_{\mathrm{V}}]$ and $\mathcal{C}_{\mathrm{A}}^{\Sigma}[f_{\mathrm{V}}]$
are, 

\begin{eqnarray}
\mathcal{C}_{\mathrm{A}}^{\partial}[f_{\mathrm{V}}] & = & 8e^{4}\int_{p^{\prime},k,k^{\prime}}w_{pk\to p^{\prime}k^{\prime},\alpha}^{1}(2\pi)^{3}\delta(k^{\prime,2})\delta(k^{2})\delta(p^{\prime,2})\nonumber \\
 &  & \qquad\qquad\times\hbar\partial_{\mu}\left\{ S^{(u),\mu\alpha}\left[f_{\mathrm{V}}^{<}(k^{\prime})f_{\mathrm{V}}^{<}(p^{\prime})f_{\mathrm{V}}^{>}(k)f_{\mathrm{V}}^{>}(p)-f_{\mathrm{V}}^{>}(k^{\prime})f_{\mathrm{V}}^{>}(p^{\prime})f_{\mathrm{V}}^{<}(k)f_{\mathrm{V}}^{<}(p)\right]\right\} ,\nonumber \\
 &  & +8e^{4}\int_{p^{\prime},k,k^{\prime}}w_{pk\to p^{\prime}k^{\prime},\mu}^{1}(2\pi)^{3}\delta(k^{\prime,2})\delta(k^{2})\delta(p^{\prime,2})\hbar S^{(u),\mu\alpha}(p)\nonumber \\
 &  & \qquad\qquad\times\left[f_{\mathrm{V}}^{<}(k^{\prime})f_{\mathrm{V}}^{<}(p^{\prime})f_{\mathrm{V}}^{>}(k)\partial_{\alpha}f_{\mathrm{V}}^{>}(p)-f_{\mathrm{V}}^{>}(k^{\prime})f_{\mathrm{V}}^{>}(p^{\prime})f_{\mathrm{V}}^{<}(k)\partial_{\alpha}f_{\mathrm{V}}^{<}(p)\right]\nonumber \\
 &  & +8e^{4}\int_{p^{\prime},k,k^{\prime}}w_{pk\to p^{\prime}k^{\prime},\mu}^{2}(2\pi)^{3}\delta(p^{\prime,2})\delta(k^{\prime,2})\delta(k^{2})\hbar S^{(u),\mu\alpha}(p^{\prime})\nonumber \\
 &  & \qquad\qquad\times\left[f_{\mathrm{V}}^{>}(p)f_{\mathrm{V}}^{>}(k)f_{\mathrm{V}}^{<}(k^{\prime})\partial_{\alpha}f_{\mathrm{V}}^{<}(p^{\prime})-f_{\mathrm{V}}^{<}(k)f_{\mathrm{V}}^{<}(p)f_{\mathrm{V}}^{>}(k^{\prime})\partial_{\alpha}f_{\mathrm{V}}^{>}(p^{\prime})\right]\nonumber \\
 &  & +8e^{4}\int_{p^{\prime},k,k^{\prime}}w_{pk\to p^{\prime}k^{\prime},\mu}^{3}(2\pi)^{3}\delta(k^{2})\delta(p^{\prime,2})\delta(k^{\prime,2})\hbar S^{(u),\mu\alpha}(k)\nonumber \\
 &  & \qquad\qquad\times\left[f_{\mathrm{V}}^{>}(p)f_{\mathrm{V}}^{<}(p^{\prime})f_{\mathrm{V}}^{<}(k^{\prime})\partial_{\alpha}f_{\mathrm{V}}^{>}(k)-f_{\mathrm{V}}^{>}(p^{\prime})f_{\mathrm{V}}^{>}(k^{\prime})f_{\mathrm{V}}^{<}(p)\partial_{\alpha}f_{\mathrm{V}}^{<}(k)\right]\nonumber \\
 &  & +8e^{4}\int_{p^{\prime},k,k^{\prime}}w_{pk\to p^{\prime}k^{\prime},\mu}^{4}(2\pi)^{3}\delta(k^{\prime,2})\delta(p^{\prime,2})\delta(k^{2})\hbar S^{(u),\mu\alpha}(k^{\prime})\nonumber \\
 &  & \qquad\qquad\times\left[f_{\mathrm{V}}^{>}(p)f_{\mathrm{V}}^{>}(k)f_{\mathrm{V}}^{<}(p^{\prime})\partial_{\alpha}f_{\mathrm{V}}^{<}(k^{\prime})-f_{\mathrm{V}}^{<}(p)f_{\mathrm{V}}^{<}(k)f_{\mathrm{V}}^{>}(p^{\prime})\partial_{\alpha}f_{\mathrm{V}}^{>}(k^{\prime})\right]\label{eq:CA=000020partial}
\end{eqnarray}
and
\begin{eqnarray}
\mathcal{C}_{\mathrm{A}}^{\Sigma}[f_{\mathrm{V}}] & = & 8e^{4}\int_{p^{\prime},k,k^{\prime}}w_{pk\to p^{\prime}k^{\prime},\mu}^{1}(2\pi)^{3}\delta(k^{\prime,2})\delta(k^{2})\delta(p^{\prime,2})\hbar S^{(u),\mu\alpha}(p)C_{\mathrm{V,\alpha}}[f_{\mathrm{V}}^{<}(p)]\nonumber \\
 &  & \qquad\qquad\times\left[f_{\mathrm{V}}^{<}(k^{\prime})f_{\mathrm{V}}^{>}(k)f_{\mathrm{V}}^{<}(p^{\prime})+f_{\mathrm{V}}^{>}(k^{\prime})f_{\mathrm{V}}^{<}(k)f_{\mathrm{V}}^{>}(p^{\prime})\right]\nonumber \\
 &  & -8e^{4}\int_{p^{\prime},k,k^{\prime}}w_{pk\to p^{\prime}k^{\prime},\mu}^{2}(2\pi)^{3}\delta(p^{\prime,2})\delta(k^{\prime,2})\delta(k^{2})\hbar S^{(u),\mu\alpha}(p^{\prime})C_{\mathrm{V,\alpha}}[f_{\mathrm{V}}^{<}(p^{\prime})]\nonumber \\
 &  & \qquad\qquad\times\left[f_{\mathrm{V}}^{>}(p)f_{\mathrm{V}}^{>}(k)f_{\mathrm{V}}^{<}(k^{\prime})+f_{\mathrm{V}}^{<}(k)f_{\mathrm{V}}^{<}(p)f_{\mathrm{V}}^{>}(k^{\prime})\right]\nonumber \\
 &  & +8e^{4}\int_{p^{\prime},k,k^{\prime}}w_{pk\to p^{\prime}k^{\prime},\mu}^{3}(2\pi)^{3}\delta(k^{2})\delta(p^{\prime,2})\delta(k^{\prime,2})\hbar S^{(u),\mu\alpha}(k)C_{\mathrm{V,\alpha}}[f_{\mathrm{V}}^{<}(k)]\nonumber \\
 &  & \qquad\qquad\times\left[f_{\mathrm{V}}^{>}(p)f_{\mathrm{V}}^{<}(p^{\prime})f_{\mathrm{V}}^{<}(k^{\prime})+f_{\mathrm{V}}^{>}(p^{\prime})f_{\mathrm{V}}^{>}(k^{\prime})f_{\mathrm{V}}^{<}(p)\right]\nonumber \\
 &  & -8e^{4}\int_{p^{\prime},k,k^{\prime}}w_{pk\to p^{\prime}k^{\prime},\mu}^{4}(2\pi)^{3}\delta(k^{\prime,2})\delta(p^{\prime,2})\delta(k^{2})\hbar S^{(u),\mu\alpha}(k^{\prime})C_{\mathrm{V},\alpha}[f_{\mathrm{V}}^{<}(k^{\prime})]\nonumber \\
 &  & \qquad\qquad\times\left[f_{\mathrm{V}}^{>}(p)f_{\mathrm{V}}^{>}(k)f_{\mathrm{V}}^{<}(p^{\prime})+f_{\mathrm{V}}^{<}(p)f_{\mathrm{V}}^{<}(k)f_{\mathrm{V}}^{>}(p^{\prime})\right]\label{eq:CA=000020Sigma}
\end{eqnarray}
where we have introduced the auxiliary scattering elements 
\begin{eqnarray}
w_{2,pk\to p^{\prime}k^{\prime}}^{\mu} & = & \frac{k^{\mu}(p\cdot k^{\prime})+k^{\prime,\mu}(p\cdot k)}{(p-p^{\prime})^{4}}+\frac{p\cdot kk^{\prime,\mu}}{(p-k^{\prime})^{2}(p-p^{\prime})^{2}},\label{eq:Auxillary_scattering_element_=00005Cmu2}\\
w_{pk\to p^{\prime}k^{\prime},\mu}^{3} & = & \frac{p_{\mu}p^{\prime}\cdot k^{\prime}-p_{\mu}^{\prime}p\cdot k^{\prime}}{(p-p^{\prime})^{4}}+\frac{p_{\mu}(p^{\prime}\cdot k^{\prime})}{(p-k^{\prime})^{2}(p-p^{\prime})^{2}},\label{eq:Auxillary_scattering_element_=00005Cmu3}\\
w_{pk\to p^{\prime}k^{\prime},\mu}^{4} & = & \frac{-p_{\mu}p^{\prime}\cdot k+p\cdot kp_{\mu}^{\prime}}{(p-p^{\prime})^{4}}+\frac{p\cdot kp_{\mu}^{\prime}}{(p-k^{\prime})^{2}(p-p^{\prime})^{2}}.\label{eq:Auxillary_scattering_element_=00005Cmu4}
\end{eqnarray}

\section{Thermodynamical functions and integrals}

We list the definitions of the thermodynamical functions and some
related integrals.

\subsection{Definition of moments \protect\label{subsec:Thermodynamical-functions-and}}

As noted in Sec.\ref{subsec:Spin-hydrodynamics-from}, spin degree
of freedom does not modify the equations of motion for viscous hydrodynamics
up to $\mathcal{O}(\partial^{1})$, 
\begin{eqnarray}
0 & = & u_{\mu}\partial_{\nu}T^{\mu\nu}=De+(e+P)\theta-\pi^{\mu\nu}\sigma_{\mu\nu},\label{eq:De=000020eq}\\
0 & = & \Delta_{\alpha}^{\mu}\partial_{\nu}T^{\alpha\nu}=(e+P)Du^{\mu}-\nabla^{\mu}P+\Delta_{\alpha}^{\mu}\partial_{\nu}\pi^{\alpha\nu},\label{eq:Du=000020eq}\\
0 & = & \partial_{\mu}j^{\mu}=Dn+n\theta+\partial_{\mu}\nu^{\mu}.\label{eq:Dn=000020eq}
\end{eqnarray}
Since $T^{[\mu\nu]}\sim\mathcal{O}(\hbar^{2}\partial^{2})$, we will
not include the conservation equation of angular momentum tensor as
explained in the context under Eq. (\ref{eq:Conservation=000020of=000020AMT}).
We then introduce the following moments \footnote{In principle, we should replace $E_{\mathbf{p}}^{n}\to E_{\mathbf{p}}^{n-2k}(E_{\mathbf{p}}^{2}-m^{2})^{k}$
in the massive case, but in the massless case the relations between
$I_{nq}$ and $I_{n0}$ is trivial, so we will focus on $I_{n0}$
in this work which can be analytically calculated.} \citep{Israel:1979wp,denicol2022microscopic},
\begin{eqnarray}
I_{nq} & = & \frac{1}{(2q+1)!!}4\int_{p}2\pi\delta(p^{2})f_{\mathrm{V,leq}}^{<}(x,p)E_{\mathbf{p}}^{n},\label{eq:Inq}\\
J_{nq} & = & \frac{1}{(2q+1)!!}4\int_{p}2\pi\delta(p^{2})f_{\mathrm{V,leq}}^{<}(x,p)f_{\mathrm{V,leq}}^{>}(x,p)E_{\mathbf{p}}^{n},\label{eq:Jnq}\\
K_{nq} & = & \frac{1}{(2q+1)!!}4\int_{p}2\pi\delta(p^{2})f_{\mathrm{V,leq}}^{<}(x,p)f_{\mathrm{V,leq}}^{>}(x,p)\left(1-2f_{\mathrm{V,leq}}^{>}(x,p)\right)E_{\mathbf{p}}^{n},\label{eq:Knq}
\end{eqnarray}
where we have dropped the sign function $\epsilon(u\cdot p)$ and
only focus on the positive-energy particles in this work. It is found
that, $n_{0}=I_{10}$, $e_{0}=I_{20}$ and $P=\frac{1}{3}I_{20}=I_{21}$.
$I_{nq},J_{nq}$ and $K_{nq}$ in Eqs.(\ref{eq:Inq}, \ref{eq:Jnq},
\ref{eq:Knq}) satisfy the following relations, 
\begin{eqnarray}
\beta J_{nq} & = & I_{n-1,q-1}+(n-2q)I_{n-1,q},\label{eq:Jnq-Inq=000020relation}\\
\beta K_{nq} & = & -J_{n-1,q-1}-(n-2q)J_{n-1,q},\label{eq:Knq-Jnq=000020relation}\\
\mathrm{d}I_{nq} & = & J_{nq}\mathrm{d}\alpha_{0}-J_{n+1,q}\mathrm{d}\beta_{0}.\label{eq:dI-J=000020relation}
\end{eqnarray}
From Eq.(\ref{eq:dI-J=000020relation}), we can derive the thermodynamical
relations,
\begin{eqnarray}
\mathrm{d}e_{0}=J_{20}\mathrm{d}\alpha_{0}-J_{30}\mathrm{d}\beta_{0} & ,\quad & \mathrm{d}n_{0}=J_{10}\mathrm{d}\alpha_{0}-J_{20}\mathrm{d}\beta_{0},\label{eq:d(e,n)-d(=00005Calpha,=00005Cbeta)}
\end{eqnarray}
and inversely, 
\begin{eqnarray}
\mathrm{d}\alpha_{0}=-\frac{J_{20}}{D_{20}}\mathrm{d}e_{0}+\frac{J_{30}}{D_{20}}\mathrm{d}n_{0} & ,\quad & \mathrm{d}\beta_{0}=-\frac{J_{10}}{D_{20}}\mathrm{d}e_{0}+\frac{J_{20}}{D_{20}}\mathrm{d}n_{0},\label{eq:d(=00005Calpha,=00005Cbeta)-d(e,n)}
\end{eqnarray}
where 
\begin{equation}
D_{nq}=J_{n+1,q}J_{n-1,q}-J_{nq}^{2}.
\end{equation}
Combining the above equations with Eqs.(\ref{eq:De=000020eq},\ref{eq:Dn=000020eq})
up to $\mathcal{O}(\partial^{1})$, we find 
\begin{eqnarray}
D\alpha_{0} & = & \frac{J_{20}(e+P)-J_{30}n}{D_{20}}\theta+\mathcal{O}(\partial^{2})=\mathcal{O}(\partial^{2}),\label{eq:D=00005Calpha}\\
D\beta_{0} & = & \frac{J_{10}(e+P)-J_{20}n}{D_{20}}\theta+\mathcal{O}(\partial^{2})=\frac{\beta}{3}\theta+\mathcal{O}(\partial^{2}).\label{eq:D=00005Cbeta}
\end{eqnarray}
We also list some other useful relations from Eqs.(\ref{eq:dI-J=000020relation},\ref{eq:Du=000020eq}),
\begin{eqnarray}
\mathrm{d}P_{0}(\alpha_{0},\beta_{0}) & = & \frac{n_{0}}{\beta_{0}}\mathrm{d}\alpha_{0}-\frac{e+P}{\beta_{0}}\mathrm{d}\beta_{0},\label{eq:dP}\\
Du_{\mu} & = & \frac{1}{\beta_{0}}h_{0}^{-1}\nabla_{\mu}\alpha_{0}-\frac{1}{\beta_{0}}\nabla_{\mu}\beta_{0}+\mathcal{O}(\partial^{2}),\label{eq:Du-D(=00005Calpha,=00005Cbeta)}
\end{eqnarray}
where $h_{0}=\frac{e_{0}+P_{0}}{n_{0}}=\frac{4}{3}\frac{I_{20}}{I_{10}}$
is the enthalpy density. 

\subsection{Definitions of some axial thermodynamical functions in Eqs.(\ref{eq:epsilon_A:=000020eqs1},
\ref{eq:epsilon_A:=000020eqs2}) \protect\label{subsec:Definitions-of-thermodynamical}}

The definition of the axial thermodynamical functions in Eqs.(\ref{eq:epsilon_A:=000020eqs1},
\ref{eq:epsilon_A:=000020eqs2}) are listed here.
\begin{eqnarray}
\alpha_{\mathrm{A},r}^{\mathrm{s},1} & = & -\alpha_{\mathrm{A},r}^{\mathrm{s},2}=-\frac{1}{3}\beta\int_{p}2\pi\delta(p^{2})E_{\mathbf{p}}^{r}f_{\mathrm{V,leq}}^{<}(p)f_{\mathrm{V,leq}}^{>}(p),\\
\alpha_{\mathrm{A},r}^{\mathrm{s},3} & = & -\int_{p}2\pi\delta(p^{2})E_{\mathbf{p}}^{r}f_{\mathrm{V,leq}}^{<}(p)f_{\mathrm{V,leq}}^{>}(p)\nonumber \\
 &  & \quad\times\left[\frac{2}{3}h_{0}^{-1}-\frac{4}{3}E_{\mathbf{p}}^{-1}+\frac{\beta}{3}(1-E_{\mathbf{p}}h_{0}^{-1})\left(1-2f_{\mathrm{V,leq}}^{>}(p)\right)\right]\\
\alpha_{\mathrm{A},r}^{\mathrm{v},1} & = & \frac{1}{3}\int_{p}2\pi\delta(p^{2})E_{\mathbf{p}}^{r+2}f_{\mathrm{V,leq}}^{<}(p)f_{\mathrm{V,leq}}^{>}(p)\left[\frac{12}{5}E_{\mathbf{p}}^{-1}-\frac{2}{5}\beta\left(1-2f_{\mathrm{V,leq}}^{>}(p)\right)\right],\\
\alpha_{\mathrm{A},r}^{\mathrm{v},2} & = & \frac{1}{3\beta}\int_{p}2\pi\delta(p^{2})E_{\mathbf{p}}^{r+1}f_{\mathrm{V,leq}}^{<}(p)f_{\mathrm{V,leq}}^{>}(p)\left(-E_{\mathbf{p}}^{-1}+h_{0}^{-1}\right),\\
\alpha_{\mathrm{A},r}^{\mathrm{ts},1} & = & 2\alpha_{A,r}^{\mathrm{ts},2}=\alpha_{A,r}^{\mathrm{ts},3}=-\frac{4}{15\beta}\int_{p}2\pi\delta(p^{2})E_{\mathbf{p}}^{r+2}f_{\mathrm{V,leq}}^{<}(p)f_{\mathrm{V,leq}}^{>}(p),\\
\alpha_{\mathrm{A},r}^{\mathrm{ts},4} & = & \frac{2}{15}\int_{p}2\pi\delta(p^{2})E_{\mathbf{p}}^{r+2}f_{\mathrm{V,leq}}^{<}(p)f_{\mathrm{V,leq}}^{>}(p)(h_{0}^{-1}+E_{\mathbf{p}}^{-1}),\\
\alpha_{\mathrm{A},r}^{\mathrm{ts},5} & = & -\frac{2}{15\beta^{2}}\int_{p}2\pi\delta(p^{2})E_{\mathbf{p}}^{r+3}f_{\mathrm{V,leq}}^{<}(p)f_{\mathrm{V,leq}}^{>}(p)\nonumber \\
 &  & \quad\times\left[E_{\mathbf{p}}^{-2}-2E_{\mathbf{p}}^{-1}h_{0}^{-1}+\beta(h_{0}^{-1}-E_{\mathbf{p}}^{-1})\left(1-2f_{\mathrm{V,leq}}^{>}(p)\right)\right]\\
\alpha_{\mathrm{A},r}^{\mathrm{tt}} & = & \frac{2}{35\beta^{2}}\int_{p}2\pi\delta(p^{2})E_{\mathbf{p}}^{r+4}f_{\mathrm{V,leq}}^{<}(p)f_{\mathrm{V,leq}}^{>}(p)\left[\left(1-2f_{\mathrm{\mathrm{V,leq}}}^{>}(p)\right)\beta-E_{\mathbf{p}}^{-1}\right].
\end{eqnarray}

\subsection{Some useful integrals \protect\label{subsec:Some-thermodynamical-integrals}}

We present some analytical results of the integrals defined in Eqs.(\ref{eq:Inq},\ref{eq:Jnq},\ref{eq:Knq}).
We first have, 
\begin{eqnarray}
 &  & I_{-1,0}=\frac{1}{\pi^{2}}\frac{\ln2}{\beta_{0}},\quad I_{00}=\frac{1}{12\beta_{0}^{2}},\quad I_{10}=\frac{3\zeta(3)}{2\pi^{2}\beta_{0}^{3}},\quad I_{20}=\frac{7\pi^{2}}{120\beta_{0}^{4}},\nonumber \\
 &  & I_{30}=\frac{45\zeta(5)}{2\pi^{2}\beta_{0}^{5}},\quad I_{40}=\frac{31\pi^{4}}{252\beta_{0}^{6}},\quad J_{-1,0}=\frac{1}{2\pi^{2}\beta_{0}},\quad K_{-1,0}=-\frac{1}{4\pi^{2}\beta_{0}},\label{eq:Thermo=000020integral-1}
\end{eqnarray}
Using the relations in Eqs.(\ref{eq:Jnq-Inq=000020relation},\ref{eq:Knq-Jnq=000020relation}),
we find, 
\begin{eqnarray}
J_{10} & = & \frac{1}{6\beta_{0}^{3}},\quad J_{20}=\frac{9\zeta(3)}{2\pi^{2}\beta_{0}^{4}},\quad K_{20}=-\frac{1}{2\beta_{0}^{4}},\quad K_{30}=-\frac{18\zeta(3)}{\pi^{2}\beta_{0}^{5}},\nonumber \\
J_{30} & = & \frac{7\pi^{2}}{30\beta_{0}^{5}},\quad J_{40}=\frac{225\zeta(5)}{2\pi^{2}\beta_{0}^{6}},\quad K_{40}=-\frac{7\pi^{2}}{6\beta_{0}^{6}}.\label{eq:Thermo=000020integral-2}
\end{eqnarray}
Accordingly, the axial thermodynamical functions are given by,
\begin{eqnarray}
\alpha_{\mathrm{A},0}^{\mathrm{s},1}=-\alpha_{\mathrm{A},0}^{\mathrm{s},2}=-\frac{\ln2}{12\pi^{2}\beta_{0}} & ,\quad & \alpha_{\mathrm{A},0}^{\mathrm{s},3}=\frac{5}{24\pi^{2}\beta_{0}}-\frac{45\zeta(3)\ln2}{7\pi^{6}\beta_{0}},\label{eq:Alpha_s123}\\
\alpha_{\mathrm{A},0}^{\mathrm{v},1}=\frac{1}{20\beta_{0}^{3}} & ,\quad & \alpha_{\mathrm{A},0}^{\mathrm{v},2}=-\frac{\ln2}{12\pi^{2}\beta_{0}^{3}}+\frac{15\zeta(3)}{56\pi^{4}\beta_{0}^{3}},\label{eq:Alpha_v12}\\
\alpha_{\mathrm{A},0}^{\mathrm{ts},1}=2\alpha_{A,0}^{\mathrm{ts},2}=\alpha_{A,0}^{\mathrm{ts},3}=-\frac{3\zeta(3)}{10\pi^{2}\beta_{0}^{5}} & ,\quad & \alpha_{\mathrm{A},0}^{\mathrm{ts},4}=\frac{81\zeta^{2}(3)}{28\pi^{6}\beta_{0}^{3}}+\frac{1}{180\beta_{0}^{3}},\label{eq:Alpha_ts1-4}\\
\alpha_{\mathrm{A},0}^{\mathrm{ts},5}=-\frac{1}{45\beta_{0}^{5}}+\frac{243\zeta^{2}(3)}{14\pi^{6}\beta_{0}^{5}} & ,\quad & \alpha_{\mathrm{A},0}^{\mathrm{tt}}=-\frac{\pi^{2}}{50\beta_{0}^{7}}.\label{eq:Alpha_ts5-tt}
\end{eqnarray}

\section{Collision integrals\protect\label{sec:Collision-integrals}}

The collision integrals are calculated using the HTL approximation
as illustrated in the end of Sec.\ref{subsec:Self-energies-in-2-2}.
We derive the collision integrals in SBE (\ref{eq:Axial=000020collision=000020int=0000201},
\ref{eq:Axial=000020collision=000020int=0000202}, \ref{eq:Axial=000020collision=000020int=0000203},
\ref{eq:Axial=000020collision=000020int=0000204}) in Appendix. \ref{subsec:Axial-collision-integrals}
and illustrate the local equilibrium condition in our specific process
and approximation scheme in Appendix.\ref{subsec:Local-equilibrium-condition}.

\subsection{Axial collision integrals\protect\label{subsec:Axial-collision-integrals}}

The axial collision integrals appeared in Eqs.(\ref{eq:phi_Ap_s,v},\ref{eq:phi_Ap_t})
can be calculated explicitly. We firstly calculate the contractors,
\begin{eqnarray}
p^{\langle\beta\rangle}(p_{\langle\beta\rangle}-p_{\langle\beta\rangle}^{\prime}) & = & -z^{\prime}|\mathbf{p}||\mathbf{q}|,\\
p^{\langle\alpha}p^{\rho\rangle}(p_{\langle\alpha}p_{\rho\rangle}-p_{\langle\alpha}^{\prime}p_{\rho\rangle}^{\prime}) & = & |\mathbf{p}|^{3}|\mathbf{q}|\frac{4}{3}z^{\prime}-|\mathbf{p}|^{2}|\mathbf{q}|^{2}(z^{\prime,2}-\frac{1}{3}),\\
p^{\langle\mu}p^{\alpha}p^{\lambda\rangle}(p_{\langle\mu}p_{\alpha}p_{\lambda\rangle}-p_{\langle\mu}^{\prime}p_{\alpha}^{\prime}p_{\lambda\rangle}^{\prime}) & = & -\frac{6}{5}|\mathbf{p}|^{5}|\mathbf{q}|z^{\prime}-3|\mathbf{p}|^{4}|\mathbf{q}|^{2}\frac{1-3z^{\prime,2}}{5},
\end{eqnarray}
with $p^{\prime}=p-q$. We decompose 
\begin{equation}
k^{\mu}=u^{\mu}E_{\mathbf{k}}+z|\mathbf{k}|\hat{q}_{\perp}^{\mu}+k^{\{\mu\}},
\end{equation}
choosing $\hat{q}_{\perp}^{\mu}$ as another reference direction (or
the $z$-direction) and $k^{\{\mu\}}=\Theta^{\mu\nu}(q)k_{\nu}$ with
$z=-\hat{q}_{\perp}^{\mu}\hat{k}_{\perp,\mu}$ \citep{Molnar:2016vvu,denicol2022microscopic},
and then obtain the following projections \citep{Yang:2020hri,Fang:2022ttm},
\begin{eqnarray}
\hat{k}_{\perp\alpha} & \to & z\hat{q}_{\perp,\alpha},\label{eq:k=000020Projection-1}\\
\hat{k}_{\perp\alpha}\hat{k}_{\perp\beta} & \to & z^{2}\hat{q}_{\perp,\alpha}\hat{q}_{\perp,\beta}+\frac{\Theta_{\alpha\beta}(q)}{2}(z^{2}-1),\label{eq:k=000020Projection-2}\\
\hat{k}_{\perp,\mu}\hat{k}_{\perp,\alpha}\hat{k}_{\perp,\beta} & \to & z^{3}\hat{q}_{\perp,\mu}\hat{q}_{\perp,\alpha}\hat{q}_{\perp,\beta}+\frac{z^{2}-1}{2}z\left(\Theta_{\mu\alpha}(q)\hat{q}_{\perp,\beta}+\Theta_{\mu\beta}(q)\hat{q}_{\perp,\alpha}+\Theta_{\alpha\beta}\hat{q}_{\perp,\mu}\right),\label{eq:k=000020Projection-3}
\end{eqnarray}
Similarly for $q\to p$ with $z^{\prime}=-\hat{p}_{\perp}^{\mu}\hat{q}_{\perp,\mu}=\cos\langle\mathbf{p},\mathbf{q}\rangle$.
To obtain the leading-log results, we only need terms of $\int_{eT}^{T}|\mathbf{q}|^{-1}\mathrm{d}|\mathbf{q}|$.
We notice that $\mathrm{d}^{4}q\sim|\mathbf{q}|^{2}\mathrm{d}|\mathbf{q}|$,
$\delta(k^{\prime,2})\sim|\mathbf{q}|^{-1}$ and the contractors are
at least of $|\mathbf{q}|^{1}$, so we at most need $q^{-4}(1+q)+\mathcal{O}(|\mathbf{q}|^{-2})$.
Such assertion is consistent with Ref. \citep{Arnold:2000dr}. Using
the tensor projections, the only possible nontrivial tensor structure
lies in $W_{pk\to p^{\prime}k^{\prime}}$,
\begin{eqnarray}
W_{pk\to p^{\prime}k^{\prime}} & = & 8e^{4}\frac{(p\cdot q)^{2}+2(p\cdot q)(p\cdot k)+2(p\cdot k)^{2}}{q^{4}}\nonumber \\
 & \to & \frac{16e^{4}|\mathbf{p}|^{2}|\mathbf{k}|^{2}}{|\mathbf{q}|^{4}(\hat{q}_{0}^{2}-1)^{2}}(-2zz^{\prime}+\frac{3-z^{2}}{2}+\frac{3z^{2}-1}{2}z^{\prime,2})+\mathcal{O}(|\mathbf{q}|^{-2}).\label{eq:W_projection}
\end{eqnarray}

We now perform an example on calculating $\mathcal{A}_{{\rm A},01}^{\mathrm{s}}$,
\begin{eqnarray}
\mathcal{A}_{\mathrm{A},01}^{\mathrm{s}} & = & \lambda^{2}\int\frac{\mathrm{d}^{4}p\mathrm{d}^{4}q}{(2\pi)^{8}}\int_{0}^{+\infty}|\mathbf{k}|\mathrm{d}|\mathbf{k}|\int_{-1}^{1}\mathrm{d}z\int\mathrm{d}k_{0}\frac{4e^{4}|\mathbf{p}||\mathbf{k}|}{|\mathbf{q}|^{4}(\hat{q}_{0}^{2}-1)^{2}}(-2zz^{\prime}+\frac{3-z^{2}}{2}+\frac{3z^{2}-1}{2}z^{\prime,2})\nonumber \\
 &  & \quad\times2\pi\delta(p^{2})\delta\left(z-(\hat{q}_{0}+\hat{q}^{2}\frac{|\mathbf{q}|}{2|\mathbf{k}|})\right)\delta(k_{0}-|\mathbf{k}|)\delta\left((p-q)^{2}\right)\nonumber \\
 &  & \quad\times f_{V,eq}^{<}(q+k)f_{V,eq}^{<}(p-q)f_{V,eq}^{>}(k)f_{V,eq}^{>}(p)\left(z^{\prime}+\frac{z^{\prime,2}-1}{2}\frac{|\mathbf{q}|}{|\mathbf{p}|}\right)\nonumber \\
 & = & \lambda^{2}\int\frac{\mathrm{d}^{4}p}{(2\pi)^{7}}\int_{eT}^{T}\mathrm{d}|\mathbf{q}|\int_{-1}^{1}\mathrm{d}z^{\prime}\int\mathrm{d}q_{0}\frac{2e^{4}}{|\mathbf{q}|^{2}(\hat{q}_{0}^{2}-1)^{2}}\left[-2\hat{q}_{0}z^{\prime}+\frac{3-z^{\prime,2}}{2}+\frac{3z^{\prime,2}-1}{2}\hat{q}_{0}^{2}\right]\nonumber \\
 &  & \quad\times2\pi\delta(p^{2})\delta\left(q_{0}-|\mathbf{q}|z^{\prime}+\frac{1-z^{\prime,2}}{2}\frac{|\mathbf{q}|^{2}}{|\mathbf{p}|}\right)f_{V,eq}^{<}(p)f_{V,eq}^{>}(p)\left(1+\beta|\mathbf{q}|\hat{q}_{0}f_{V,eq}^{>}(p)\right)\nonumber \\
 &  & \quad\times\left(1+z^{\prime}\frac{|\mathbf{q}|}{|\mathbf{p}|}\right)\left(\frac{\pi^{2}}{6\beta^{3}}-|\mathbf{q}|\hat{q}_{0}\frac{\pi^{2}+12\ln2}{12\beta^{2}}\right)\left(z^{\prime}+\frac{z^{\prime,2}-1}{2}\frac{|\mathbf{q}|}{|\mathbf{p}|}\right)\nonumber \\
 &  & +\lambda^{2}\int\frac{\mathrm{d}^{4}p}{(2\pi)^{7}}\int_{eT}^{T}\mathrm{d}|\mathbf{q}|\int_{-1}^{1}\mathrm{d}z^{\prime}\int\mathrm{d}q_{0}\frac{e^{4}}{|\mathbf{q}|(\hat{q}_{0}^{2}-1)}\left[(3z^{\prime,2}-1)z^{\prime}\hat{q}_{0}-2z^{\prime,2}\right]\nonumber \\
 &  & \quad\times(2\pi)\delta(p^{2})\delta(q_{0}-|\mathbf{q}|z^{\prime})f_{V,eq}^{<}(p)f_{V,eq}^{>}(p)\frac{\ln2}{\beta^{2}}\nonumber \\
 & = & -\lambda^{2}e^{4}\ln\frac{1}{e}\int\frac{\mathrm{d}^{4}p}{(2\pi)^{7}}(2\pi)\delta(p^{2})f_{V,eq}^{<}(p)f_{V,eq}^{>}(p)\left[\frac{\pi^{2}}{6\beta^{2}}\left(1-2f_{V,eq}^{>}(p)\right)+\frac{\pi^{2}}{3\beta^{3}}\frac{1}{|\mathbf{p}|}\right]\nonumber \\
 & = & -\frac{\lambda^{2}e^{4}\ln\frac{1}{e}}{192\pi\beta^{3}}\left(\beta K_{00}+2J_{-1,0}\right).\label{eq:A_01=000020s}
\end{eqnarray}
where we have used,
\begin{eqnarray}
\delta(k^{2}) & = & \frac{\delta(k_{0}-|\mathbf{k}|)}{2|\mathbf{k}|},\label{eq:delta(k2)}\\
\delta\left((q+k)^{2}\right) & = & \frac{1}{2|\mathbf{q}||\mathbf{k}|}\delta\left(z-\frac{q^{2}+2q_{0}k_{0}}{2|\mathbf{q}||\mathbf{k}|}\right).\label{eq:delta(q+k)^2}
\end{eqnarray}
and 
\begin{eqnarray}
\delta\left((p-q)^{2}\right) & = & \frac{1}{2|\mathbf{p}|}\delta\left(q_{0}-|\mathbf{q}|z^{\prime}+\frac{1-z^{\prime,2}}{2}\frac{|\mathbf{q}|^{2}}{|\mathbf{p}|}\right)\left(1+z^{\prime}\frac{|\mathbf{q}|}{|\mathbf{p}|}\right)+\mathcal{O}(|\mathbf{q}|^{2}),\label{eq:delta(p-q)^2}\\
|\mathbf{p}-\mathbf{q}| & = & |\mathbf{p}|\left(1-\frac{|\mathbf{q}|z^{\prime}}{|\mathbf{p}|}+\frac{(1-z^{\prime,2})}{2}\frac{|\mathbf{q}|^{2}}{|\mathbf{p}|^{2}}\right)+\mathcal{O}(\frac{|\mathbf{q}|^{3}}{|\mathbf{p}|^{3}}).\label{eq:|p-q|=000020expansion}
\end{eqnarray}
Different from the self-energies computed in our previous work \citep{Fang:2022ttm},
which are divergent both collinearly and infraredly, the collision
integral is divergence free as expected because it will directly appear
in the physical quantities.

Following the similar calculations, we can get,
\begin{eqnarray}
\mathcal{A}_{\mathrm{A},00}^{\mathrm{v}}=\frac{\lambda^{2}e^{4}\ln\frac{1}{e}}{576\pi\beta^{2}}K_{10} & ,\quad & \mathcal{A}_{\mathrm{A},00}^{\mathrm{ts}}=-\frac{\lambda^{2}e^{4}\ln\frac{1}{e}}{720\pi\beta^{2}}K_{30}\label{eq:A_00=000020v,ts}\\
\mathcal{A}_{\mathrm{A},00}^{\mathrm{tt}}=\frac{\lambda^{2}e^{4}\ln\frac{1}{e}}{1120\pi\beta^{2}}K_{50} & .\label{eq:A_00=000020t}
\end{eqnarray}

\subsection{Local equilibrium condition for quantum kinetic theory\protect\label{subsec:Local-equilibrium-condition}}

We will show that the local equilibrium conditions for the QKT in
M{\o}ller scattering process is simply Eq.(\ref{eq:Leq_condition=000020HTL}),
$\omega_{\mu\nu}^{s}=\Omega_{\mu\nu}.$ 

The Wigner functions in local equilibrium with spin chemical potential
are, 
\begin{eqnarray}
\mathcal{V}_{\mathrm{leq}}^{<,\mu}(x,p) & = & 2\pi\delta(p^{2})p^{\mu}f_{\mathrm{V},\mathrm{leq}}^{<}(x,p),\label{eq:Wigner-V=000020Leq}\\
\mathcal{A}_{\mathrm{leq}}^{<,\mu}(x,p) & = & 2\pi\hbar\delta(p^{2})f_{\mathrm{V,leq}}^{<}(p)f_{\mathrm{V,leq}}^{>}(p)\left[S_{(u)}^{\mu\alpha}(p)p_{\nu}(\omega_{\alpha}^{s,\;\;\nu}-\Omega_{\alpha}^{\;\;\nu}-\xi_{\alpha}^{\nu})\right.\nonumber \\
 &  & \quad\qquad\left.+\frac{1}{4}\epsilon^{\mu\rho\alpha\beta}p_{\rho}\omega_{\alpha\beta}^{s}+S_{(u)}^{\mu\alpha}(p)\nabla_{\alpha}\alpha_{0}\right],\label{eq:Wigner-A=000020Leq_spin=000020chemical=000020potential}
\end{eqnarray}
where we have defined the thermal shear $\xi_{\mu\nu}=\partial_{(\mu}\beta u_{\nu)}$.
From Eq.(\ref{eq:V=000020BE=000020Moller=000020process}), we find
that the conditions in Eq.(\ref{eq:Wigner-V=000020Leq}) are sufficient
to make the vector collision kernel zero, $\mathcal{C}_{\mathrm{V}}=0$.
For $\mathcal{C}_{\mathrm{A}}$, we use Eqs.(\ref{eq:Vector=000020SE},
\ref{eq:Axial=000020SE}, \ref{eq:Wigner-V=000020Leq}, \ref{eq:Wigner-A=000020Leq_spin=000020chemical=000020potential})
to get, 
\begin{eqnarray}
\mathcal{C}_{\mathrm{A,leq}} & = & -8\hbar e^{4}\int_{p^{\prime},k,k^{\prime}}w_{pk\to p^{\prime}k^{\prime},\mu}^{1,\perp}(2\pi)^{3}\delta(k^{\prime,2})\delta(k^{2})\delta(p^{\prime,2})\nonumber \\
 &  & \quad\qquad\times f_{\mathrm{V,leq}}^{>}(p)f_{\mathrm{V,leq}}^{>}(k)f_{\mathrm{V,leq}}^{<}(k^{\prime})f_{\mathrm{V,leq}}^{<}(p^{\prime})S_{(u)}^{\mu\alpha}(p)\left[p_{\nu}(\omega_{\alpha}^{s,\;\;\nu}-\Omega_{\alpha}^{\;\;\nu}-\xi_{\alpha}^{\nu})+\nabla_{\alpha}\alpha_{0}\right]\nonumber \\
 &  & +8\hbar e^{4}\int_{p^{\prime},k,k^{\prime}}w_{pk\to p^{\prime}k^{\prime},\mu}^{2,\perp}(2\pi)^{3}\delta(k^{2})\delta(k^{\prime,2})\delta(p^{\prime2})\nonumber \\
 &  & \quad\qquad\times f_{\mathrm{V,leq}}^{<}(k^{\prime})f_{\mathrm{V,leq}}^{>}(p)f_{\mathrm{V,leq}}^{>}(k)f_{\mathrm{V,leq}}^{<}(p^{\prime})S_{(u)}^{\mu\alpha}(p^{\prime})\left[p_{\nu}^{\prime}(\omega_{\alpha}^{s,\;\;\nu}-\Omega_{\alpha}^{\;\;\nu}-\xi_{\alpha}^{\nu})+\nabla_{\alpha}\alpha_{0}\right]\nonumber \\
 &  & -8\hbar e^{4}\int_{p^{\prime},k,k^{\prime}}w_{pk\to p^{\prime}k^{\prime},\mu}^{3,\perp}(2\pi)^{3}\delta(k^{2})\delta(p^{\prime,2})\delta(k^{\prime,2})\nonumber \\
 &  & \quad\qquad\times f_{\mathrm{V,leq}}^{>}(p)f_{\mathrm{V,leq}}^{>}(k)f_{\mathrm{V,leq}}^{<}(p^{\prime})f_{\mathrm{V,leq}}^{<}(k^{\prime})S_{(u)}^{\mu\alpha}(k)\left[k_{\nu}(\omega_{\alpha}^{s,\;\;\nu}-\Omega_{\alpha}^{\;\;\nu}-\xi_{\alpha}^{\nu})+\nabla_{\alpha}\alpha_{0}\right]\nonumber \\
 &  & +8\hbar e^{4}\int_{p^{\prime},k,k^{\prime}}w_{pk\to p^{\prime}k^{\prime},\mu}^{4,\perp}(2\pi)^{3}\delta(k^{2})\delta(p^{\prime,2})\delta(k^{\prime2})\nonumber \\
 &  & \quad\qquad\times f_{\mathrm{V,leq}}^{>}(p)f_{\mathrm{V,leq}}^{>}(k)f_{\mathrm{V,leq}}^{<}(p^{\prime})f_{\mathrm{V,leq}}^{<}(k^{\prime})S_{(u)}^{\mu\alpha}(k^{\prime})\left[k_{\nu}^{\prime}(\omega_{\alpha}^{s,\;\;\nu}-\Omega_{\alpha}^{\;\;\nu}-\xi_{\alpha}^{\nu})+\nabla_{\alpha}\alpha_{0}\right]\nonumber \\
 &  & +\mathcal{C}_{\mathrm{A,leq}}^{\omega^{s}},
\end{eqnarray}
where $w_{pk\to p^{\prime}k^{\prime},\mu}^{i}$ are defined in Eqs.(\ref{eq:Auxillary_scattering_element_=00005Cmu1},
\ref{eq:Auxillary_scattering_element_=00005Cmu2}, \ref{eq:Auxillary_scattering_element_=00005Cmu3},
\ref{eq:Auxillary_scattering_element_=00005Cmu4}) and we have used
$w_{pk\to p^{\prime}k^{\prime},\mu}^{i,\perp}=\Delta_{\mu}^{\alpha}w_{pk\to p^{\prime}k^{\prime},\alpha}^{i}$.
Here,
\begin{eqnarray}
\mathcal{C}_{\mathrm{A,leq}}^{\omega^{s}} & = & 8\hbar e^{4}\int_{p^{\prime},k,k^{\prime}}\left[\frac{(p\cdot k^{\prime})(p_{\rho}^{\prime}k_{\mu}+p_{\mu}^{\prime}k_{\rho}-k_{\mu}^{\prime}p_{\rho}-p_{\mu}k_{\rho}^{\prime})+(p\cdot k)(p_{\rho}^{\prime}k_{\mu}^{\prime}+p_{\mu}^{\prime}k_{\rho}^{\prime}-p_{\mu}k_{\rho}-p_{\rho}k_{\mu})}{(p-p^{\prime})^{4}}\right.\nonumber \\
 &  & \qquad\qquad\qquad\left.+\frac{(p\cdot k)(p_{\rho}^{\prime}k_{\mu}^{\prime}+p_{\mu}^{\prime}k_{\rho}^{\prime})-(p_{\rho}k_{\mu}+p_{\mu}k_{\rho})(p^{\prime}\cdot k^{\prime})}{(p-k^{\prime})^{2}(p-p^{\prime})^{2}}\right]\frac{1}{4}\epsilon^{\mu\rho\alpha\beta}\omega_{\alpha\beta}^{s}\nonumber \\
 &  & \quad\qquad\times(2\pi)^{4}\delta(p^{2})\delta(k^{2})\delta(k^{\prime,2})\delta(p^{\prime2})f_{\mathrm{V,leq}}^{<}(k^{\prime})f_{\mathrm{V,leq}}^{>}(p)f_{\mathrm{V,leq}}^{>}(k)f_{\mathrm{V,leq}}^{<}(p^{\prime}),\nonumber \\
 & = & 0.
\end{eqnarray}

Then using the projection from $k_{\mu}$ on $q_{\mu}$ in Eqs.(\ref{eq:k=000020Projection-1},
\ref{eq:k=000020Projection-2}, \ref{eq:k=000020Projection-3}) and
from $q_{\mu}$ on $p_{\mu}$, we can show that, all the coefficient
of $\xi_{\alpha}^{\nu}$ and $\nabla_{\alpha}\alpha_{0}$ are zero.
Now we are left with,
\begin{eqnarray}
\mathcal{C}_{\mathrm{A,leq}} & = & 8\hbar e^{4}S_{(u)}^{\alpha\mu}(p)(\omega_{\alpha\mu}^{s}-\Omega_{\alpha\mu})\int\frac{\mathrm{d}^{4}q\mathrm{d}^{4}k}{(2\pi)^{5}}\delta(k^{2})\delta\left((k+q)^{2}\right)\delta\left((p-q)^{2}\right)\nonumber \\
 &  & \quad\qquad\times f_{\mathrm{V,leq}}^{<}(q+k)f_{\mathrm{V,leq}}^{>}(p)f_{\mathrm{V,leq}}^{>}(k)f_{\mathrm{V,leq}}^{<}(p-q)L(p,q,k),
\end{eqnarray}
where we have denoted the integral kernel, 
\begin{eqnarray}
L(p,q,k) & = & \frac{z^{\prime,2}-1}{2}\frac{|\mathbf{p}||\mathbf{q}|}{E_{\mathbf{p}-\mathbf{q}}}\left(\frac{|\mathbf{q}||\mathbf{k}|^{2}(z^{2}-1)}{q^{4}}+|\mathbf{p}||\mathbf{q}||\mathbf{k}|\frac{(\hat{q}_{0}-z^{\prime})z+(1-z^{\prime}z)}{q^{4}}\right.\nonumber \\
 &  & \quad\left.+|\mathbf{p}||\mathbf{k}|^{2}\frac{2(z-z^{\prime}\frac{3z^{2}-1}{2})}{q^{4}}-\frac{|\mathbf{q}|+|\mathbf{k}|z}{q^{2}}\right)\nonumber \\
 &  & +\left(\frac{|\mathbf{p}||\mathbf{q}|(\hat{q}_{0}-z^{\prime})}{q^{4}}+\frac{1}{q^{2}}\right)|\mathbf{p}||\mathbf{k}|(\frac{3z^{2}-1}{2}\frac{z^{\prime,2}-1}{2}+\frac{z^{2}-1}{2})\nonumber \\
 &  & +\frac{z^{2}-1}{2}\frac{|\mathbf{p}||\mathbf{q}||\mathbf{k}|}{q^{4}}\left(z\frac{3z^{\prime,2}-1}{2}|\mathbf{k}|+z^{\prime}(-|\mathbf{q}|(\hat{q}_{0}-z^{\prime})-|\mathbf{k}|)\right)\nonumber \\
 &  & +\left(|\mathbf{k}|z+|\mathbf{q}|\right)\frac{z^{\prime,2}-1}{2}\left(\frac{-|\mathbf{p}||\mathbf{q}|(\hat{q}_{0}-z^{\prime})}{q^{4}}-\frac{1}{q^{2}}\right)\frac{|\mathbf{p}||\mathbf{q}|}{E_{\mathbf{q}+\mathbf{k}}}\nonumber \\
 &  & +\left(|\mathbf{k}|(\frac{3z^{2}-1}{2}\frac{z^{\prime,2}-1}{2}+\frac{z^{2}-1}{2})+|\mathbf{q}|z\frac{z^{\prime,2}-1}{2}\right)\left(\frac{-|\mathbf{p}||\mathbf{q}|(\hat{q}_{0}-z^{\prime})}{q^{4}}-\frac{1}{q^{2}}\right)\frac{|\mathbf{p}||\mathbf{k}|}{E_{\mathbf{q}+\mathbf{k}}}\nonumber \\
 &  & +\frac{z^{2}-1}{2}\frac{|\mathbf{q}||\mathbf{k}|^{2}}{E_{\mathbf{q}+\mathbf{k}}}\left((\frac{3z^{\prime,2}-1}{2}z-z^{\prime})\frac{|\mathbf{p}||\mathbf{k}|}{q^{4}}+\frac{z^{\prime,2}-1}{2}\frac{|\mathbf{p}||\mathbf{q}|}{q^{4}}+\frac{z^{\prime}}{q^{2}}\right).
\end{eqnarray}
Now let us keep the results up to leading-log order, insert the delta
functions (\ref{eq:delta(k2)}, \ref{eq:delta(q+k)^2}, \ref{eq:delta(p-q)^2}),
and integrate them over $z$ and $|\mathbf{k}|$. The $\mathcal{C}_{\mathrm{A,leq}}$
becomes, 
\begin{eqnarray}
\mathcal{C}_{\mathrm{A,leq}} & = & \hbar e^{4}S_{(u)}^{\alpha\rho}(p)(\omega_{\alpha\rho}^{s}-\Omega_{\alpha\rho})\frac{\int_{eT}^{T}\mathrm{d}|\mathbf{q}|\int_{-1}^{1}\mathrm{d}z^{\prime}\int\mathrm{d}q_{0}}{(2\pi)^{3}}\delta\left(q_{0}-|\mathbf{q}|z^{\prime}+\frac{1-z^{\prime,2}}{2}\frac{|\mathbf{q}|^{2}}{|\mathbf{p}|}\right)f_{\mathrm{V,leq}}^{>}(p)f_{\mathrm{V,leq}}^{<}(p)\nonumber \\
 &  & \quad\times\frac{1}{|\mathbf{q}|(\hat{q}_{0}^{2}-1)}\left\{ \frac{z^{\prime,2}-1}{2}\frac{1-z^{\prime}\hat{q}_{0}}{\hat{q}_{0}^{2}-1}\frac{\ln2}{\beta^{2}}+\frac{z^{\prime,2}-1}{2}\frac{1}{|\mathbf{p}|}\frac{\pi^{2}}{6\beta^{3}}\right.\nonumber \\
 &  & \quad\qquad+\frac{1}{|\mathbf{q}|}\left(1+\beta|\mathbf{q}|\hat{q}_{0}f_{\mathrm{V,leq}}^{>}(p)+z^{\prime}\frac{|\mathbf{q}|}{|\mathbf{p}|}\right)\left(\frac{\pi^{2}}{6\beta^{3}}-\frac{|\mathbf{q}|\hat{q}_{0}}{\beta^{2}}(\frac{\pi^{2}}{12}+\ln2)\right)\nonumber \\
 &  & \quad\qquad\quad\times\left(-z^{\prime}+\hat{q}_{0}\frac{3z^{\prime,2}-1}{2}+(\hat{q}_{0}-z^{\prime}\frac{3\hat{q}_{0}^{2}-1}{2})(z^{\prime,2}-1)(1+z^{\prime}\frac{|\mathbf{q}|}{|\mathbf{p}|})\right)\nonumber \\
 &  & \quad\qquad\left.+\frac{1}{2}\left((2\hat{q}_{0}^{2}-1)\frac{(3z^{\prime,2}-1)}{2}+\frac{3(z^{\prime,2}-1)}{2}-\hat{q}_{0}z^{\prime}(3z^{\prime,2}-2)\right)\frac{\ln2}{\beta^{2}}\right\} ,
\end{eqnarray}
After further integrating $\mathcal{C}_{\mathrm{A,leq}}$ over $q_{0},z^{\prime},$
and $|\mathbf{q}|$, we arrive at,
\begin{eqnarray}
\mathcal{C}_{\mathrm{A,leq}} & = & \frac{\hbar e^{4}\ln\frac{1}{e}}{(2\pi)^{3}}f_{\mathrm{V,leq}}^{>}(p)f_{\mathrm{V,leq}}^{<}(p)S_{(u)}^{\mu\alpha}(p)(\omega_{\mu\alpha}^{s}-\Omega_{\mu\alpha})\frac{\pi^{2}}{6\beta^{3}|\mathbf{p}|}.
\end{eqnarray}

Therefore, with the distribution functions defined in Eqs. (\ref{eq:fA=000020local=000020eq},
\ref{eq:fV=000020local=000020eq}) in local equilibrium, the corresponding
collision kernels $\mathcal{C}_{\mathrm{A,leq}}$ must vanish. We
only need 
\begin{eqnarray}
\omega_{\mu\alpha}^{s}-\Omega_{\mu\alpha} & = & 0,
\end{eqnarray}
to hold. 

\bibliographystyle{h-physrev}
\bibliography{qkt-ref}

\begin{thebibliography}{100}

\bibitem{PhysRev.6.239}
S.~J. Barnett,
\newblock Phys. Rev. {\bf 6}, 239 (1915).

\bibitem{1915KNAB...18..696E}
A.~{Einstein} and W.~J. {de Haas},
\newblock Koninklijke Nederlandse Akademie van Wetenschappen Proceedings Series B Physical Sciences {\bf 18}, 696 (1915).

\bibitem{Liang:2004ph}
Z.-T. Liang and X.-N. Wang,
\newblock Phys. Rev. Lett. {\bf 94}, 102301 (2005), nucl-th/0410079,
\newblock [Erratum: Phys.Rev.Lett. 96, 039901 (2006)].

\bibitem{Liang:2004xn}
Z.-T. Liang and X.-N. Wang,
\newblock Phys. Lett. B {\bf 629}, 20 (2005), nucl-th/0411101.

\bibitem{Gao:2007bc}
J.-H. Gao {\em et~al.},
\newblock Phys. Rev. C {\bf 77}, 044902 (2008), 0710.2943.

\bibitem{STAR:2017ckg}
STAR, L.~Adamczyk {\em et~al.},
\newblock Nature {\bf 548}, 62 (2017), 1701.06657.

\bibitem{STAR:2019erd}
STAR, J.~Adam {\em et~al.},
\newblock Phys. Rev. Lett. {\bf 123}, 132301 (2019), 1905.11917.

\bibitem{ALICE:2019aid}
ALICE, S.~Acharya {\em et~al.},
\newblock Phys. Rev. Lett. {\bf 125}, 012301 (2020), 1910.14408.

\bibitem{STAR:2020xbm}
STAR, J.~Adam {\em et~al.},
\newblock Phys. Rev. Lett. {\bf 126}, 162301 (2021), 2012.13601.

\bibitem{STAR:2022fan}
STAR, M.~S. Abdallah {\em et~al.},
\newblock Nature {\bf 614}, 244 (2023), 2204.02302.

\bibitem{ALICE:2023jad}
ALICE, S.~Acharya {\em et~al.},
\newblock Phys. Rev. Lett. {\bf 131}, 042303 (2023).

\bibitem{Sheng:2019kmk}
X.-L. Sheng, L.~Oliva, and Q.~Wang,
\newblock Phys. Rev. D {\bf 101}, 096005 (2020), 1910.13684.

\bibitem{Sheng:2020ghv}
X.-L. Sheng, Q.~Wang, and X.-N. Wang,
\newblock Phys. Rev. D {\bf 102}, 056013 (2020), 2007.05106.

\bibitem{Sheng:2022ffb}
X.-L. Sheng, L.~Oliva, Z.-T. Liang, Q.~Wang, and X.-N. Wang,
\newblock (2022), 2206.05868.

\bibitem{Sheng:2022wsy}
X.-L. Sheng, L.~Oliva, Z.-T. Liang, Q.~Wang, and X.-N. Wang,
\newblock Phys. Rev. Lett. {\bf 131}, 042304 (2023), 2205.15689.

\bibitem{Sheng:2023urn}
X.-L. Sheng, S.~Pu, and Q.~Wang,
\newblock Phys. Rev. C {\bf 108}, 054902 (2023), 2308.14038.

\bibitem{Muller:2021hpe}
B.~M\"uller and D.-L. Yang,
\newblock Phys. Rev. D {\bf 105}, L011901 (2022), 2110.15630,
\newblock [Erratum: Phys.Rev.D 106, 039904 (2022)].

\bibitem{Kumar:2023ghs}
A.~Kumar, B.~M\"uller, and D.-L. Yang,
\newblock Phys. Rev. D {\bf 108}, 016020 (2023), 2304.04181.

\bibitem{Karpenko:2016jyx}
I.~Karpenko and F.~Becattini,
\newblock Eur. Phys. J. C {\bf 77}, 213 (2017), 1610.04717.

\bibitem{Becattini:2017gcx}
F.~Becattini and I.~Karpenko,
\newblock Phys. Rev. Lett. {\bf 120}, 012302 (2018), 1707.07984.

\bibitem{Xie:2017upb}
Y.~Xie, D.~Wang, and L.~P. Csernai,
\newblock Phys. Rev. C {\bf 95}, 031901 (2017), 1703.03770.

\bibitem{Pang:2016igs}
L.-G. Pang, H.~Petersen, Q.~Wang, and X.-N. Wang,
\newblock Phys. Rev. Lett. {\bf 117}, 192301 (2016), 1605.04024.

\bibitem{Li:2017slc}
H.~Li, L.-G. Pang, Q.~Wang, and X.-L. Xia,
\newblock Phys. Rev. C {\bf 96}, 054908 (2017), 1704.01507.

\bibitem{Wei:2018zfb}
D.-X. Wei, W.-T. Deng, and X.-G. Huang,
\newblock Phys. Rev. C {\bf 99}, 014905 (2019), 1810.00151.

\bibitem{Ryu:2021lnx}
S.~Ryu, V.~Jupic, and C.~Shen,
\newblock Phys. Rev. C {\bf 104}, 054908 (2021), 2106.08125.

\bibitem{Shi:2017wpk}
S.~Shi, K.~Li, and J.~Liao,
\newblock Phys. Lett. B {\bf 788}, 409 (2019), 1712.00878.

\bibitem{Fu:2020oxj}
B.~Fu, K.~Xu, X.-G. Huang, and H.~Song,
\newblock Phys. Rev. C {\bf 103}, 024903 (2021), 2011.03740.

\bibitem{Sun:2017xhx}
Y.~Sun and C.~M. Ko,
\newblock Phys. Rev. {\bf C96}, 024906 (2017), 1706.09467.

\bibitem{Wu:2022mkr}
X.-Y. Wu, C.~Yi, G.-Y. Qin, and S.~Pu,
\newblock Phys. Rev. C {\bf 105}, 064909 (2022), 2204.02218.

\bibitem{Alzhrani:2022dpi}
S.~Alzhrani, S.~Ryu, and C.~Shen,
\newblock Phys. Rev. C {\bf 106}, 014905 (2022), 2203.15718.

\bibitem{Becattini:2013fla}
F.~Becattini, V.~Chandra, L.~Del~Zanna, and E.~Grossi,
\newblock Annals Phys. {\bf 338}, 32 (2013), 1303.3431.

\bibitem{Fang:2016vpj}
R.-h. Fang, L.-g. Pang, Q.~Wang, and X.-n. Wang,
\newblock Phys. Rev. C {\bf 94}, 024904 (2016), 1604.04036.

\bibitem{Becattini:2020ngo}
F.~Becattini and M.~A. Lisa,
\newblock Ann. Rev. Nucl. Part. Sci. {\bf 70}, 395 (2020), 2003.03640.

\bibitem{Gao:2020vbh}
J.-H. Gao, G.-L. Ma, S.~Pu, and Q.~Wang,
\newblock Nucl. Sci. Tech. {\bf 31}, 90 (2020), 2005.10432.

\bibitem{Hidaka:2022dmn}
Y.~Hidaka, S.~Pu, Q.~Wang, and D.-L. Yang,
\newblock Prog. Part. Nucl. Phys. {\bf 127}, 103989 (2022), 2201.07644.

\bibitem{Becattini:2022zvf}
F.~Becattini,
\newblock Rept. Prog. Phys. {\bf 85}, 122301 (2022), 2204.01144.

\bibitem{Becattini:2024uha}
F.~Becattini {\em et~al.},
\newblock (2024), 2402.04540.

\bibitem{Xia:2018tes}
X.-L. Xia, H.~Li, Z.-B. Tang, and Q.~Wang,
\newblock Phys. Rev. C {\bf 98}, 024905 (2018), 1803.00867.

\bibitem{Liu:2021uhn}
S.~Y.~F. Liu and Y.~Yin,
\newblock JHEP {\bf 07}, 188 (2021), 2103.09200.

\bibitem{Fu:2021pok}
B.~Fu, S.~Y.~F. Liu, L.~Pang, H.~Song, and Y.~Yin,
\newblock Phys. Rev. Lett. {\bf 127}, 142301 (2021), 2103.10403.

\bibitem{Becattini:2021suc}
F.~Becattini, M.~Buzzegoli, and A.~Palermo,
\newblock Phys. Lett. B {\bf 820}, 136519 (2021), 2103.10917.

\bibitem{Becattini:2021iol}
F.~Becattini, M.~Buzzegoli, G.~Inghirami, I.~Karpenko, and A.~Palermo,
\newblock Phys. Rev. Lett. {\bf 127}, 272302 (2021), 2103.14621.

\bibitem{Hidaka:2017auj}
Y.~Hidaka, S.~Pu, and D.-L. Yang,
\newblock Phys. Rev. D {\bf 97}, 016004 (2018), 1710.00278.

\bibitem{Yi:2021ryh}
C.~Yi, S.~Pu, and D.-L. Yang,
\newblock Phys. Rev. C {\bf 104}, 064901 (2021), 2106.00238.

\bibitem{Fu:2022myl}
B.~Fu, L.~Pang, H.~Song, and Y.~Yin,
\newblock (2022), 2201.12970.

\bibitem{STAR:2023eck}
STAR, M.~Abdulhamid {\em et~al.},
\newblock Phys. Rev. Lett. {\bf 131}, 202301 (2023), 2303.09074.

\bibitem{Yi:2024kwu}
C.~Yi, X.-Y. Wu, J.~Zhu, S.~Pu, and G.-Y. Qin,
\newblock (2024), 2408.04296.

\bibitem{Florkowski:2017ruc}
W.~Florkowski, B.~Friman, A.~Jaiswal, and E.~Speranza,
\newblock Phys. Rev. C {\bf 97}, 041901 (2018), 1705.00587.

\bibitem{Florkowski:2018fap}
W.~Florkowski, A.~Kumar, and R.~Ryblewski,
\newblock Prog. Part. Nucl. Phys. {\bf 108}, 103709 (2019), 1811.04409.

\bibitem{Florkowski:2018myy}
W.~Florkowski, E.~Speranza, and F.~Becattini,
\newblock Acta Phys. Polon. B {\bf 49}, 1409 (2018), 1803.11098.

\bibitem{Hattori:2019lfp}
K.~Hattori, M.~Hongo, X.-G. Huang, M.~Matsuo, and H.~Taya,
\newblock Phys. Lett. B {\bf 795}, 100 (2019), 1901.06615.

\bibitem{Hongo:2021ona}
M.~Hongo, X.-G. Huang, M.~Kaminski, M.~Stephanov, and H.-U. Yee,
\newblock JHEP {\bf 11}, 150 (2021), 2107.14231.

\bibitem{Hongo:2022izs}
M.~Hongo, X.-G. Huang, M.~Kaminski, M.~Stephanov, and H.-U. Yee,
\newblock (2022), 2201.12390.

\bibitem{Cao:2022aku}
Z.~Cao, K.~Hattori, M.~Hongo, X.-G. Huang, and H.~Taya,
\newblock PTEP {\bf 2022}, 071D01 (2022), 2205.08051.

\bibitem{Li:2020eon}
S.~Li, M.~A. Stephanov, and H.-U. Yee,
\newblock Phys. Rev. Lett. {\bf 127}, 082302 (2021), 2011.12318.

\bibitem{Fukushima:2020qta}
K.~Fukushima and S.~Pu,
\newblock Lect. Notes Phys. {\bf 987}, 381 (2021), 2001.00359.

\bibitem{Fukushima:2020ucl}
K.~Fukushima and S.~Pu,
\newblock Phys. Lett. B {\bf 817}, 136346 (2021), 2010.01608.

\bibitem{Wang:2021ngp}
D.-L. Wang, S.~Fang, and S.~Pu,
\newblock Phys. Rev. D {\bf 104}, 114043 (2021), 2107.11726.

\bibitem{Wang:2021wqq}
D.-L. Wang, X.-Q. Xie, S.~Fang, and S.~Pu,
\newblock Phys. Rev. D {\bf 105}, 114050 (2022), 2112.15535.

\bibitem{Xie:2023gbo}
X.-Q. Xie, D.-L. Wang, C.~Yang, and S.~Pu,
\newblock (2023), 2306.13880.

\bibitem{She:2021lhe}
D.~She, A.~Huang, D.~Hou, and J.~Liao,
\newblock Sci. Bull. {\bf 67}, 2265 (2022), 2105.04060.

\bibitem{Bhadury:2020puc}
S.~Bhadury, W.~Florkowski, A.~Jaiswal, A.~Kumar, and R.~Ryblewski,
\newblock Phys. Lett. B {\bf 814}, 136096 (2021), 2002.03937.

\bibitem{Bhadury:2020cop}
S.~Bhadury, W.~Florkowski, A.~Jaiswal, A.~Kumar, and R.~Ryblewski,
\newblock Phys. Rev. D {\bf 103}, 014030 (2021), 2008.10976.

\bibitem{Bhadury:2022ulr}
S.~Bhadury, W.~Florkowski, A.~Jaiswal, A.~Kumar, and R.~Ryblewski,
\newblock Phys. Rev. Lett. {\bf 129}, 192301 (2022), 2204.01357.

\bibitem{Biswas:2023qsw}
R.~Biswas, A.~Daher, A.~Das, W.~Florkowski, and R.~Ryblewski,
\newblock Phys. Rev. D {\bf 108}, 014024 (2023), 2304.01009.

\bibitem{Garbiso:2020puw}
M.~Garbiso and M.~Kaminski,
\newblock JHEP {\bf 12}, 112 (2020), 2007.04345.

\bibitem{Montenegro:2017lvf}
D.~Montenegro, L.~Tinti, and G.~Torrieri,
\newblock Phys. Rev. D {\bf 96}, 076016 (2017), 1703.03079.

\bibitem{Montenegro:2017rbu}
D.~Montenegro, L.~Tinti, and G.~Torrieri,
\newblock Phys. Rev. D {\bf 96}, 056012 (2017), 1701.08263,
\newblock [Addendum: Phys.Rev.D 96, 079901 (2017)].

\bibitem{Peng:2021ago}
H.-H. Peng, J.-J. Zhang, X.-L. Sheng, and Q.~Wang,
\newblock Chin. Phys. Lett. {\bf 38}, 116701 (2021), 2107.00448.

\bibitem{Weickgenannt:2022zxs}
N.~Weickgenannt, D.~Wagner, E.~Speranza, and D.~H. Rischke,
\newblock Phys. Rev. D {\bf 106}, 096014 (2022), 2203.04766.

\bibitem{Weickgenannt:2022qvh}
N.~Weickgenannt, D.~Wagner, E.~Speranza, and D.~H. Rischke,
\newblock Phys. Rev. D {\bf 106}, L091901 (2022), 2208.01955.

\bibitem{Weickgenannt:2022zhj}
N.~Weickgenannt,
\newblock {\em {Dissipative spin hydrodynamics from quantum field theory}},
\newblock PhD thesis, Goethe U., Frankfurt (main), 2022.

\bibitem{Weickgenannt:2023btk}
N.~Weickgenannt,
\newblock (2023), 2307.13561.

\bibitem{Gallegos:2020otk}
A.~D. Gallegos and U.~G\"ursoy,
\newblock JHEP {\bf 11}, 151 (2020), 2004.05148.

\bibitem{Gallegos:2021bzp}
A.~D. Gallegos, U.~G\"ursoy, and A.~Yarom,
\newblock SciPost Phys. {\bf 11}, 041 (2021), 2101.04759.

\bibitem{Shi:2020htn}
S.~Shi, C.~Gale, and S.~Jeon,
\newblock Phys. Rev. C {\bf 103}, 044906 (2021), 2008.08618.

\bibitem{Hu:2021lnx}
J.~Hu,
\newblock Phys. Rev. D {\bf 103}, 116015 (2021), 2101.08440.

\bibitem{Hu:2021pwh}
J.~Hu,
\newblock Phys. Rev. D {\bf 105}, 076009 (2022), 2111.03571.

\bibitem{Hu:2022azy}
J.~Hu,
\newblock Phys. Rev. C {\bf 107}, 024915 (2023), 2209.10979.

\bibitem{Becattini:2023ouz}
F.~Becattini, A.~Daher, and X.-L. Sheng,
\newblock Phys. Lett. B {\bf 850}, 138533 (2024), 2309.05789.

\bibitem{Wagner:2024fhf}
D.~Wagner, M.~Shokri, and D.~H. Rischke,
\newblock (2024), 2405.00533.

\bibitem{Ren:2024pur}
X.~Ren, C.~Yang, D.-L. Wang, and S.~Pu,
\newblock (2024), 2405.03105.

\bibitem{Wang:2024afv}
D.-L. Wang, L.~Yan, and S.~Pu,
\newblock (2024), 2408.03781.

\bibitem{Gao:2019znl}
J.-H. Gao and Z.-T. Liang,
\newblock Phys. Rev. D {\bf 100}, 056021 (2019), 1902.06510.

\bibitem{Hattori:2019ahi}
K.~Hattori, Y.~Hidaka, and D.-L. Yang,
\newblock Phys. Rev. D {\bf 100}, 096011 (2019), 1903.01653.

\bibitem{Weickgenannt:2019dks}
N.~Weickgenannt, X.-L. Sheng, E.~Speranza, Q.~Wang, and D.~H. Rischke,
\newblock Phys. Rev. D {\bf 100}, 056018 (2019), 1902.06513.

\bibitem{Weickgenannt:2020aaf}
N.~Weickgenannt, E.~Speranza, X.-l. Sheng, Q.~Wang, and D.~H. Rischke,
\newblock Phys. Rev. Lett. {\bf 127}, 052301 (2021), 2005.01506.

\bibitem{Yang:2020hri}
D.-L. Yang, K.~Hattori, and Y.~Hidaka,
\newblock JHEP {\bf 07}, 070 (2020), 2002.02612.

\bibitem{Liu:2020flb}
Y.-C. Liu, K.~Mameda, and X.-G. Huang,
\newblock Chin. Phys. C {\bf 44}, 094101 (2020), 2002.03753,
\newblock [Erratum: Chin.Phys.C 45, 089001 (2021)].

\bibitem{Weickgenannt:2021cuo}
N.~Weickgenannt, E.~Speranza, X.-l. Sheng, Q.~Wang, and D.~H. Rischke,
\newblock Phys. Rev. D {\bf 104}, 016022 (2021), 2103.04896.

\bibitem{Sheng:2021kfc}
X.-L. Sheng, N.~Weickgenannt, E.~Speranza, D.~H. Rischke, and Q.~Wang,
\newblock Phys. Rev. D {\bf 104}, 016029 (2021), 2103.10636.

\bibitem{Wang:2021qnt}
Z.~Wang and P.~Zhuang,
\newblock (2021), 2105.00915.

\bibitem{Weickgenannt:2022jes}
N.~Weickgenannt, D.~Wagner, and E.~Speranza,
\newblock Phys. Rev. D {\bf 105}, 116026 (2022), 2204.01797.

\bibitem{Wang:2019moi}
Z.~Wang, X.~Guo, S.~Shi, and P.~Zhuang,
\newblock Phys. Rev. D {\bf 100}, 014015 (2019), 1903.03461.

\bibitem{Huang:2020wrr}
A.~Huang {\em et~al.},
\newblock Phys. Rev. D {\bf 103}, 056025 (2021), 2007.02858.

\bibitem{Lin:2021mvw}
S.~Lin,
\newblock Phys. Rev. D {\bf 105}, 076017 (2022), 2109.00184.

\bibitem{Fang:2022ttm}
S.~Fang, S.~Pu, and D.-L. Yang,
\newblock Phys. Rev. D {\bf 106}, 016002 (2022), 2204.11519.

\bibitem{Wang:2022yli}
Z.~Wang,
\newblock Phys. Rev. D {\bf 106}, 076011 (2022), 2205.09334.

\bibitem{Lin:2022tma}
S.~Lin and Z.~Wang,
\newblock JHEP {\bf 12}, 030 (2022), 2206.12573.

\bibitem{Sheng:2022ssd}
X.-L. Sheng, Q.~Wang, and D.~H. Rischke,
\newblock Phys. Rev. D {\bf 106}, L111901 (2022), 2202.10160.

\bibitem{Wagner:2022amr}
D.~Wagner, N.~Weickgenannt, and D.~H. Rischke,
\newblock Phys. Rev. D {\bf 106}, 116021 (2022), 2210.06187.

\bibitem{Li:2019qkf}
S.~Li and H.-U. Yee,
\newblock Phys. Rev. D {\bf 100}, 056022 (2019), 1905.10463.

\bibitem{Das:2022azr}
A.~Das, W.~Florkowski, A.~Kumar, R.~Ryblewski, and R.~Singh,
\newblock (2022), 2203.15562.

\bibitem{Yang:2021fea}
D.-L. Yang,
\newblock JHEP {\bf 06}, 140 (2022), 2112.14392.

\bibitem{Fang:2023bbw}
S.~Fang, S.~Pu, and D.-L. Yang,
\newblock (2023), 2311.15197.

\bibitem{Huang:2020kik}
X.-G. Huang, P.~Mitkin, A.~V. Sadofyev, and E.~Speranza,
\newblock JHEP {\bf 10}, 117 (2020), 2006.03591.

\bibitem{Hattori:2020gqh}
K.~Hattori, Y.~Hidaka, N.~Yamamoto, and D.-L. Yang,
\newblock JHEP {\bf 02}, 001 (2021), 2010.13368.

\bibitem{Mameda:2022ojk}
K.~Mameda, N.~Yamamoto, and D.-L. Yang,
\newblock Phys. Rev. D {\bf 105}, 096019 (2022), 2203.08449.

\bibitem{Son:2012wh}
D.~T. Son and N.~Yamamoto,
\newblock Phys. Rev. Lett. {\bf 109}, 181602 (2012), 1203.2697.

\bibitem{Son:2012zy}
D.~T. Son and N.~Yamamoto,
\newblock Phys. Rev. D {\bf 87}, 085016 (2013), 1210.8158.

\bibitem{Stephanov:2012ki}
M.~A. Stephanov and Y.~Yin,
\newblock Phys. Rev. Lett. {\bf 109}, 162001 (2012), 1207.0747.

\bibitem{Gao:2012ix}
J.-H. Gao, Z.-T. Liang, S.~Pu, Q.~Wang, and X.-N. Wang,
\newblock Phys. Rev. Lett. {\bf 109}, 232301 (2012), 1203.0725.

\bibitem{Chen:2012ca}
J.-W. Chen, S.~Pu, Q.~Wang, and X.-N. Wang,
\newblock Phys. Rev. Lett. {\bf 110}, 262301 (2013), 1210.8312.

\bibitem{Chen:2014cla}
J.-Y. Chen, D.~T. Son, M.~A. Stephanov, H.-U. Yee, and Y.~Yin,
\newblock Phys. Rev. Lett. {\bf 113}, 182302 (2014), 1404.5963.

\bibitem{Chen:2015gta}
J.-Y. Chen, D.~T. Son, and M.~A. Stephanov,
\newblock Phys. Rev. Lett. {\bf 115}, 021601 (2015), 1502.06966.

\bibitem{Hidaka:2016yjf}
Y.~Hidaka, S.~Pu, and D.-L. Yang,
\newblock Phys. Rev. D {\bf 95}, 091901 (2017), 1612.04630.

\bibitem{Huang:2017tsq}
A.~Huang, Y.~Jiang, S.~Shi, J.~Liao, and P.~Zhuang,
\newblock Phys. Lett. B {\bf 777}, 177 (2018), 1703.08856.

\bibitem{Wang:2020pej}
Z.~Wang, X.~Guo, and P.~Zhuang,
\newblock Eur. Phys. J. C {\bf 81}, 799 (2021), 2009.10930.

\bibitem{Yamamoto:2023okm}
N.~Yamamoto and D.-L. Yang,
\newblock (2023), 2308.08257.

\bibitem{Lin:2024zik}
S.~Lin and Z.~Wang,
\newblock (2024), 2406.10003.

\bibitem{Xiao:2009rm}
D.~Xiao, M.-C. Chang, and Q.~Niu,
\newblock Rev. Mod. Phys. {\bf 82}, 1959 (2010), 0907.2021.

\bibitem{RevModPhys.82.1539}
N.~Nagaosa, J.~Sinova, S.~Onoda, A.~H. MacDonald, and N.~P. Ong,
\newblock Rev. Mod. Phys. {\bf 82}, 1539 (2010).

\bibitem{sinova2015spin}
J.~Sinova, S.~O. Valenzuela, J.~Wunderlich, C.~Back, and T.~Jungwirth,
\newblock Reviews of modern physics {\bf 87}, 1213 (2015).

\bibitem{DeGroot:1980dk}
S.~R. De~Groot,
\newblock {\em {Relativistic Kinetic Theory. Principles and Applications}} (, 1980).

\bibitem{denicol2022microscopic}
G.~S. Denicol and D.~H. Rischke,
\newblock {\em Microscopic Foundations of Relativistic Fluid Dynamics} (Springer, 2022).

\bibitem{Blaizot:2001nr}
J.-P. Blaizot and E.~Iancu,
\newblock Phys. Rept. {\bf 359}, 355 (2002), hep-ph/0101103.

\bibitem{Blaizot:1996az}
J.-P. Blaizot and E.~Iancu,
\newblock Phys. Rev. D {\bf 55}, 973 (1997), hep-ph/9607303.

\bibitem{Arnold:2000dr}
P.~B. Arnold, G.~D. Moore, and L.~G. Yaffe,
\newblock JHEP {\bf 11}, 001 (2000), hep-ph/0010177.

\bibitem{Peskin:1995ev}
M.~E. Peskin and D.~V. Schroeder,
\newblock {\em {An Introduction to quantum field theory}} (Addison-Wesley, Reading, USA, 1995).

\bibitem{Blaizot:1999xk}
J.-P. Blaizot and E.~Iancu,
\newblock Nucl. Phys. B {\bf 557}, 183 (1999), hep-ph/9903389.

\bibitem{Elze:1986qd}
H.~T. Elze, M.~Gyulassy, and D.~Vasak,
\newblock Nucl. Phys. {\bf B276}, 706 (1986).

\bibitem{Vasak:1987um}
D.~Vasak, M.~Gyulassy, and H.~T. Elze,
\newblock Annals Phys. {\bf 173}, 462 (1987).

\bibitem{Itzykson:1980rh}
C.~Itzykson and J.~B. Zuber,
\newblock {\em {Quantum Field Theory}}International Series In Pure and Applied Physics (McGraw-Hill, New York, 1980).

\bibitem{Kovtun:2012rj}
P.~Kovtun,
\newblock J. Phys. A {\bf 45}, 473001 (2012), 1205.5040.

\bibitem{Becattini:2012tc}
F.~Becattini,
\newblock Phys. Rev. Lett. {\bf 108}, 244502 (2012), 1201.5278.

\bibitem{Liu:2020dxg}
S.~Y.~F. Liu and Y.~Yin,
\newblock Phys. Rev. D {\bf 104}, 054043 (2021), 2006.12421.

\bibitem{Bellac:2011kqa}
M.~L. Bellac,
\newblock {\em {Thermal Field Theory}}Cambridge Monographs on Mathematical Physics (Cambridge University Press, 2011).

\bibitem{Deng:2012pc}
W.-T. Deng and X.-G. Huang,
\newblock Phys. Rev. C {\bf 85}, 044907 (2012), 1201.5108.

\bibitem{Mrowczynski:1992hq}
S.~Mrowczynski and U.~W. Heinz,
\newblock Annals Phys. {\bf 229}, 1 (1994).

\bibitem{Arnold:2002zm}
P.~B. Arnold, G.~D. Moore, and L.~G. Yaffe,
\newblock JHEP {\bf 01}, 030 (2003), hep-ph/0209353.

\bibitem{Arnold:2003zc}
P.~B. Arnold, G.~D. Moore, and L.~G. Yaffe,
\newblock JHEP {\bf 05}, 051 (2003), hep-ph/0302165.

\bibitem{Cooper:1974mv}
F.~Cooper and G.~Frye,
\newblock Phys. Rev. D {\bf 10}, 186 (1974).

\bibitem{Andronic:2017pug}
A.~Andronic, P.~Braun-Munzinger, K.~Redlich, and J.~Stachel,
\newblock Nature {\bf 561}, 321 (2018), 1710.09425.

\bibitem{Arnold:2007pg}
P.~B. Arnold,
\newblock Int. J. Mod. Phys. E {\bf 16}, 2555 (2007), 0708.0812.

\bibitem{Yang:2018lew}
D.-L. Yang,
\newblock Phys. Rev. D {\bf 98}, 076019 (2018), 1807.02395.

\bibitem{landau2003fluid}
L.~Landau and E.~Lifshits,
\newblock {\em Fluid Mechanics}Teoreticheskaia fizika (Butterworth-Heinemann, 2003).

\bibitem{Israel:1979wp}
W.~Israel and J.~M. Stewart,
\newblock Annals Phys. {\bf 118}, 341 (1979).

\bibitem{chapman1990mathematical}
S.~Chapman and T.~Cowling,
\newblock {\em The Mathematical Theory of Non-uniform Gases: An Account of the Kinetic Theory of Viscosity, Thermal Conduction and Diffusion in Gases}Cambridge Mathematical Library (Cambridge University Press, 1990).

\bibitem{Kockel:1959auz}
W.~Franz,
\newblock {\em {Max-Planck-Festschrift 1958}} (VEB Deutscher Verlag der Wissenschaften, Berlin, 1959).

\bibitem{israel1963relativistic}
W.~Israel,
\newblock Journal of Mathematical Physics {\bf 4}, 1163 (1963).

\bibitem{cercignani2002relativistic}
C.~Cercignani, G.~M. Kremer, C.~Cercignani, and G.~M. Kremer,
\newblock {\em Relativistic boltzmann equation} (Springer, 2002).

\bibitem{Denicol:2012cn}
G.~S. Denicol, H.~Niemi, E.~Molnar, and D.~H. Rischke,
\newblock Phys. Rev. D {\bf 85}, 114047 (2012), 1202.4551,
\newblock [Erratum: Phys.Rev.D 91, 039902 (2015)].

\bibitem{dyakonov2017spin}
M.~I. Dyakonov and A.~Khaetskii,
\newblock {\em Spin physics in semiconductors} (Springer, 2017).

\bibitem{Yang:2017sdk}
Y.-G. Yang, R.-H. Fang, Q.~Wang, and X.-N. Wang,
\newblock Phys. Rev. C {\bf 97}, 034917 (2018), 1711.06008.

\bibitem{Speranza:2020ilk}
E.~Speranza and N.~Weickgenannt,
\newblock Eur. Phys. J. A {\bf 57}, 155 (2021), 2007.00138.

\bibitem{Liu:2021nyg}
Y.-C. Liu and X.-G. Huang,
\newblock Sci. China Phys. Mech. Astron. {\bf 65}, 272011 (2022), 2109.15301.

\bibitem{Palermo:2023cup}
A.~Palermo and F.~Becattini,
\newblock Eur. Phys. J. Plus {\bf 138}, 547 (2023), 2304.02276.

\bibitem{ANDERSON1974466}
J.~Anderson and H.~Witting,
\newblock Physica {\bf 74}, 466 (1974).

\bibitem{Bjorken:1982qr}
J.~D. Bjorken,
\newblock Phys. Rev. D {\bf 27}, 140 (1983).

\bibitem{BAYM198418}
G.~Baym,
\newblock Physics Letters B {\bf 138}, 18 (1984).

\bibitem{Ebihara:2017suq}
S.~Ebihara, K.~Fukushima, and S.~Pu,
\newblock Phys. Rev. D {\bf 96}, 016016 (2017), 1705.08611.

\bibitem{Molnar:2016vvu}
E.~Molnar, H.~Niemi, and D.~H. Rischke,
\newblock Phys. Rev. D {\bf 93}, 114025 (2016), 1602.00573.

\end{thebibliography}

\end{document}